\documentclass[11pt]{revtex4-2}

\usepackage{graphicx}
\usepackage{dcolumn}
\usepackage{color}
\usepackage{bm}
\usepackage{amsmath}
\usepackage{amsfonts}
\usepackage{amssymb}
\usepackage{array}
\usepackage{multirow}
\begin{document}
	\title{Wormhole generating function in $f(R,T)$ gravity}
	
	\author{Bikram Ghosh$^1$\footnote {bikramghosh13@gmail.com}}
	\author{Saugata Mitra$^1$\footnote {saugatamitra20@gmail.com}}
	
	\affiliation{$^1$Department of Mathematics,
		 Ramakrishna Mission Vidyamandira, Howrah-711202, West Bengal, India}

	
	\begin{abstract} In this present work, we have studied the traversable wormhole geometries in $f(R,T)$ gravity theory, where $R$ denotes the Ricci scalar and $T$ is the trace of the energy-momentum tensor. Firstly, two new shape functions are obtained for some assumed generating function. Also, some new generating functions are obtained in wormhole geometry for some well known shape functions and redshift functions. Energy conditions are examined in each wormhole solution and it is found that a particular type of wormhole satisfies all the energy conditions in a region.
	\end{abstract}

	\maketitle
	
\section{Introduction}  
Wormholes are hypothetical tunnels which connect two different points of the Universe. The term `wormhole' was first used by Misner and Wheeler \cite{MW1}. Einstein and Rosen described the structure of the wormhole mathematically \cite{rER} whereas the study of wormholes became popular after the work of Morris and Thorne \cite{r2}. In $4D$ space-time, the line element of static spherically symmetric wormholes is given by \cite{r2, r1, Y1}
\begin{equation} \label{eq1} ds^2=-e^{2\phi(r)}dt^2+\left[1-\frac{b(r)}{r}\right]^{-1}dr^2+r^2(d\theta^2+sin^2\theta d\phi^2),
\end{equation}
where $\phi(r)$ is the gravitational redshift function and $b(r)$ is the shape function. The radial co-ordinate `$r$' decreases from infinity to a minimum value $r_0$, where $b(r_0)=r_0$ and then again increases from $r_0$ back to infinity. The formation of an event horizon should be avoided for a traversable wormhole \cite{r3}, which are identified as the surfaces with infinite redshift  at the horizon, so that $\phi(r)$ must be finite everywhere for traversability. The shape function $b(r)$ must satisfy the following conditions for obtaining wormhole solutions and for flaring out from the throat \cite{r4,r5,1}:
\begin{eqnarray}\
b(r_0)&=&r_0,\label{eqc2}\\
\frac{b}{r}&\leq&1 ,\label{eqc3}\\
\frac{b-b^\prime r}{b^2}&>&0.\label{eqc4}
\end{eqnarray} 
Also for asymptotically flatness property we must have, $\frac{b(r)}{r}\rightarrow 0$ and $\phi (r)\rightarrow \phi_0 $ (constant) as $r \rightarrow \infty$ \cite{r2, Y1}.

Modified gravity theory is an answer to the accelerated expansion of the Universe and $f(R,T)$ gravity model is one of these modified gravity theories. In $f(R,T)$ gravity theory,  Einstein-Hilbert action is modified by replacing  $R$ (Ricci scalar) by $f(R,T)$, which is an arbitrary function of  R and trace of energy momentum tensor ($T$) \cite{fRT1}. Researchers are analyzing various studies under the context of this gravity theory. In the background of $f(R,T)$ gravity, dynamical  in/stability of celestial compact system has been estimated \cite{fRT2}. Different models of $f(R,T)$ gravity theory are studied with observational constraints in Lyra-Geometry \cite{fRT3}. Godani \cite{fRT4}, determined deceleration and Hubble parameter in terms of redshift and the age of the Universe is also estimated using various supernovae data in the background of this gravity theory. Tiwari et al., \cite{fRT5}, studied deceleration, Hubble and jerk parameters under LRS Bianchi type-I cosmological model in $f(R,T)$ gravity. Bianchi type -III cosmological models are studied in the presence of cosmological constants in this scenario \cite{fRT6}. There are also various topics which are studied like, study of $f(R,T)$ gravity, with the interaction between dark energy and dark matter \cite{fRT7};  with the restriction of conservation of matter \cite{fRT8}; study of energy conditions considering perfect fluid \cite{fRT9}.

Study of wormholes in $f(R,T)$ gravity theory is not new. Charged wormholes are studied in \cite{WH1}, wormhole solutions are obtained by analytical approach in \cite{1}.  Sahoo et al., obtained shape function using the relation $p_r=\omega \rho$, where $p_r$ is the radial pressure and $\rho $ is the energy density; also they have investigated the energy conditions \cite{2}. Mandal et al., studied the geometrical behavior of wormholes under anisotropic and isotropic cases considering the shape function $b(r)=r_0^me^{r_0-r}r^{1-m}$ \cite{3}. In \cite{4}, wormhole solutions are obtained considering different relation between the radial and transverse pressure and in \cite{5} wormhole solutions are obtained considering energy density $\rho$ as a function of Ricci scalar and its derivative with respect to the radial coordinate. Noether symmetry is also applied to obtaining wormhole solutions \cite{6}. The simplest form of $f(R,T)$ is $f(R,T)=R+\lambda T$, which is used in various papers \cite{1, 2, 3, 4, 8, 15, 16, 17, 18, 19}.
\par 
Harko et al. \cite{harko2013} found that this is the extra curvature terms of $f(R)$ gravity which support the wormhole geometries while the matter satisfies all the energy conditions. Capozziello et al. \cite{Cap2012} discussed the possibility for the existence of wormholes in hybrid metric-Palatini gravity by exploring general conditions to violate the null energy condition at the throat. They also studied some particular examples to support their investigation by using redshift function, potential as well as shape functions etc. Many authors have worked on the existence of wormholes and energy conditions in various interesting scenarios \cite{Bej2017}-\cite{Ghashti}. Alvarenga et al. \cite{Alva} tested particular $f(R,T)$ gravity models which satisfy the Energy conditions (which are worked out via the Raychaudhuri equation for expansion) and found stable power-law and
de-Sitter solutions for some values of the input parameters. Also many researchers \cite{you2017a}-\cite{MSetall} demonstrated the inhomogeneity factors of matter density for self-gravitating celestial stars evolving in the background of $f(R,T)$ gravity and imperfect fluid configurations.

\par 
Recently, obtaining generating functions is an important tool to find wormhole solutions \cite{20, 21, 22} in Einstein gravity as well as in modified gravity theory. Herrera et.al., \cite{20} discovered that there are two generating functions which describe all static spherically symmetric anisotropic perfect fluid solutions.  Using this notion Rahaman et. al., \cite{21} found that the generating function associated with redshift function is always positive and decreasing in nature in the context of Einstein gravity. The second generating function plays an important role to check the violation of null energy condition as it relates to matter distribution. The main motivation of this paper is to check the prescription, provided by Rahaman et. al., \cite{21} in the $f(R,T)$ gravity scenario. In this paper, the necessary field equations on $f(R,T)$ have been discussed in section \ref{sec2}. In section \ref{sec3}, the way of obtaining wormhole generating functions has been shown and also wormhole solutions are obtained in this section. Some new generating functions have been presented in section \ref{secG}.  Energy conditions are described and examined in section \ref{sec5}. In section \ref{sec6}, wormhole embedding diagrams have been studied. The paper ends with a brief discussion in section \ref{sec7}.

\section{Field equations on $f(R,T)$ gravity}\label{sec2}
In $f(R,T)$ gravity theory the action is given by \cite{fRT1,1,4}:
\begin{equation}\label{action}
S=\frac{1}{16\pi} \int \left[f(R,T)+L_m\right]\sqrt{-g}d^4x,
\end{equation}
where $f(R,T)$ is an arbitrary function of Ricci scalar $(R)$ and $T$ is the trace term of the energy momentum function. In equation $(\ref{action})$, $g$ is the metric determinant, $L_m$ is the matter Lagarangian density and $c=1=G$ is considered.

The energy momentum tensor is given by, 
\begin{equation}\label{T}
T_{ij}=-\frac{2}{\sqrt{-g}}\left[\frac{\partial (\sqrt{-g}L_m)}{\partial g^{ij}}-\frac{\partial }{\partial x^k} \frac{\partial (\sqrt{-g}L_m)}{\partial (\frac{\partial g^{ij}}{\partial x^k})} \right].
\end{equation}
Considering $L_m=-P$ ( where $P$ is the total pressure), the energy momentum tensor reduces to \cite{1,5},
\begin{equation}
T_{ij}=(\rho+p_t)u_i u_j +p_tg_{ij}+(p_r-p_t)\xi_i \xi_j,
\end{equation}
where $\xi_i$ is a space like vector which is orthogonal to $u_i$ such that $u^iu_i=-1$, $\xi^i \xi_i=1$; $\rho$ is the energy density, $P=\frac{p_r+2p_t}{3}$ and $p_r$, $p_t$ are radial pressure and transverse pressure, respectively.

By varying the action ($\ref{action}$) with respect to the metric $g_{ij}$ we get \cite{1},
\begin{equation}\label{FE}
f_R R_{\mu \nu} -\frac{1}{2}fg_{\mu \nu} +(g_{\mu \nu} \Box -\nabla_\mu \nabla_\nu)f_R=8\pi T_{\mu \nu} +f_T(T_{\mu \nu }+Pg_{\mu \nu}),
\end{equation}
where $f_R\equiv \frac{\partial f}{\partial R}$, $R_{\mu \nu}$ is the Ricci tensor, $\Box$ is the D'Alembert operator, $\nabla_\mu$ is the covariant derivative and  $f_T\equiv \frac{\partial f}{\partial T}$.
Also covariant derivative of energy momentum tensor reduces to \cite{1} (considering $T=\rho-3P$), 
\begin{equation}
\nabla^\mu T_{\mu \nu}=-\frac{f_T}{f_T+8\pi}\left[(T_{\mu \nu}+Pg_{\mu \nu})\nabla^\mu \ln f_T+\frac{1}{2} g_{\mu \nu} \nabla^{\mu } (\rho -P)\right].
\end{equation}
Considering $f(R,T)=R+2\lambda T$ (where $\lambda$ is constant) \cite{1,2,4}, Einstein tensor $G_{\mu \nu}$ becomes \cite{1},
\begin{equation}\label{ET}
G_{\mu \nu}=8\pi T_{\mu \nu}+2\lambda \left[T_{\mu \nu} +(\rho -P)g_{\mu \nu}\right].
\end{equation}
Hence, the field equations are given by,
\begin{eqnarray}\label{f1}
&&\frac{b^\prime}{r^2}=8\pi \rho+2\lambda \left( 2\rho -\frac{p_r+2p_t}{3}\right) ,  \\\label{f2}
&&\frac{1}{r} \left[\frac{b}{r^2}+2\phi^\prime (\frac{b}{r}-1)\right]=-8\pi p_r+2\lambda \left( \rho -\frac{4p_r+2p_t}{3} \right), \\\label{f3}\nonumber
&&\frac{1}{2r} \left[\frac{1}{r}(\phi^\prime b+b^\prime-\frac{b}{r})+2(\phi^{\prime \prime} -(\phi^{\prime })^2)b-\phi^{\prime}(2-b^\prime)\right]-\phi^{\prime \prime}-(\phi^{\prime })^2\\
&&~~~~~~~~~=-8\pi p_t+2\lambda \left( \rho -\frac{p_r+5p_t}{3} \right). 
\end{eqnarray}
Now solving equations (\ref{f1})--(\ref{f3}) we get,
\begin{eqnarray}\label{rho}
\rho&=&\frac{6B(2\pi+\lambda)+\lambda(2D+C)}{6(5\lambda^2+16\lambda\pi+16\pi^2)},\\\label{rp}
p_r&=&\frac{(6D-12C+3B)\lambda^2+\lambda\pi(8D-44C+12B)-48\pi^2C}{6(5\lambda^2+16\lambda\pi+16\pi^2)(\lambda+4\pi)},\\\label{tp}
p_t&=&\frac{(-9D+3C+3B)\lambda^2+\lambda\pi(4C-40D+12B)-48\pi^2D}{6(5\lambda^2+16\lambda\pi+16\pi^2)(\lambda+4\pi)},
\end{eqnarray}
where
\begin{eqnarray}
	B&=&\frac{b^\prime}{r^2},~
	C=\frac{1}{r} \left[\frac{b}{r^2}+2\phi^\prime (\frac{b}{r}-1)\right]\text{and}\nonumber\\
	D&=&\frac{1}{2r} \left[\frac{1}{r}(\phi^\prime b+b^\prime-\frac{b}{r})+2(\phi^{\prime \prime} -(\phi^{\prime })^2)b-\phi^{\prime}(2-b^\prime)\right]-\phi^{\prime \prime}-(\phi^{\prime })^2.\nonumber
\end{eqnarray}
 
\section{Obtaining Generating functions and a new shape function corresponding these}
\label{sec3}
A mechanism was showed by Herrera et al.\cite{20} to obtain all static spherically symmetric solutions of locally anisotropic fluids. To generate all possible solutions, this mechanism needs two types of functions, known as wormhole generating functions.
Using equation (\ref{rp}) and (\ref{tp}) we obtain the result,
\begin{equation}
2(\lambda+4\pi)[p_r-p_t]=\left(1-\frac{b}{r}\right)\left(\frac{\phi^\prime}{r}-\phi^{\prime\prime}-{\phi^\prime}^2+\frac{1}{r^2}\right)+\frac{1}{2}\left(\frac{b^\prime}{r}-\frac{b}{r^2}\right)\left(\phi^\prime+\frac{1}{r}\right)-\frac{1}{r^2}.
\end{equation}
Now, we define a function $H(r)$ by 
\begin{equation}\label{hr}
H(r)=2(\lambda+4\pi)[p_r-p_t]=\left(1-\frac{b}{r}\right)\left(\frac{\phi^\prime}{r}-\phi^{\prime\prime}-{\phi^\prime}^2+\frac{1}{r^2}\right)+\frac{1}{2}\left(\frac{b^\prime}{r}-\frac{b}{r^2}\right)\left(\phi^\prime+\frac{1}{r}\right)-\frac{1}{r^2}.
\end{equation}
Let us introduce a new variable function $G(r)$ by
\begin{equation}\label{G}
e^{\phi(r)}=\exp\left(\int\left(2G(r)-\frac{1}{r}\right)dr\right).
 \end{equation}
Then we have $\phi^\prime=2G(r)-\frac{1}{r}$ {\it i.e.,} $G(r)=\frac{1}{2}(\phi^\prime+\frac{1}{r})$.
Considering $1-\frac{b}{r}=v(r)$, equation (\ref{hr}) reduces 
\begin{equation}\label{id}
\left(H(r)+\frac{1}{r^2}\right)=2v(r)\left[-G^\prime-2G^2+3\frac{G}{r}-\frac{1}{r^2}\right]+v^\prime(r)G(r).
\end{equation}
After solving the differential equation (\ref{id}) for the variable $v$, we get the solution
\begin{equation}
v(r)=\frac{G^2}{r^6}e^{\int\left(\frac{2}{Gr^2}+4G\right)dr}\times\left[\int\Big\{\frac{r^6}{G^3}\left(H(r)+\frac{1}{r^2}\right)e^{-\int\left(\frac{2}{Gr^2}+4G\right)dr}\Big\}dr+C_1\right],
\end{equation} 
where $C_1$ is an arbitrary constant.
Hence we obtain the $b(r)$ as follows:
\begin{equation}\label{b}
b(r)=r-\frac{G^2}{r^5}e^{\int\left(\frac{2}{Gr^2}+4G\right)dr}\times\left[\int\Big\{\frac{r^6}{G^3}\left(H(r)+\frac{1}{r^2}\right)e^{-\int\left(\frac{2}{Gr^2}+4G\right)dr}\Big\}dr+C_1\right].
\end{equation}
From equation (\ref{b}), it is clear that a shape function can be obtained by choosing two functions $G$ and $H$ provided it satisfies all other conditions to be a shape function.
\par 
Now, we will obtain a new shape function by assuming generating functions
 {$G=1/r$} and $H=-r$.
Putting the values of $G$, $H$ in (\ref{b}) and using the throat condition $b(r_0)=r_0$, we get
\begin{equation}\label{obs}
b(r)=r-\frac{1}{r^6}\left(\frac{1}{17}(r_0^{17}-r^{17})+\frac{1}{14}(r^{14}-r_0^{14})\right)^6,
\end{equation}
and from equation (\ref{G}) we obtain $\phi(r)=\ln r$.
\begin{figure}[htb!]
	\centering
	\begin{minipage}{.45\textwidth}
		\centering
		\includegraphics[width=.9\linewidth]{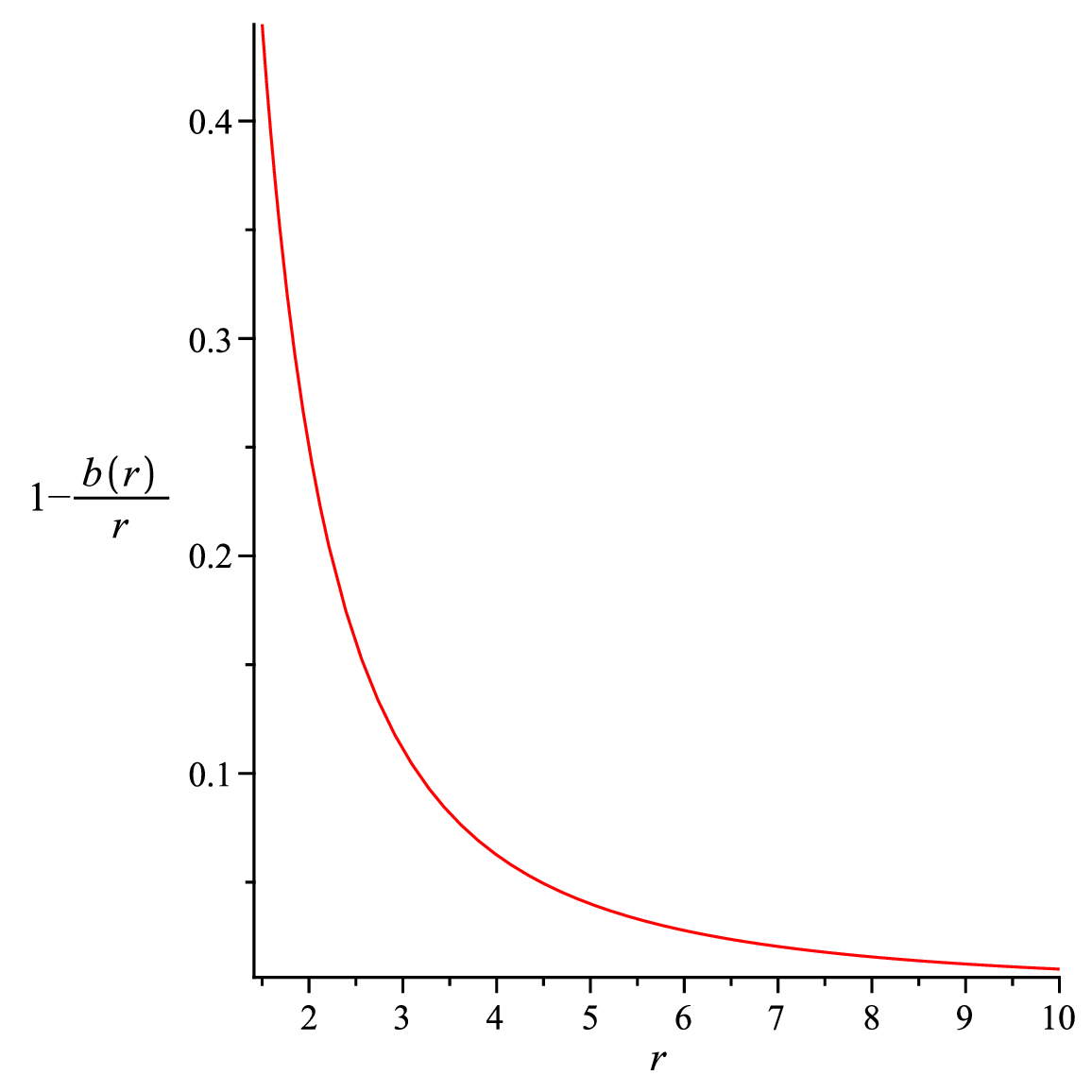}
		\centering (a)
	\end{minipage}
	\begin{minipage}{.45\textwidth}
		\centering
		\includegraphics[width=.9\linewidth]{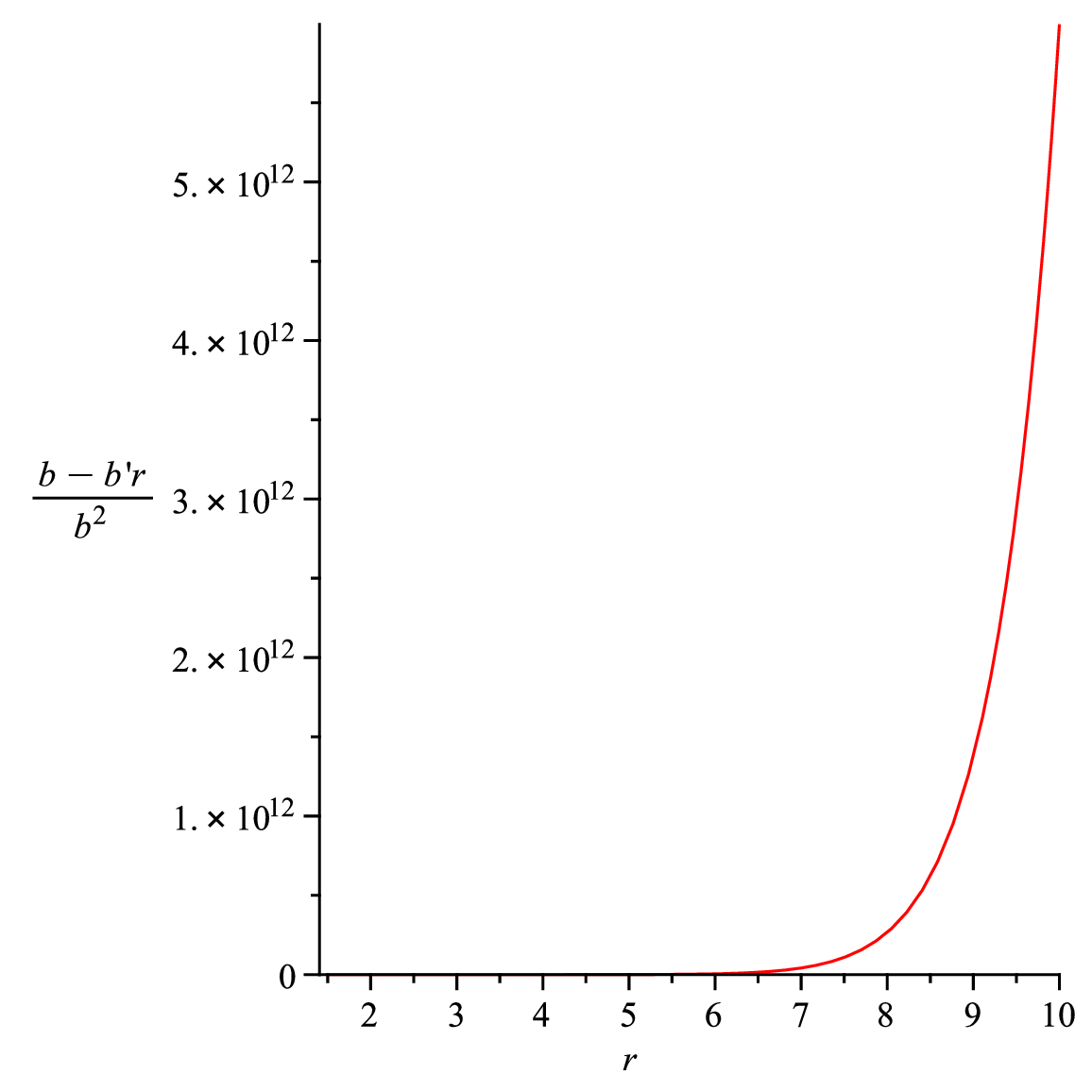}
		\centering (b)
	\end{minipage}
\caption{Behavior of $\left(1-\frac{b(r)}{r}\right)$(a) and $\frac{b-b'r}{b^2}$(b) versus `$r$' for the obtained new shape function (\ref{obs}) with $r_0=1.5$.}\label{figure1}
\end{figure}
\begin{figure}[htb!]
	
	\centering
	\begin{minipage}{.45\textwidth}
		\centering
		\includegraphics[width=.8\linewidth]{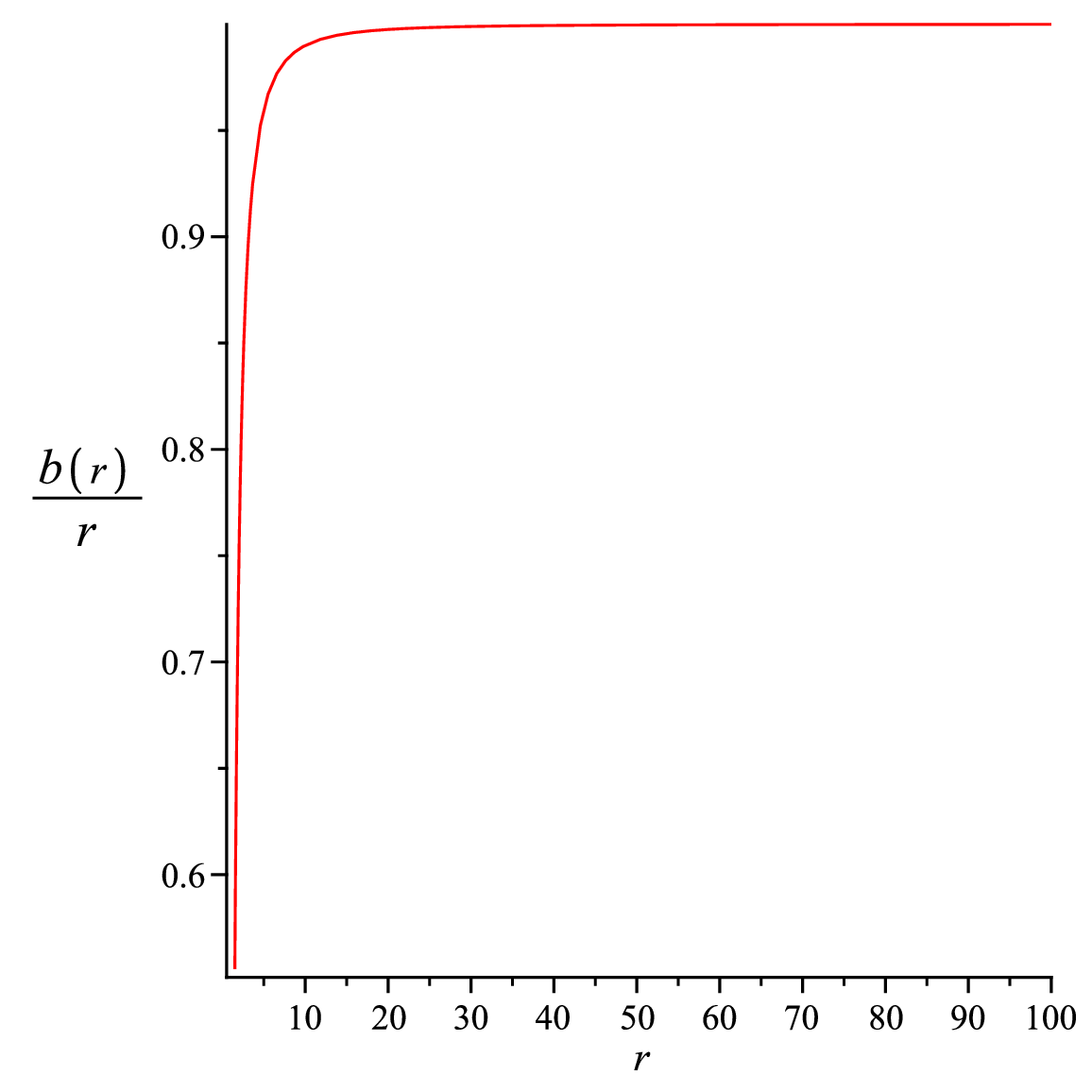}
	\end{minipage}
	\caption{Behavior of $\frac{b(r)}{r}$ versus `$r$' for the obtained new shape function (\ref{obs}) with $r_0=1.5$.}\label{figure2}
\end{figure}
From figure (\ref{figure1}), it is clearly shown that the new shape function obeys  all the required conditions to be a shape function. This type of wormhole is not asymptotically flat (from figure (\ref{figure2})) and we will have to use junction conditions. 
\par
\begin{figure}[htb!]
\centering
\begin{minipage}{.45\textwidth}
	\centering
	\includegraphics[width=.9\linewidth]{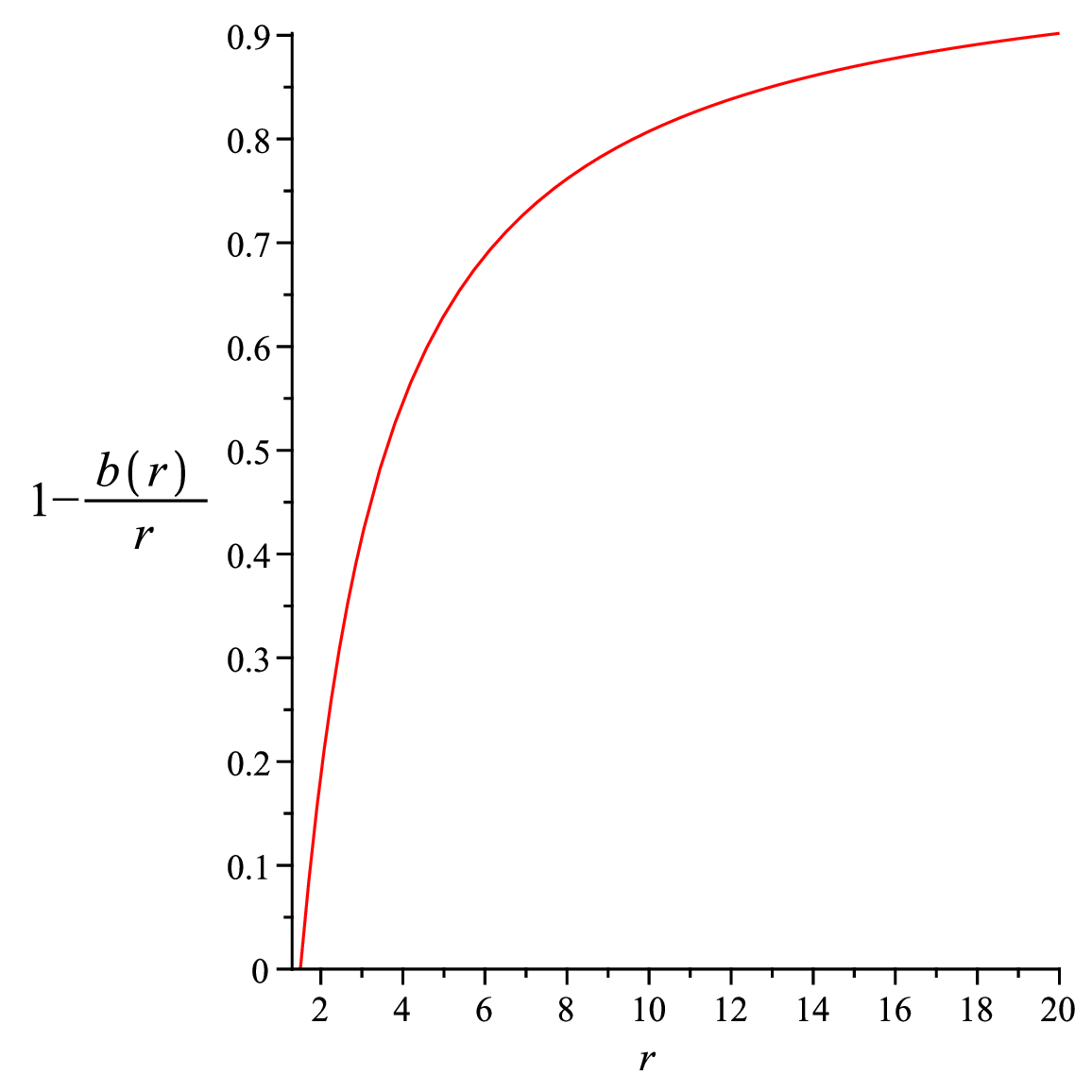}
	\centering (a)
\end{minipage}
\begin{minipage}{.45\textwidth}
	\centering
	\includegraphics[width=.9\linewidth]{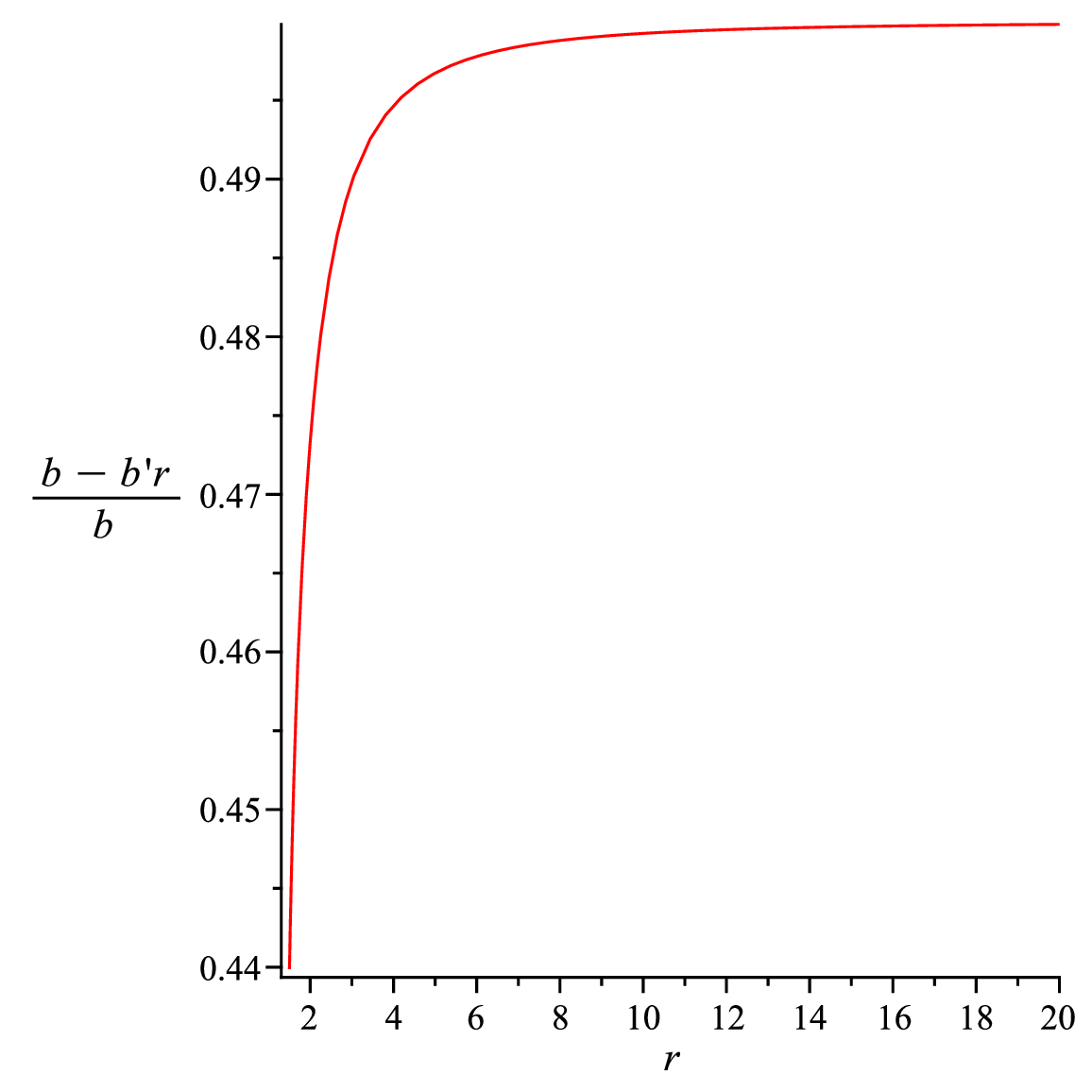}
	\centering (b)
\end{minipage}
\caption{Behavior of
	$\left(1-\frac{b(r)}{r}\right)$(a) and $\frac{b-b'r}{b^2}$(b) versus `$r$' for the obtained new shape function (\ref{obs2}) with $r_0=1.5$.}\label{figure11}
\end{figure}

\begin{figure}[htb!]
	\centering
	\begin{minipage}{.45\textwidth}
		\centering
		\includegraphics[width=.9\linewidth]{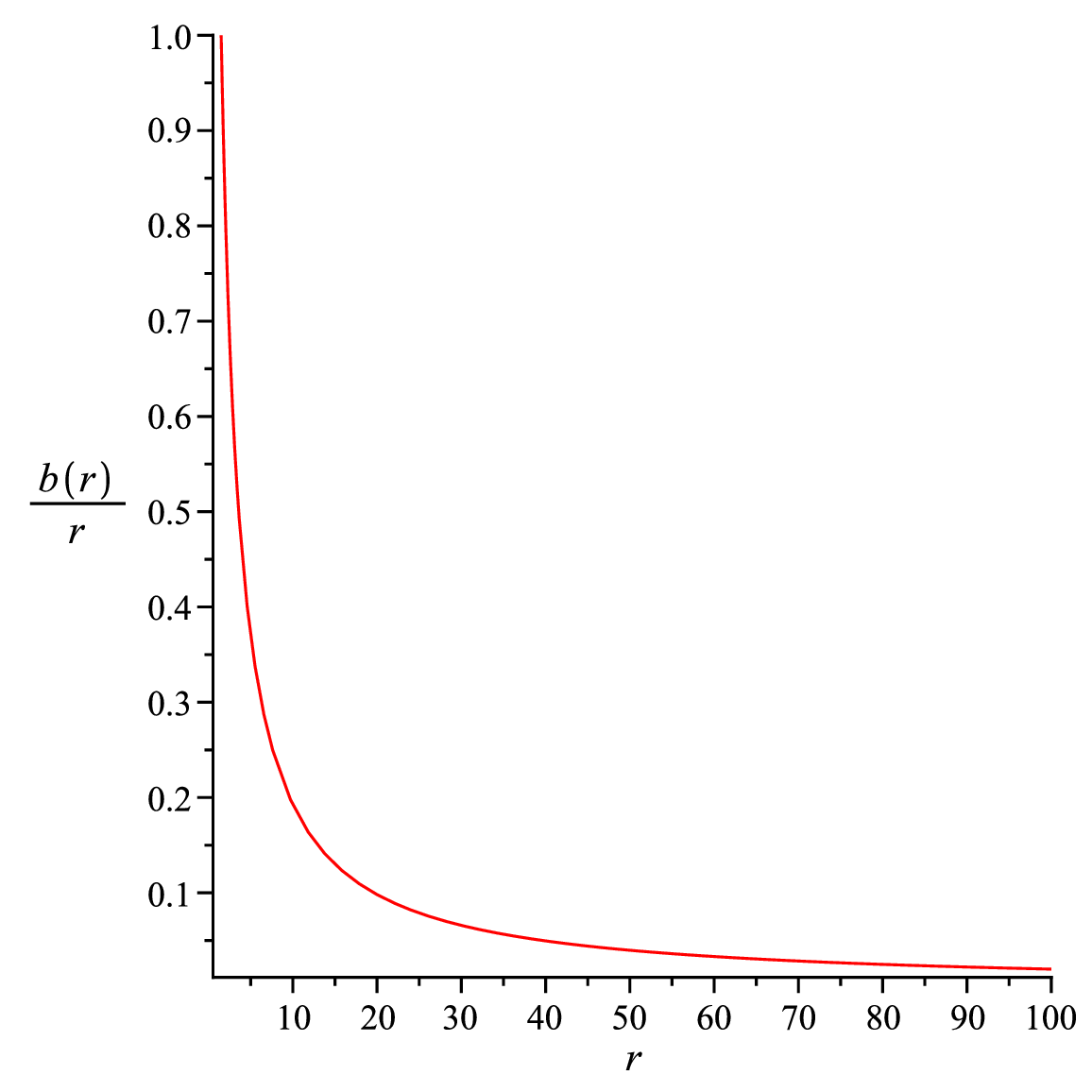}
	\end{minipage}
	\caption{Behavior of $\frac{b(r)}{r}$ versus `$r$' for the obtained new shape function (\ref{obs2}) with $r_0=1.5$.}\label{figure12}
\end{figure}

 
Again, we will obtain another new shape function by assuming generating functions
{$G=\dfrac{1}{2r}$} and $H=-\dfrac{1}{r^3}\left(1+\dfrac{1}{r^{13}}\right)$.
Putting the values of $G$, $H$ in (\ref{b}) and using the throat condition $b(r_0)=r_0$, we get
\begin{equation}\label{obs2}
	b(r)=r-\frac{1}{2r}\left\{2(r^2-r_0^2)-4(r-r_0)+\frac{1}{3}\left(\frac{1}{r^{12}}-\frac{1}{r_0^{12}}\right)\right\}
\end{equation}
and from equation (\ref{G}) we obtain $\phi(r)=\text{Constant}$ (say, $\phi_0$). From figure (\ref{figure11}), it is clearly shown that the new shape function obeys  all the required conditions to be a shape function. Because of finite redshift function and since $\frac{b(r)}{r}\rightarrow 0$ as $r\rightarrow\infty$ (see figure (\ref{figure12})) so this represents an asymptotically flat wormhole.

\begin{figure}[!]
	\centering
	\begin{minipage}{.45\textwidth}
		\centering
		\includegraphics[width=.6\linewidth]{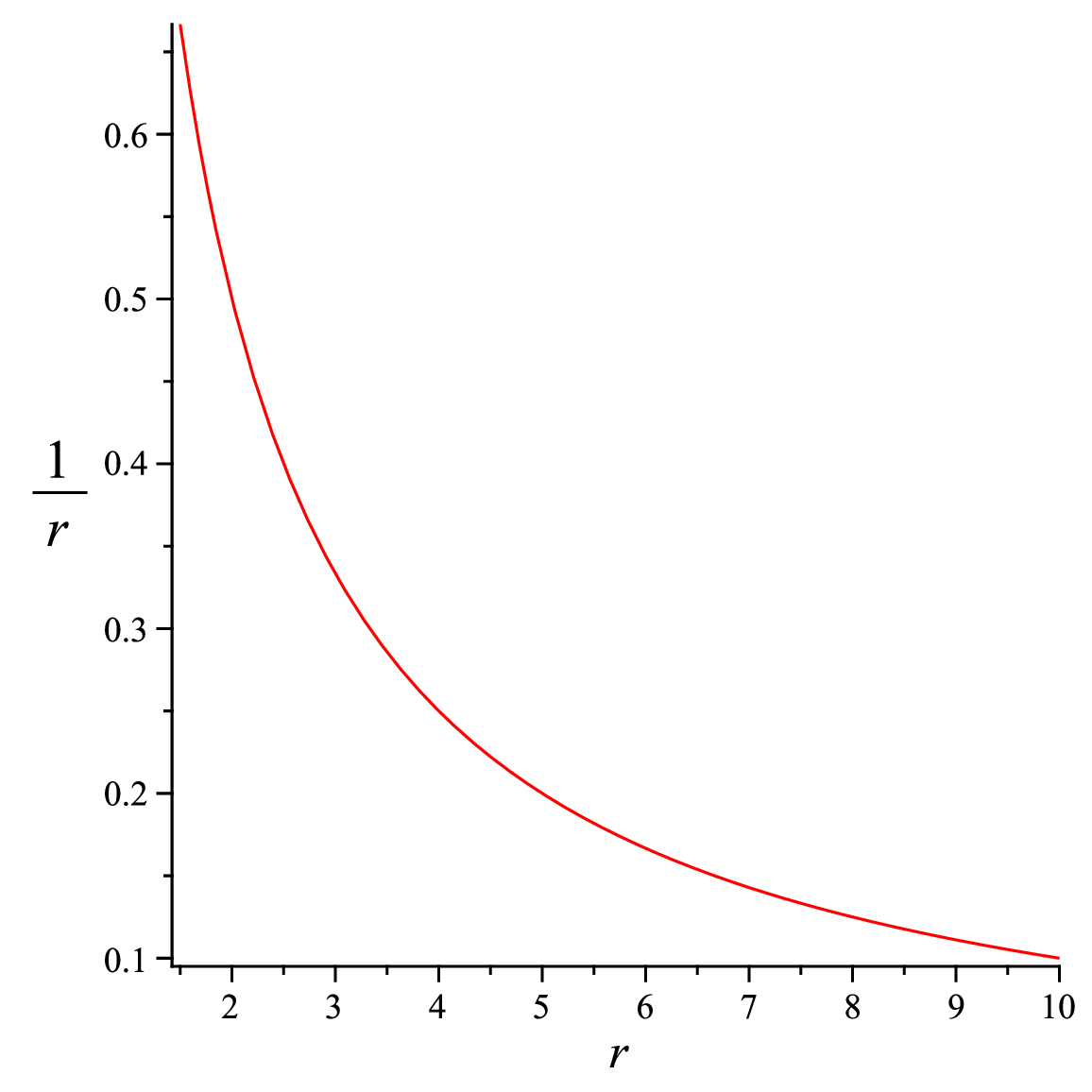}
		\centering (a)
	\end{minipage}
	\begin{minipage}{.45\textwidth}
		\centering
		\includegraphics[width=.6\linewidth]{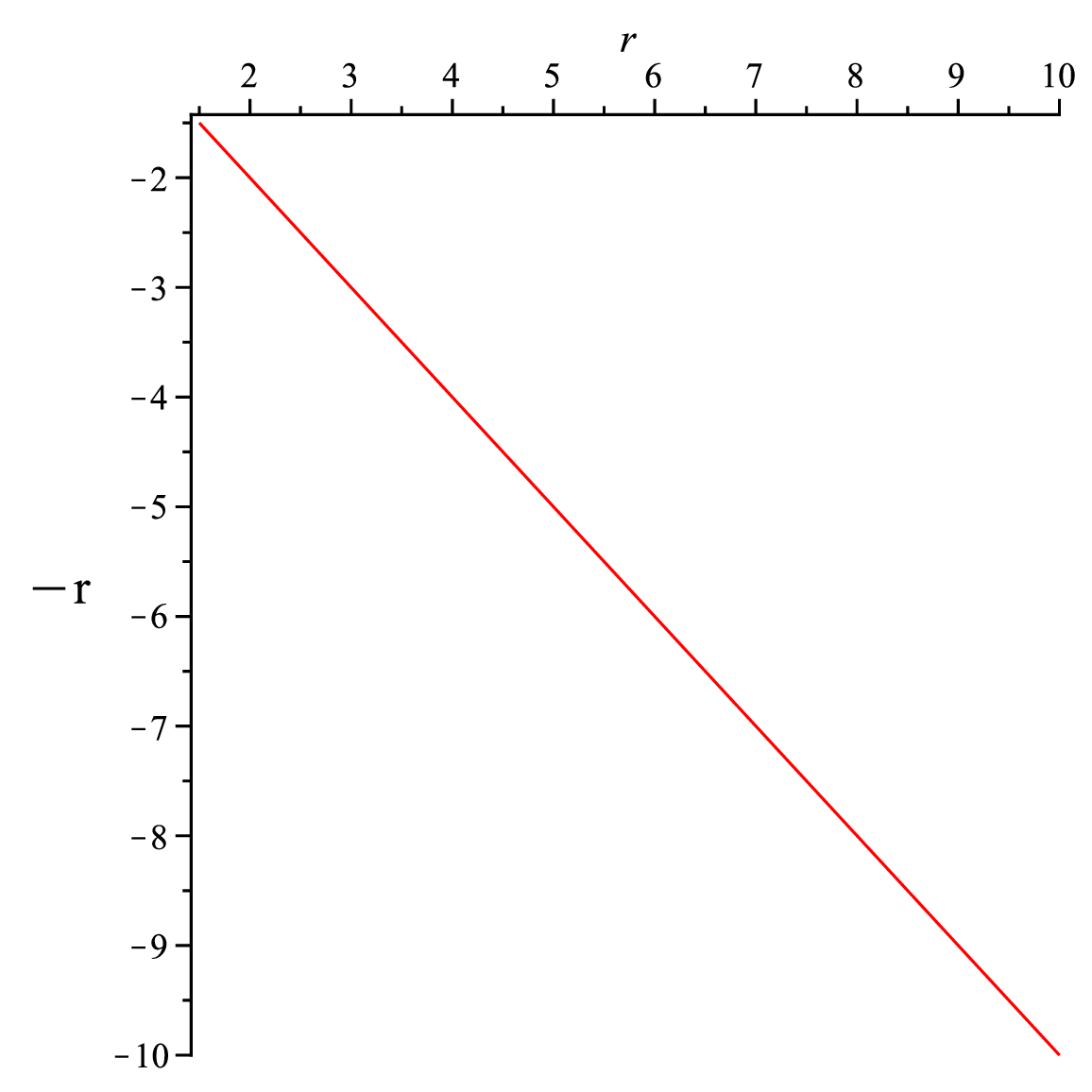}
		\centering (b)
	\end{minipage}
	\centering
	\begin{minipage}{.45\textwidth}
		\centering
		\includegraphics[width=.6\linewidth]{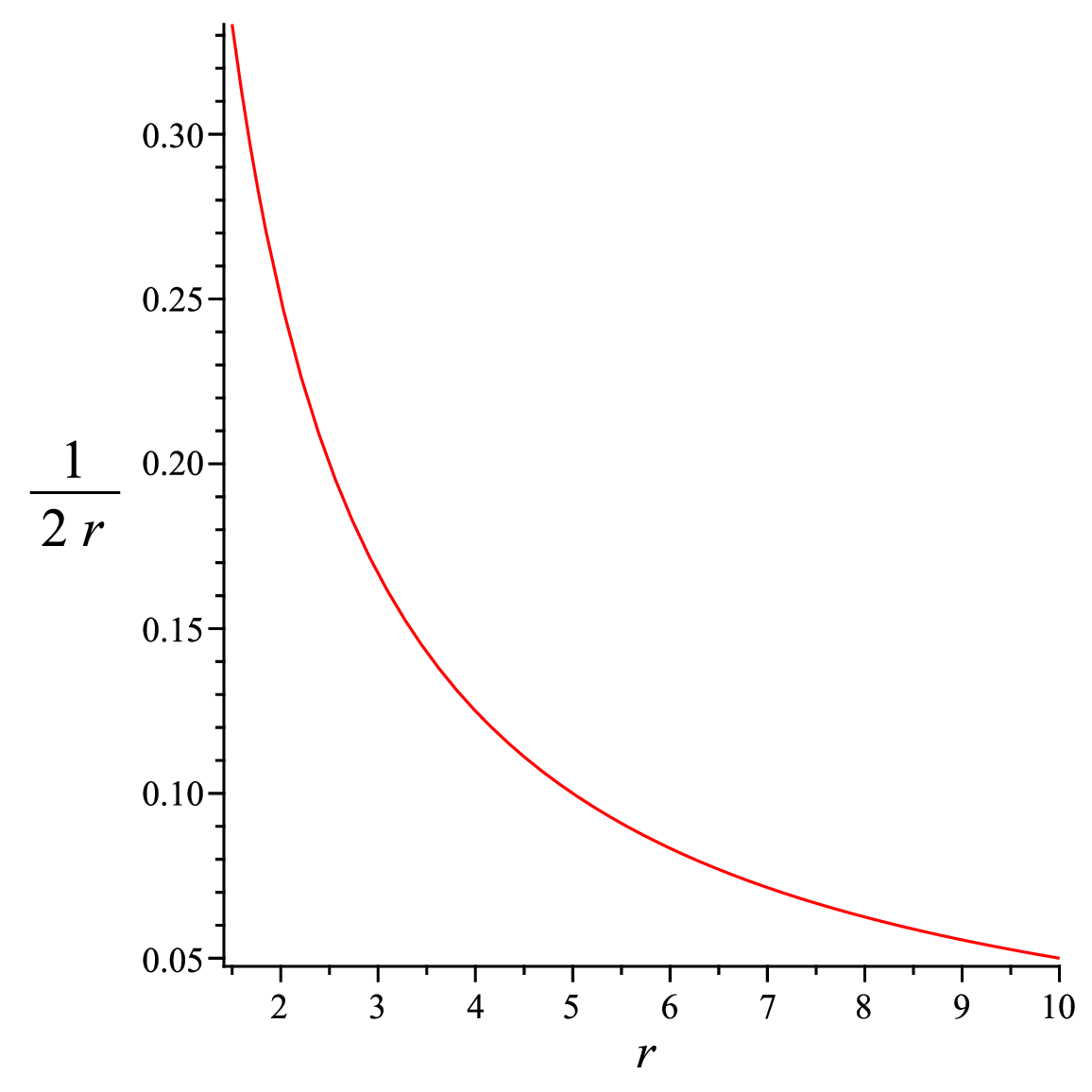}
		\centering (c)
	\end{minipage}
	\begin{minipage}{.45\textwidth}
		\centering
		\includegraphics[width=.7\linewidth]{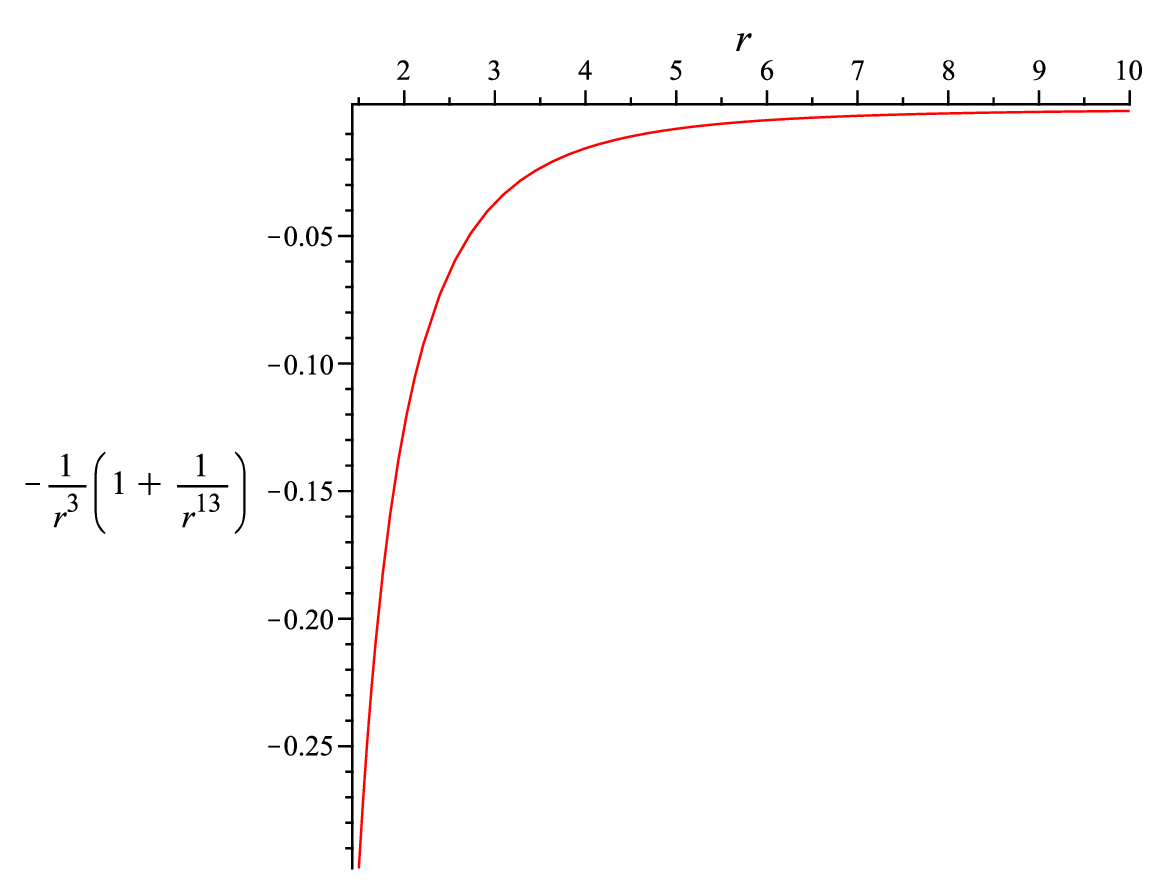}
		\centering (d)
	\end{minipage}
	\caption{Behavior of the assumed generating functions $G(r)$((a) and (c)) and $H(r)$((b) and (d)).}
	\label{figgen}
\end{figure}

\section{Generating functions corresponding to known shape and redshift functions}\label{secG}
In this section, we are interested to find some generating functions corresponding to some known redshift and shape functions which are used in different literature.
\subsection{$\phi(r)=j\ln(\frac{r}{r_0})$,  $j$ is an arbitrary real number.}
Here we will consider the redshift function $\phi(r)=j\ln(\frac{r}{r_0})$ \cite{rd1}. Using equation (\ref{G}), we have found the generating function  
\begin{equation}
G(r)=\frac{1}{2r}(1+j).
\end{equation}
\subsection{$\phi(r)=e^{-\frac{r_0}{r}}$}
In this frame, we consider the redshift function $\phi(r)=e^{-\frac{r_0}{r}}$\cite{rp}. This redshift function obeys the characteristics of a wormhole. For this model we obtain the generating function $G(r)$ by 
\begin{equation}
G(r)=\frac{r_0e^{-r_0/r}}{r^2}+\frac{1}{2r}.
\end{equation}
\subsection{$\phi(r)=\ln\sqrt{1+\frac{\gamma^2}{r^2}}$, $\gamma$ is an arbitrary constant}
Here we use the redshift function $\phi(r)=\ln\sqrt{1+\frac{\gamma^2}{r^2}}$ \cite{rp}, for this model we obtain the generating function $G(r)$ by 
\begin{equation}
G(r)=-\frac{1}{2}\left(1+\frac{\gamma^2}{r^2}\right)^{-1}+\frac{1}{2r}.
\end{equation} 
We have summarized the new generating functions $G(r)$ corresponding to the redshift function in the following table (\ref{Table:T1}).
\begin{table}[!htb]
	\centering
\caption{Generating function $G(r)$ corresponding to redshift functions:}
\begin{tabular}{|c|c|}\hline
	{\bfseries Redshift functions $\phi(r)$} & {\bfseries Generating function $G(r)$}
	\\ \hline
	\text{$\phi(r)=j\ln(\frac{r}{r_0})$,  $j$ is an } & \text{$G(r)=\frac{1}{2r}(1+j)$ }      \\  
	\text{ arbitrary real number} & \text{} \\ \hline
	$\phi(r)=e^{-\frac{r_0}{r}}$   & $G(r)=\frac{r_0e^{-r_0/r}}{r^2}+\frac{1}{2r}$
	\\
	\hline $\phi(r)=\ln\sqrt{1+\frac{\gamma^2}{r^2}}$
	& $G(r)=-\frac{1}{2}\left(1+\frac{\gamma^2}{r^2}\right)^{-1}+\frac{1}{2r}$ \\
	$\gamma$ is an arbitrary constant &  
	\\ 
	\hline
	
\end{tabular}
\label{Table:T1}
\end{table}
\par 

To obtain the generating function $H(r)$, we will consider three cases with different shape functions $(1).~  b(r)=r_0e^{1-\frac{r}{r_0}}$, $(2).~ b(r)=r\frac{\ln(r+1)}{\ln(r_0+1)}$ and $(3).~  b(r)=r_0\frac{a^r}{a^r_0}$, $a\in(0, 1)$ for each of the  the above redshift functions (A), (B) and (C), respectively;
$${\bf I}.~ b(r)=r_0e^{1-\frac{r}{r_0}}~ \&~ \phi(r)=j\ln\left(\frac{r}{r_0}\right) $$
Let us consider the shape function $b(r)=r_0e^{1-\frac{r}{r_0}}$\cite{b1} 
and using equation (\ref{hr}) we get,
\begin{eqnarray}\label{h1}
H(r)&=&\frac{1}{r^2}(2j-j^2)+\frac{r_0^2e^{1-\frac{r_0}{r}}}{2r^4}(1+j)-\frac{r_0e^{1-\frac{r_0}{r}}}{2r^3}(3j-2j^2+3).
\end{eqnarray}
  $${\bf II}.~b(r)=r\frac{\ln(r+1)}{\ln(r_0+1)}~\&~ \phi(r)=j\ln\left(\frac{r}{r_0}\right)$$ 
Here, we consider the shape function $b(r)=r\frac{\ln(r+1)}{\ln(r_0+1)}$\cite{b2}. For this shape function, we obtain the generating function $H(r)$ by
\begin{eqnarray}
H(r)&=&\frac{1}{r^2}\left(1-\frac{\ln(r+1)}{\ln(r_0+1)}\right)(2j-j^2+1)+\frac{1}{2r\ln(r+1)\ln(r_0+1)}(1+j)-\frac{1}{r^2}.
\end{eqnarray}
$${\bf III}.~ b(r)=r_0\frac{a^r}{a^r_0}, ~a\in(0, 1)~ \&~ \phi(r)=j\ln\left(\frac{r}{r_0}\right)$$
The form of the shape function $b(r)=r_0\frac{a^r}{a^r_0}$\cite{b3} gives a wormhole solution provided $a\in(0,1)$. For the above shape function we get $H(r)$ as follows:
\begin{eqnarray}
H(r)&=&\frac{j}{r^2}(2-j)-\frac{r_0a^{r-r_0}}{2r^3}(3+5j-2j^2)+\frac{r_0a^{r-r_0}\ln a }{2r^2}(j+1).
\end{eqnarray}

\begin{figure}[!]
	\centering
	\begin{minipage}{.45\textwidth}
		\centering
		\includegraphics[width=.6\linewidth]{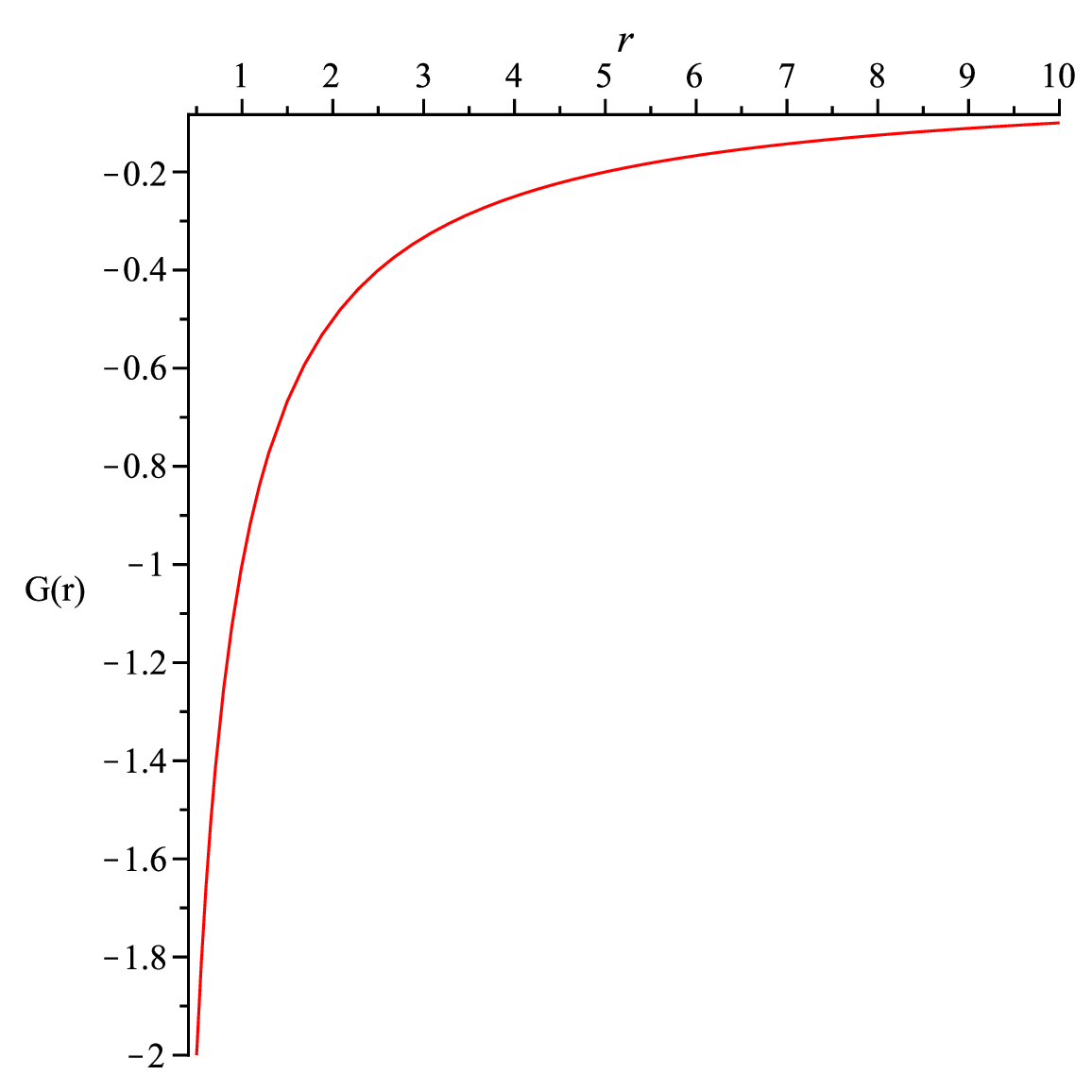}
		\centering (a)
	\end{minipage}
	\begin{minipage}{.45\textwidth}
		\centering
		\includegraphics[width=.6\linewidth]{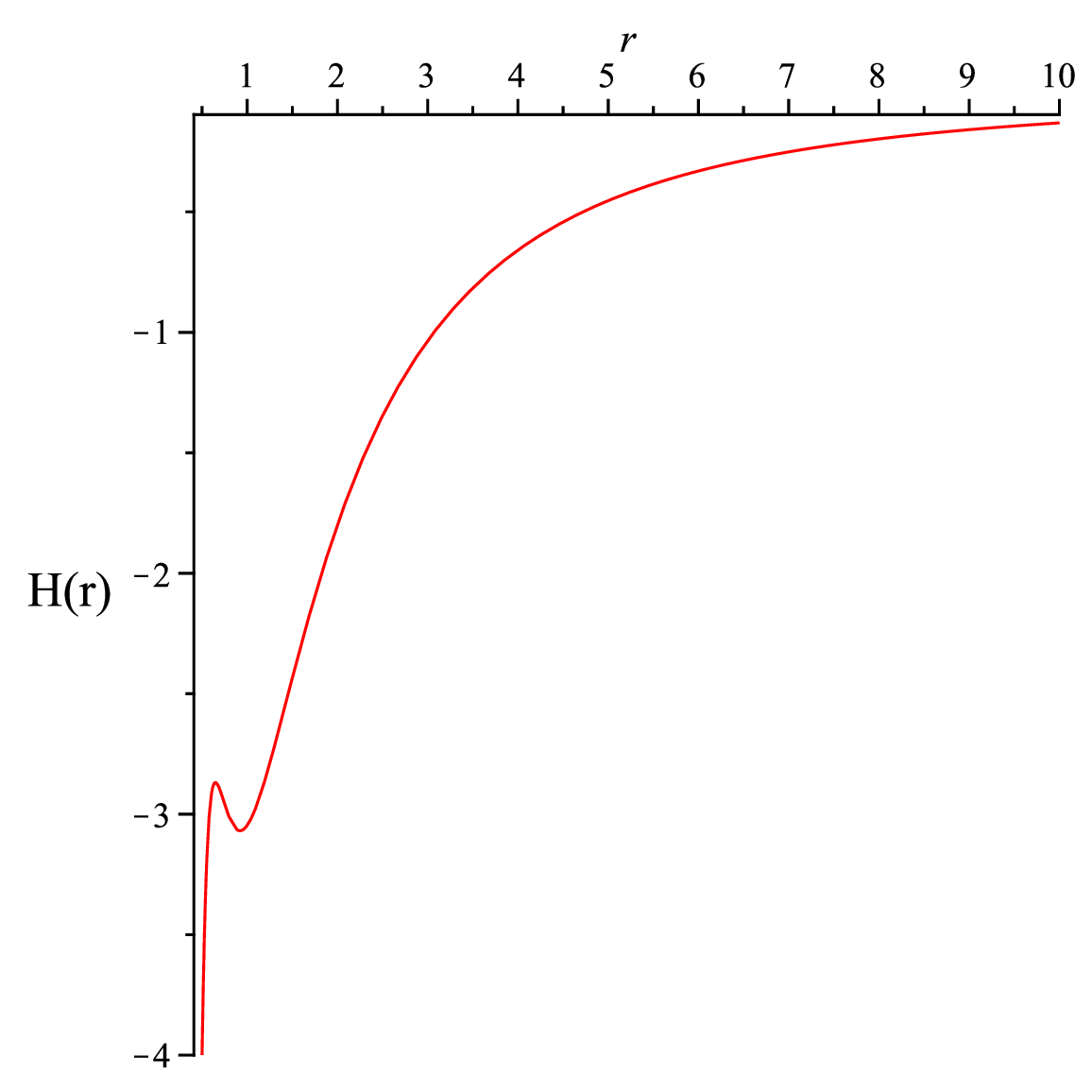}
		\centering (b)
	\end{minipage}
	\centering
	\begin{minipage}{.45\textwidth}
		\centering
		\includegraphics[width=.6\linewidth]{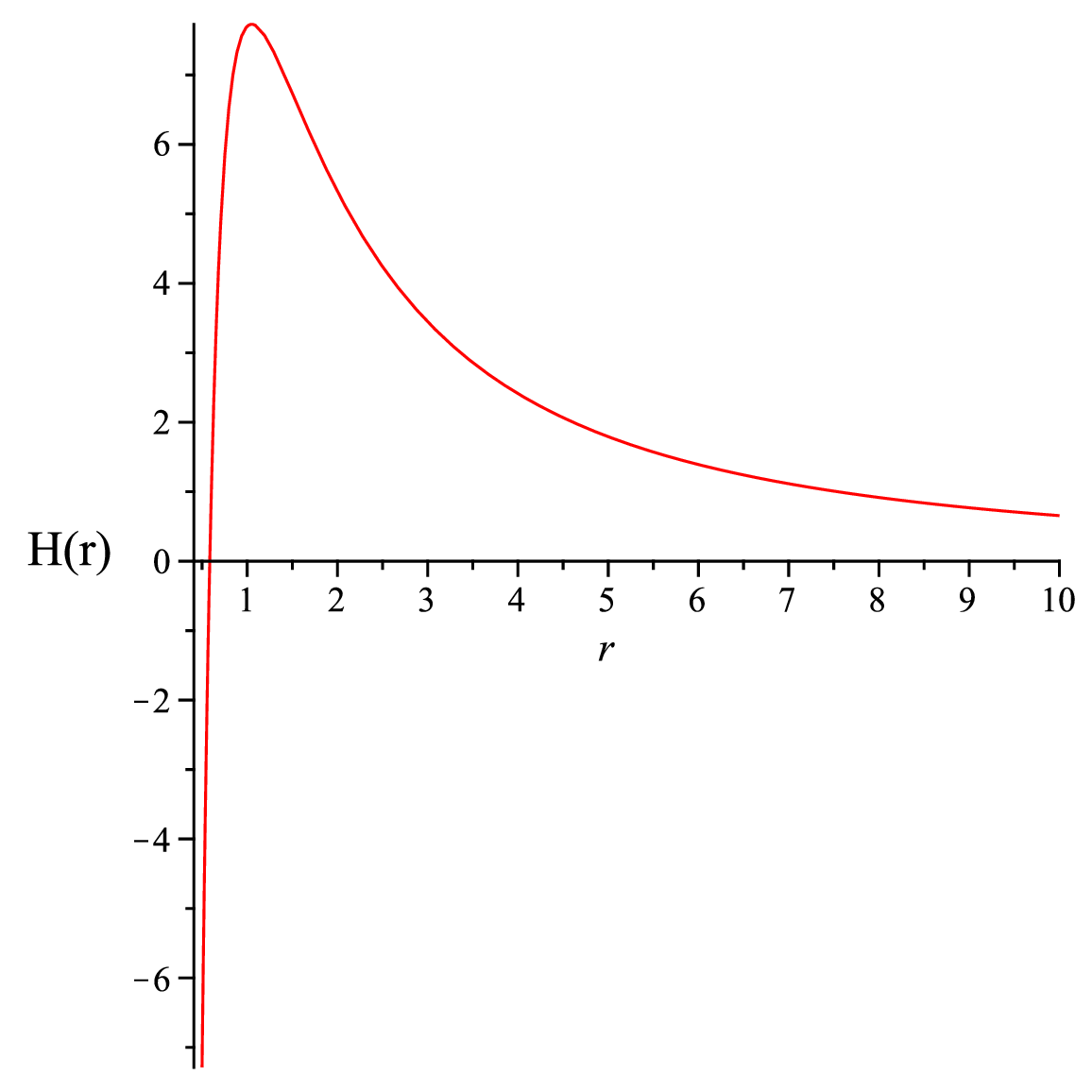}
		\centering (c)
	\end{minipage}
	\begin{minipage}{.45\textwidth}
		\centering
		\includegraphics[width=.6\linewidth]{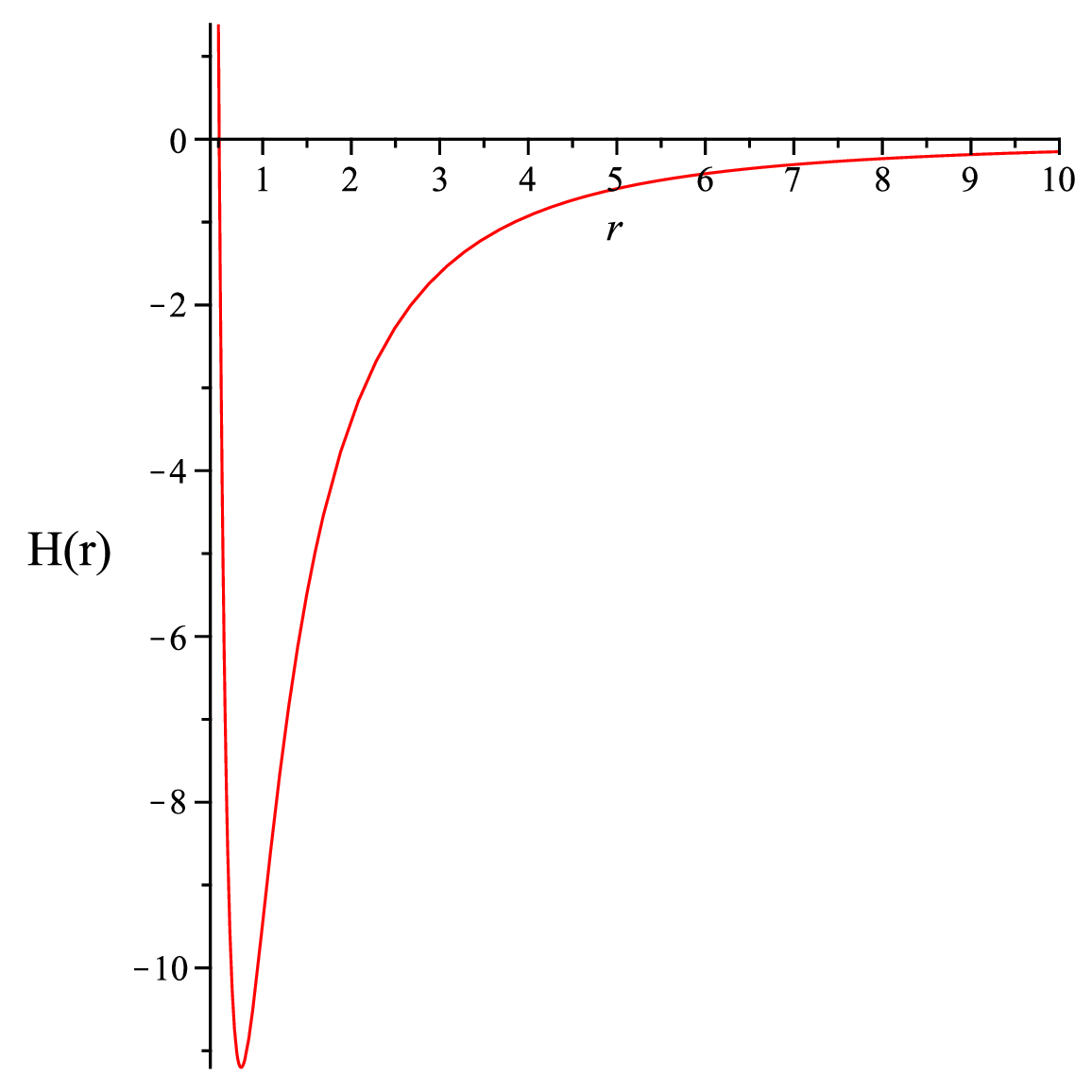}
		\centering (d)
	\end{minipage}
	\caption{Diagram of generating functions $G(r)$(a) for the redshift function $\phi(r)=j\ln(\frac{r}{r_0})$, and $H(r)$((b), (c)  and (d) for  the shape functions 1, 2 and 3 respectively with the same redshift function) when $r_0=0.5$, $j=-3$ and $a=0.5$.}
	\label{fig3}
\end{figure}

 $${\bf IV}.~b(r)=r_0e^{1-\frac{r}{r_0}} ~\&~ \phi(r)=e^{\frac{-r_0}{r}}$$
Using the combinations of $b(r)=r_0e^{1-\frac{r}{r_0}}$ and $\phi(r)=e^{\frac{-r_0}{r}}$, we obtain the generating function $H(r)$ (from equation \ref{hr}) as follows
\begin{eqnarray}\nonumber
H(r)&=&\left(1-\frac{r_0e^{-\frac{r_0}{r}}}{r}\right)\left\{\frac{e^{-\frac{r_0}{r}}}{r^4}(3r_0r-r_0^2)-\frac{r_0^2e^{-\frac{2r_0}{r}}}{r^4}+\frac{1}{r^2}\right\}\\
&~&+\frac{e^{1-\frac{r_0}{r}}}{2r^5}(r_0^2-rr_0)(r_0e^{-\frac{r_0}{r}}+r)-\frac{1}{r^2}.
\end{eqnarray}
$${\bf V}. ~b(r)=r\frac{\ln(r+1)}{\ln(r_0+1)} ~\&~ \phi(r)=e^{\frac{-r_0}{r}}$$
For this $b(r)=r\frac{\ln(r+1)}{\ln(r_0+1)}$, equation (\ref{hr}) gives  
\begin{eqnarray}\nonumber
H(r)&=&\left(1-\frac{\ln(r+1)}{\ln(r_0+1)}\right)\left\{\frac{e^{-\frac{r_0}{r}}}{r^4}(3r_0r-r_0^2)-\frac{r_0^2e^{-\frac{2r_0}{r}}}{r^4}+\frac{1}{r^2}\right\}\\
&&+\frac{1}{2r(r+1)\ln(r+1)}(r_0e^{-\frac{r_0}{r}}+r)-\frac{1}{r^2}.
\end{eqnarray}

$${\bf VI}.~b(r)=r_0\frac{a^r}{a^r_0}, ~a\in(0, 1) ~\&~ \phi(r)=e^{\frac{-r_0}{r}}$$
Here, we consider the form $b(r)=r_0\frac{a^r}{a^r_0}$ and using equation (\ref{hr}) we get
\begin{eqnarray}\nonumber
H(r)&=&\left(1-\frac{r_0a^{r-r_0}}{r}\right)\left\{\frac{e^{-\frac{r_0}{r}}}{r^4}(3r_0r-r_0^2)-\frac{r_0^2e^{-\frac{2r_0}{r}}}{r^4}+\frac{1}{r^2}\right\}\\
&&+\frac{r_0a^{r-r_0}}{2r^4}(r\ln a-1)(r_0e^{-\frac{r_0}{r}}+r)-\frac{1}{r^2}.
\end{eqnarray}
\begin{figure}[!]
	\centering
	\begin{minipage}{.45\textwidth}
		\centering
		\includegraphics[width=.6\linewidth]{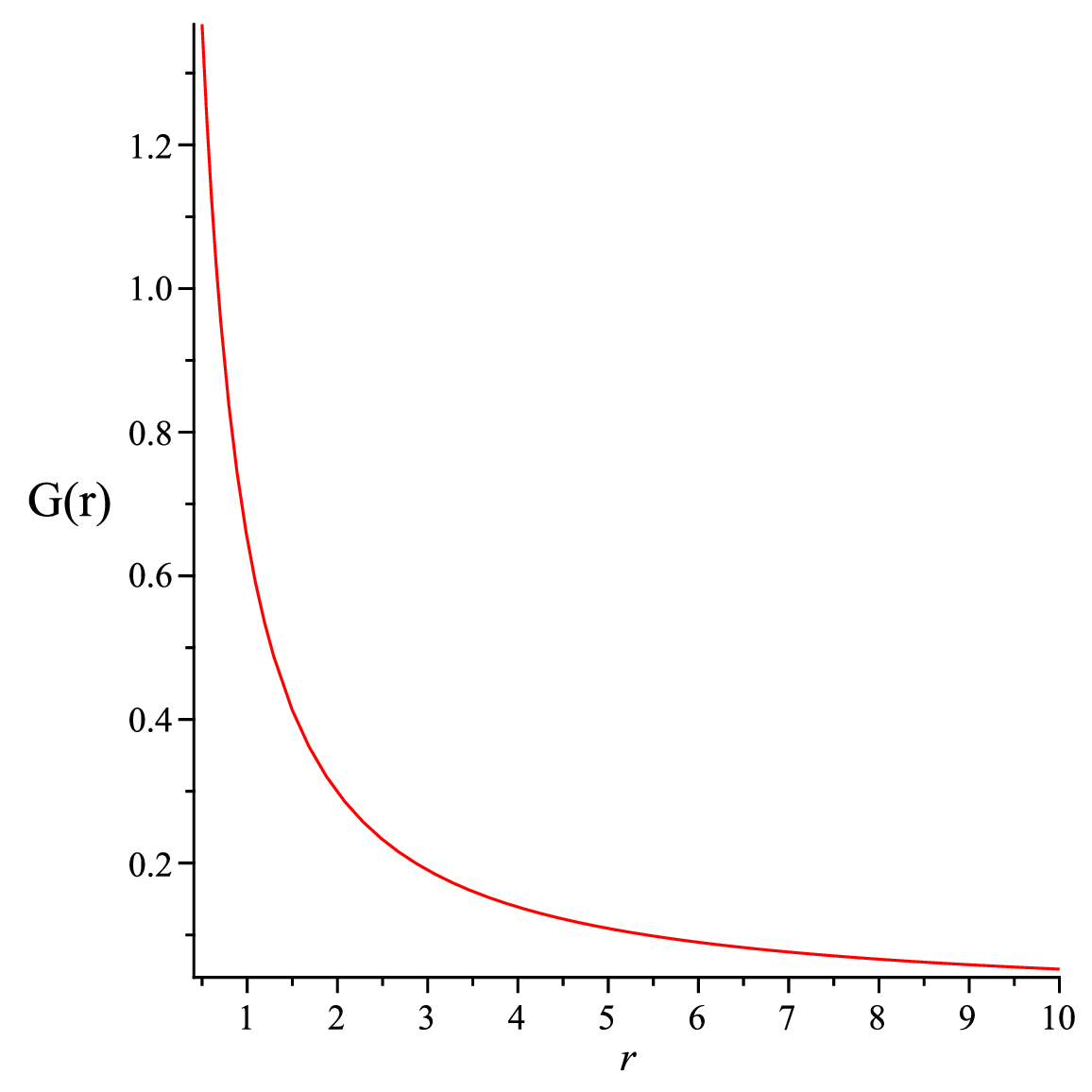}
		\centering (a)
	\end{minipage}
	\begin{minipage}{.45\textwidth}
		\centering
		\includegraphics[width=.6\linewidth]{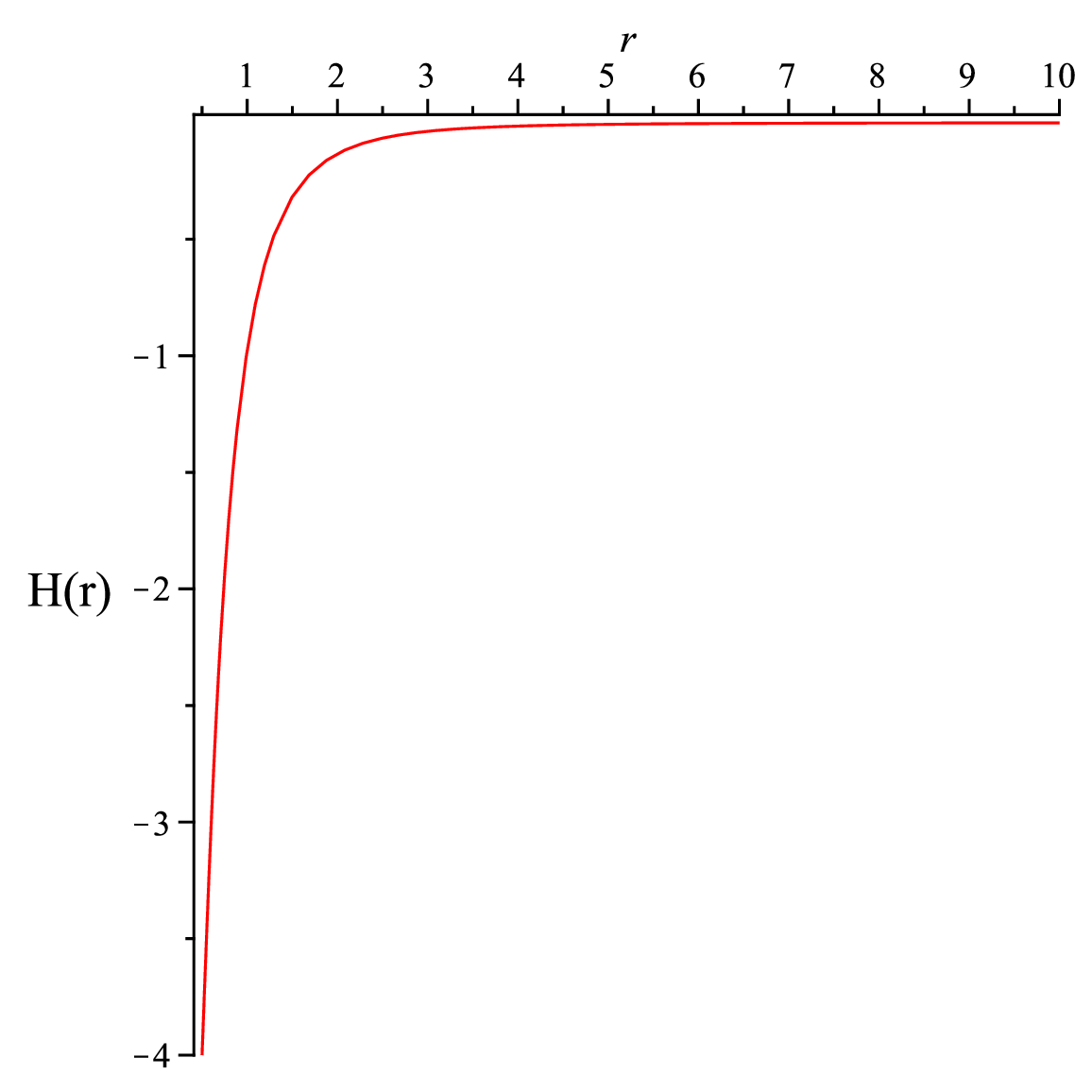}
		\centering (b)
	\end{minipage}
	\centering
	\begin{minipage}{.45\textwidth}
		\centering
		\includegraphics[width=.6\linewidth]{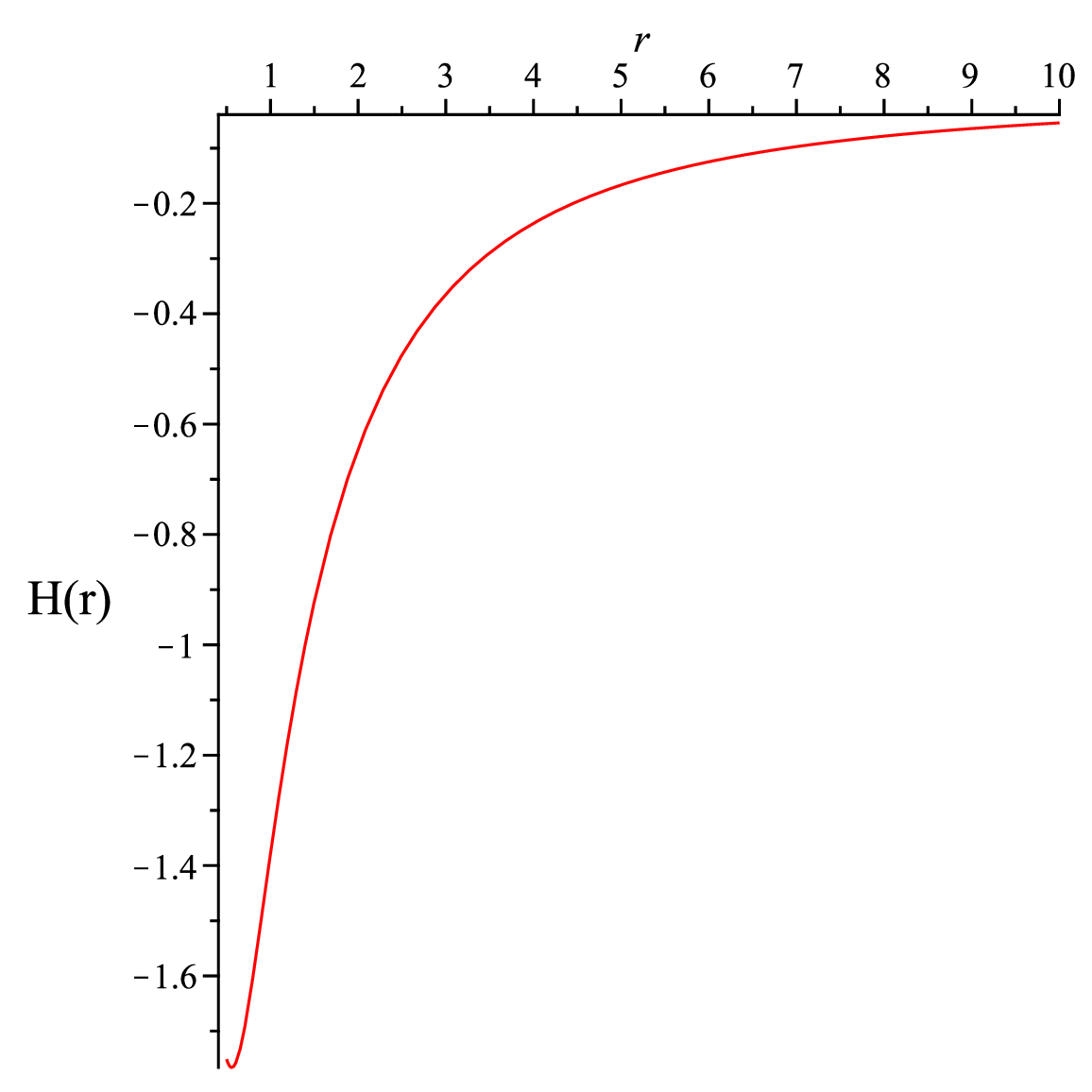}
		\centering (c)
	\end{minipage}
	\begin{minipage}{.45\textwidth}
		\centering
		\includegraphics[width=.6\linewidth]{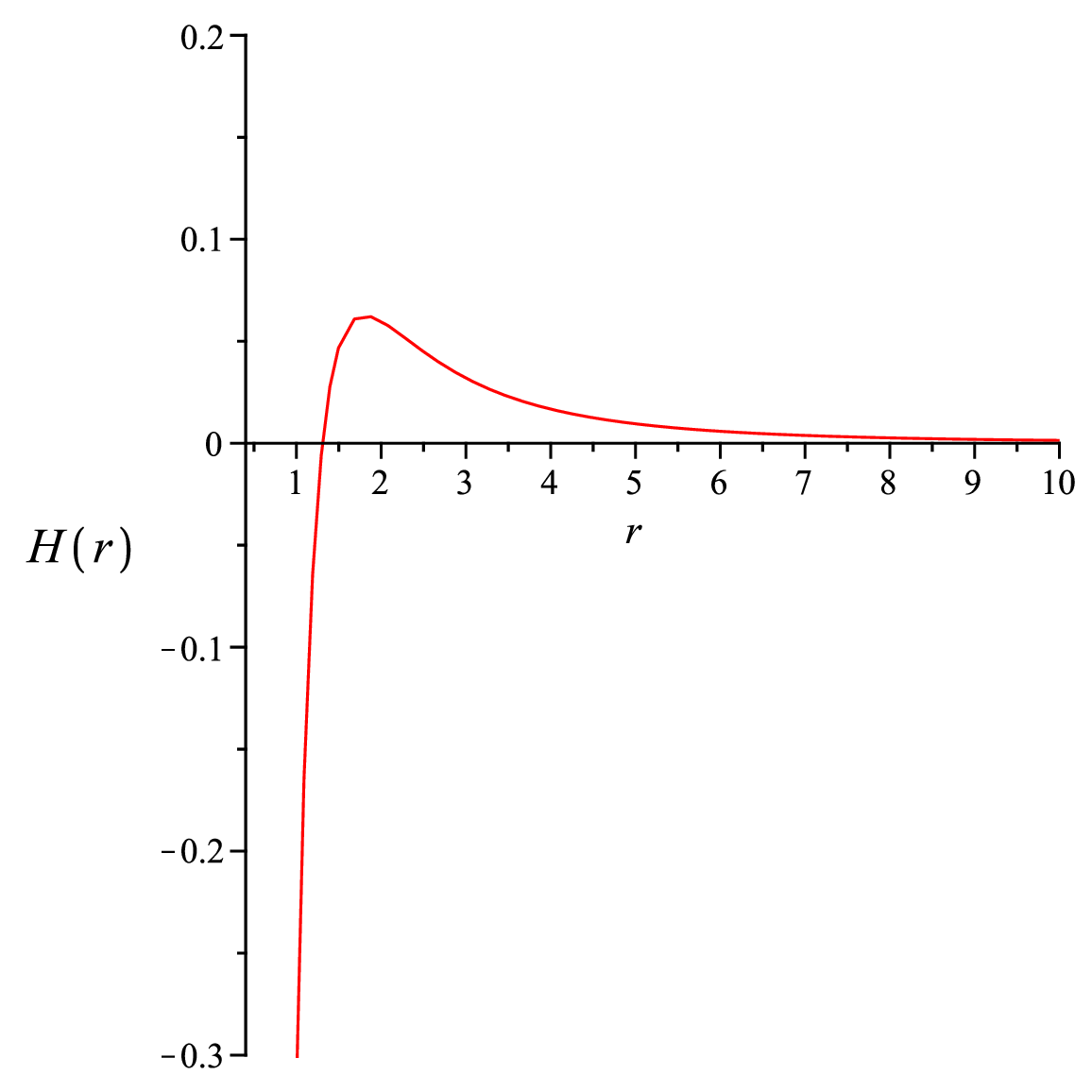}
		\centering (d)
	\end{minipage}
	\caption{Diagram of generating functions $G(r)$(a) for the redshift function $\phi(r)=e^{\frac{-r_0}{r}}$ and $H(r)$((b), (c)  and (d) for  the shape functions 1, 2 and 3 respectively with the same redshift function) when $r_0=0.5$, $a=0.5$}
	\label{fig4}
\end{figure}
$${\bf VII}. ~b(r)=r_0e^{1-\frac{r}{r_0}} ~\&~ \phi(r)=\ln\sqrt{1+\frac{\gamma^2}{r^2}}$$
The above consideration of $b(r)$ and $\phi(r)$ gives 
\begin{eqnarray}\nonumber
H(r)&=&\left(1-\frac{r_0e^{1-\frac{r_0}{r}}}{r}\right)\left\{-\frac{4\gamma^2}{r^4}\left(1+\frac{\gamma^2}{r^2}\right)^{-1}+\frac{\gamma^2}{r^6}\left(1+\frac{\gamma^2}{r^2}\right)^{-2}+\frac{1}{r^2}\right\}\\
&&+\frac{e^{1-\frac{r_0}{r}}(r_0^2-r_0r)}{r^3}\left\{-\frac{\gamma^2}{2r^3}\left(1+\frac{\gamma^2}{r^2}\right)^{-1}+\frac{1}{2r}\right\}-\frac{1}{r^2}.
\end{eqnarray}
$${\bf VIII}.~ b(r)=r\frac{\ln(r+1)}{\ln(r_0+1)}~\&~ \phi(r)=\ln\sqrt{1+\frac{\gamma^2}{r^2}}$$
Considering $ b(r)=r\frac{\ln(r+1)}{\ln(r_0+1)}$ and $\phi(r)=\ln\sqrt{1+\frac{\gamma^2}{r^2}}$, from equation (\ref{hr}) we get
\begin{eqnarray}\nonumber
H(r)&=&\left(1-\frac{\ln(r+1)}{\ln(r_0+1)}\right)\left\{-\frac{4\gamma^2}{r^4}\left(1+\frac{\gamma^2}{r^2}\right)^{-1}+\frac{\gamma^2}{r^6}\left(1+\frac{\gamma^2}{r^2}\right)^{-2}+\frac{1}{r^2}\right\}\\
&&+\frac{1}{(r+1)\ln(r_0+1)}\left\{-\frac{\gamma^2}{2r^3}\left(1+\frac{\gamma^2}{r^2}\right)^{-1}+\frac{1}{2r}\right\}-\frac{1}{r^2}.
\end{eqnarray}
$${\bf IX}. ~b(r)=r_0\frac{a^r}{a^r_0},~ a\in(0, 1)~ \& ~\phi(r)=\ln\sqrt{1+\frac{\gamma^2}{r^2}}$$
Now, we will consider the pair $b(r)=r_0\frac{a^r}{a^r_0}$ and $\phi(r)=\ln\sqrt{1+\frac{\gamma^2}{r^2}}$. Considering these, from equation (\ref{hr}) we get 
\begin{eqnarray}\nonumber
H(r)&=&\left(1-\frac{r_0a^{r-r_0}}{r}\right)\left\{-\frac{4\gamma^2}{r^4}\left(1+\frac{\gamma^2}{r^2}\right)^{-1}+\frac{\gamma^2}{r^6}\left(1+\frac{\gamma^2}{r^2}\right)^{-2}+\frac{1}{r^2}\right\}\\
&&+\frac{r_0a^{r-r_0}(rr_0\ln(a)-1
	)}{r^2}\left\{-\frac{\gamma^2}{2r^3}\left(1+\frac{\gamma^2}{r^2}\right)^{-1}+\frac{1}{2r}\right\}-\frac{1}{r^2}.
\end{eqnarray}
\begin{figure}[!]
	\centering
	\begin{minipage}{.45\textwidth}
		\centering
		\includegraphics[width=.6\linewidth]{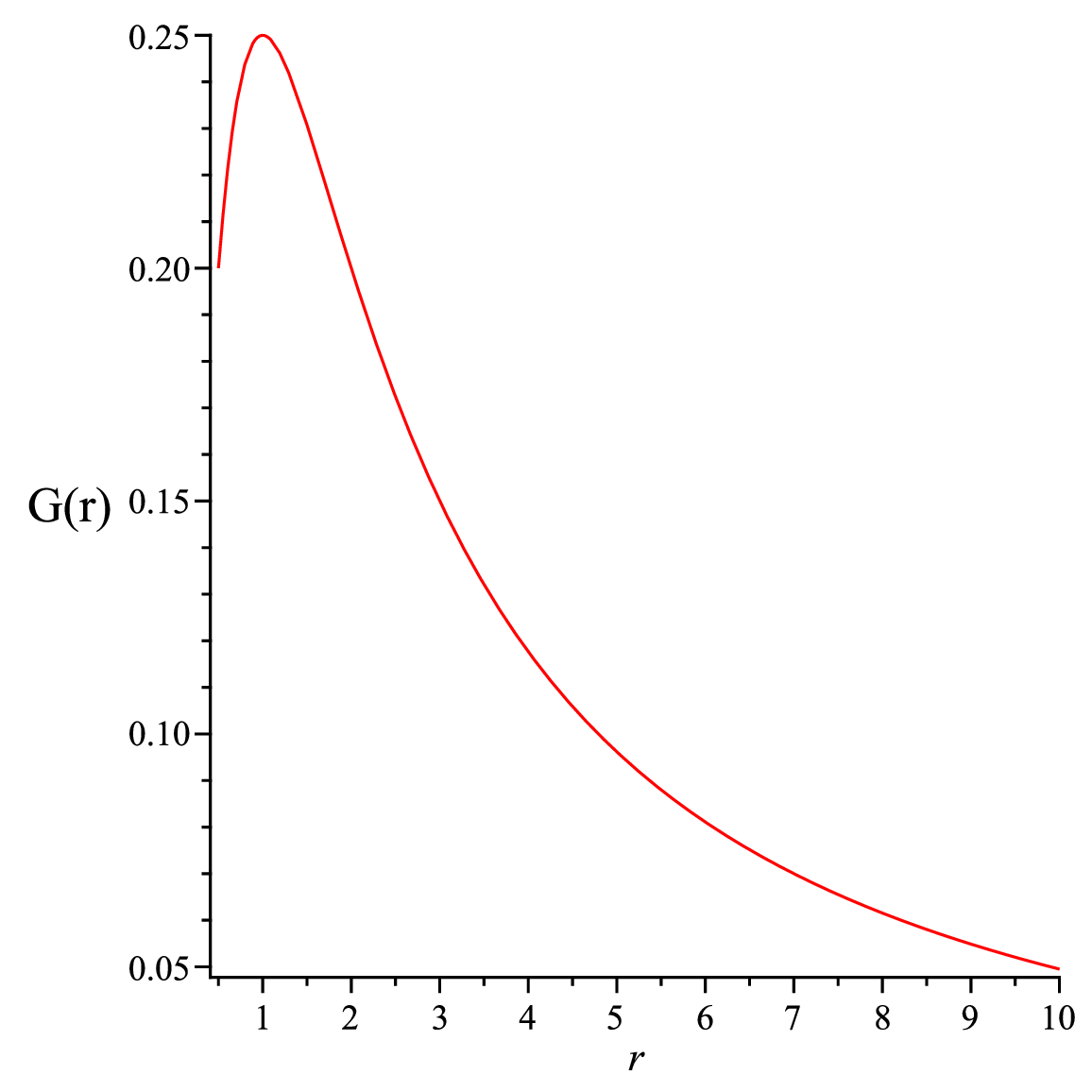}
		\centering (a)
	\end{minipage}
	\begin{minipage}{.45\textwidth}
		\centering
		\includegraphics[width=.6\linewidth]{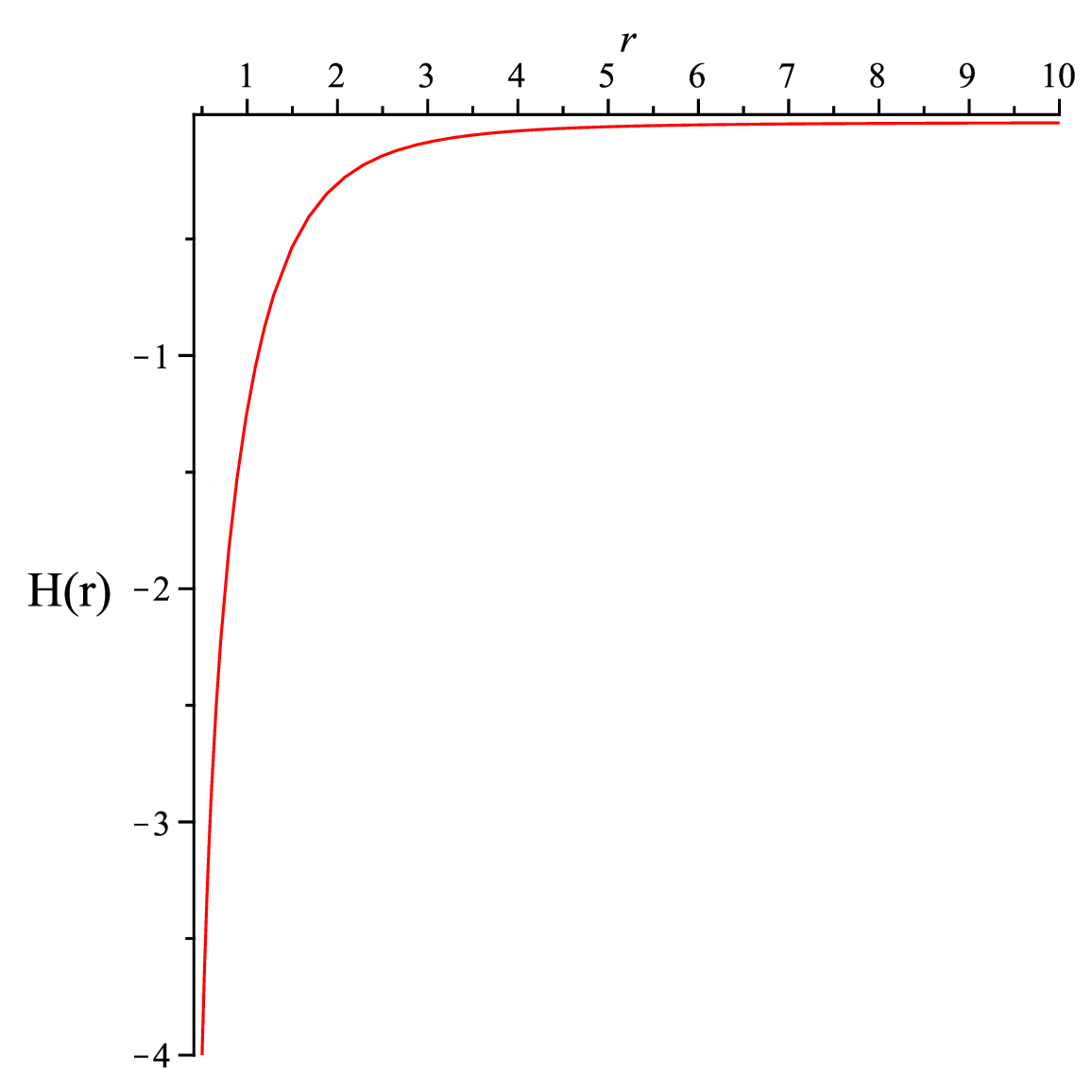}
		\centering (b)
	\end{minipage}
	\centering
	\begin{minipage}{.45\textwidth}
		\centering
		\includegraphics[width=.6\linewidth]{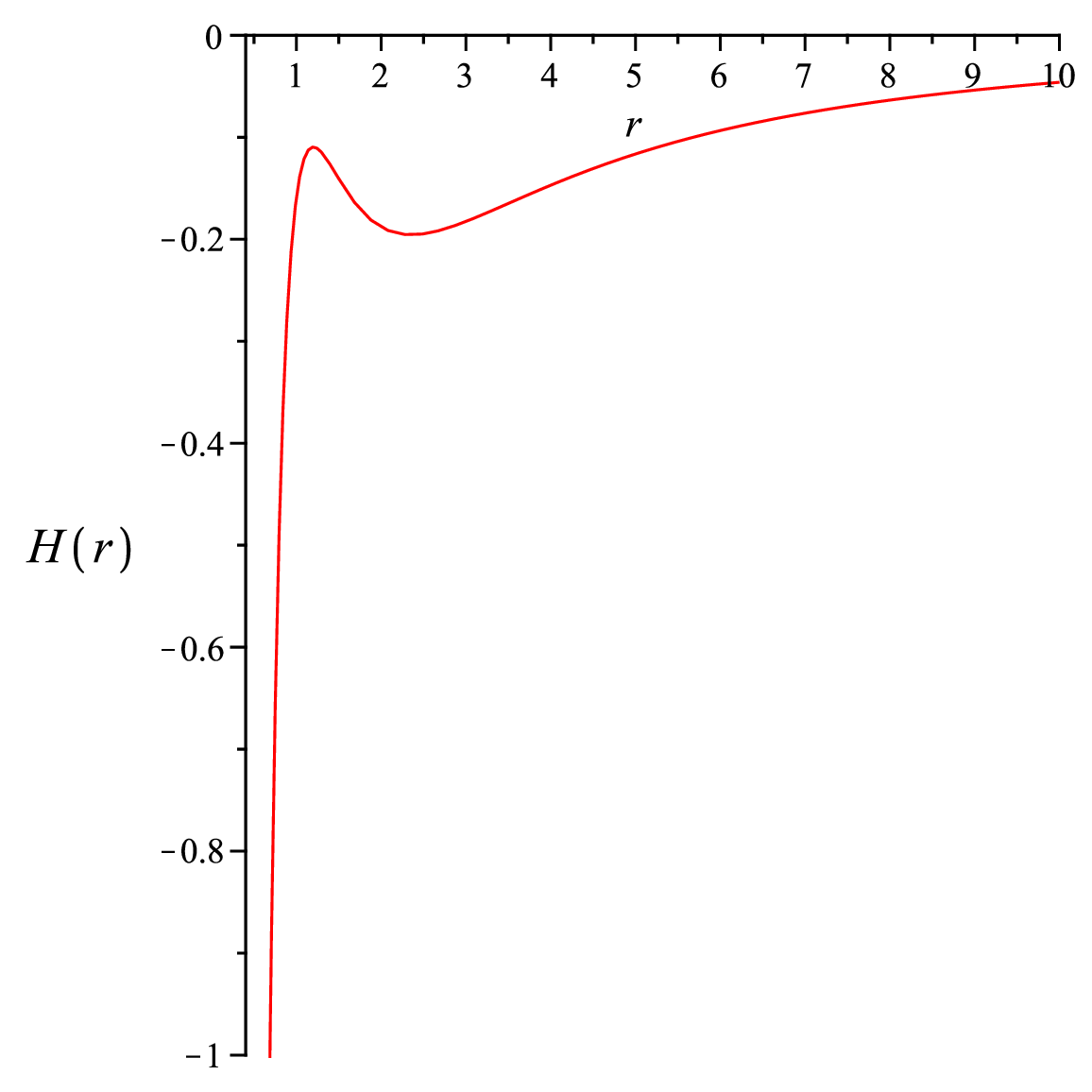}
		\centering (c)
	\end{minipage}
	\begin{minipage}{.45\textwidth}
		\centering
		\includegraphics[width=.6\linewidth]{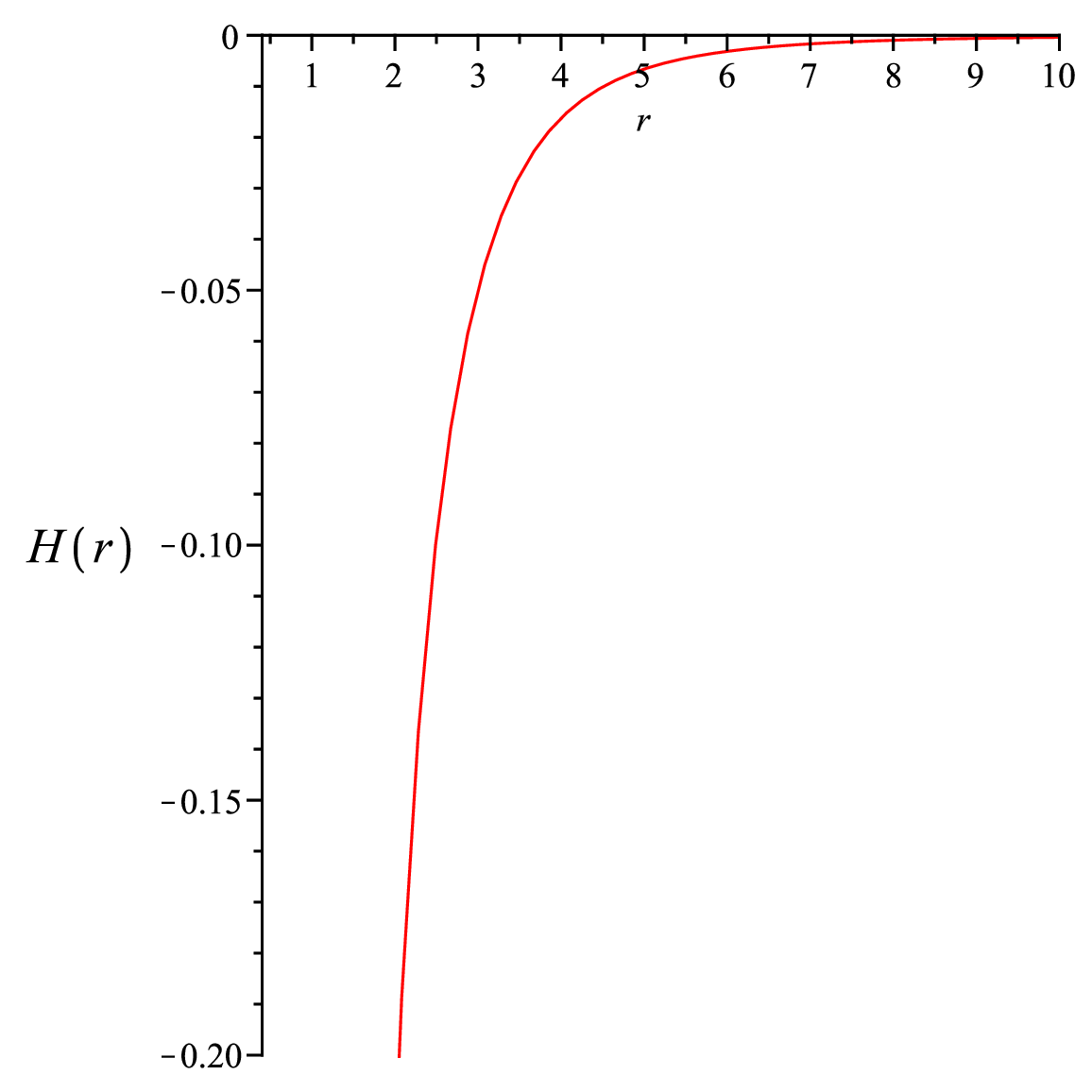}
		\centering (d)
	\end{minipage}
	\caption{Diagram of generating functions $G(r)$(a) for the redshift function $\phi(r)=\ln\sqrt{1+\frac{\gamma^2}{r^2}}$ and $H(r)$((b), (c)  and (d) for  the shape functions 1, 2 and 3 respectively with the same redshift function) when $r_0=0.5$, $\gamma=1$ and $a=0.5$}
	\label{fig5}
\end{figure}
\begin{figure}
\begin{minipage}{.35\textwidth}
	\includegraphics[width=.6\linewidth]{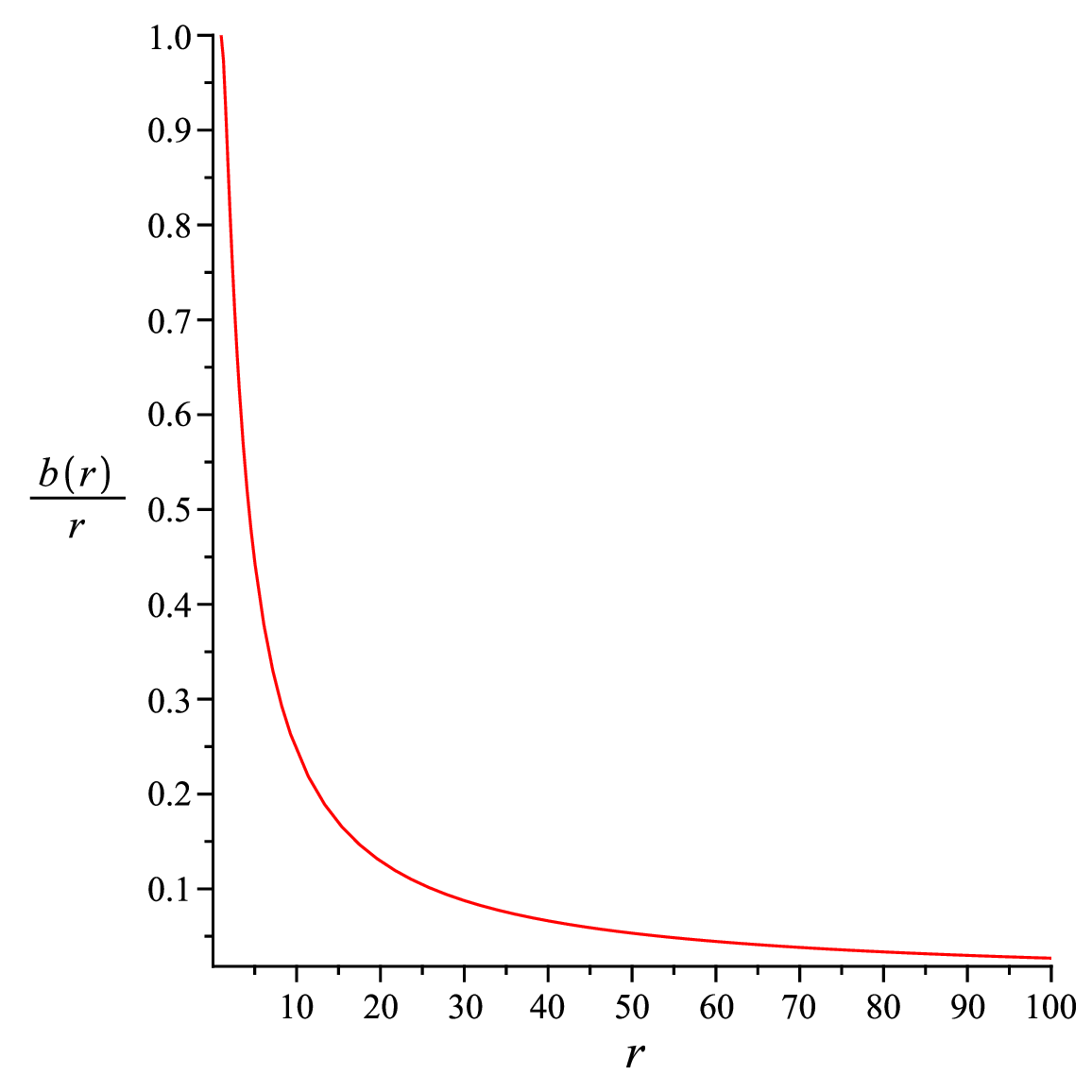}
	(a)
\end{minipage}
\begin{minipage}{.35\textwidth}
	\centering
	\includegraphics[width=.6\linewidth]{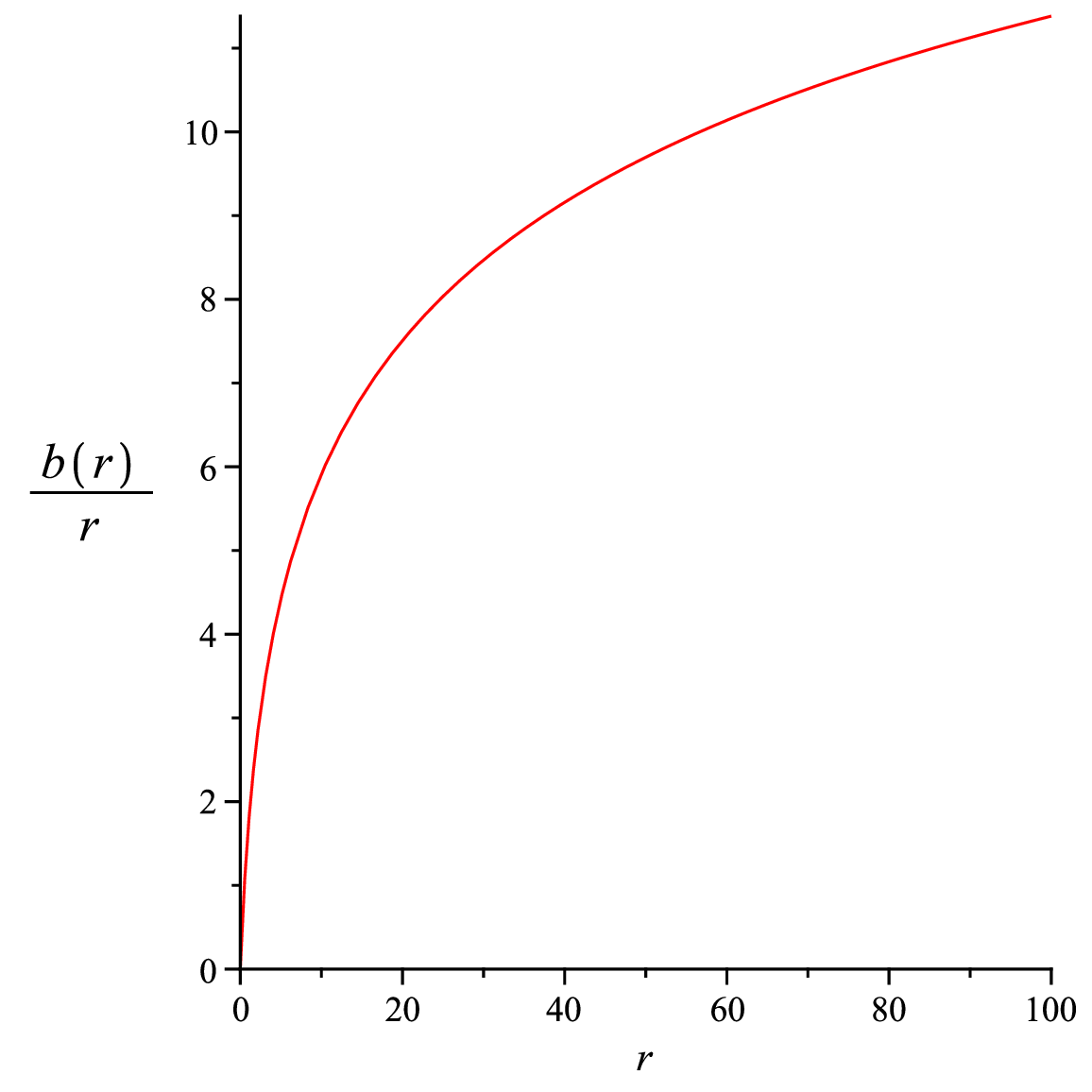}
	\centering (b)
\end{minipage}
\begin{minipage}{.35\textwidth}
	\centering
	\includegraphics[width=.6\linewidth]{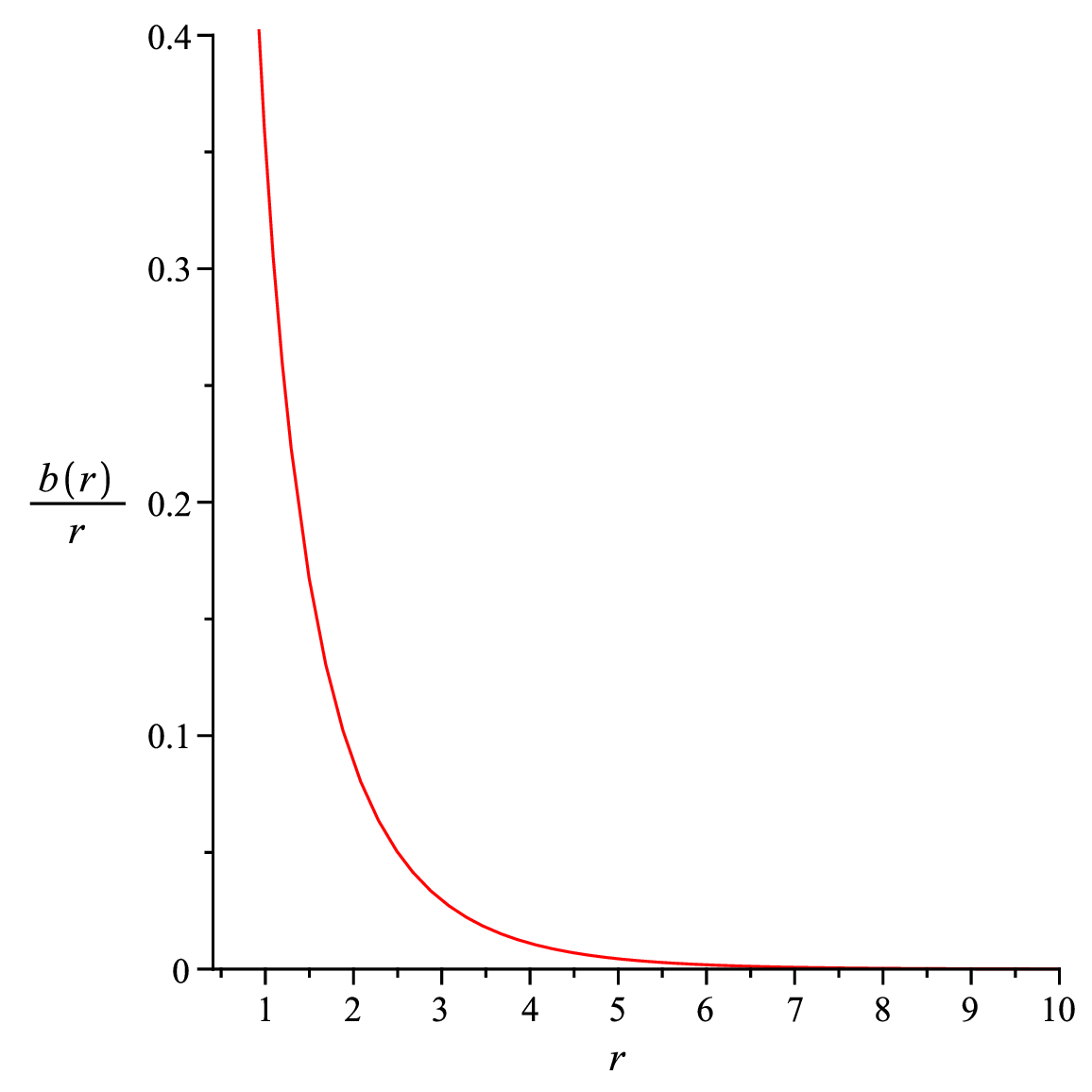}
	\centering (c)
\end{minipage}
\caption{Variation of $\frac{b(r)}{r}$ ($(a)$ for $b(r)=r_0e^{1-\frac{r}{r_0}},~r_0=1$, $(b)$ for $b(r)=r\frac{\ln(r+1)}{\ln(r_0+1)},~r_0=1$ and $(c)$ for $b(r)=r_0\frac{a^r}{a_0^r}$, $a=0.5$, $r_0=0.5$  ) with radial co-ordinate `$r$'.}
\label{figss}
\end{figure}
\begin{figure}[htb!]
	\begin{minipage}{.35\textwidth}
		\includegraphics[width=.6\linewidth]{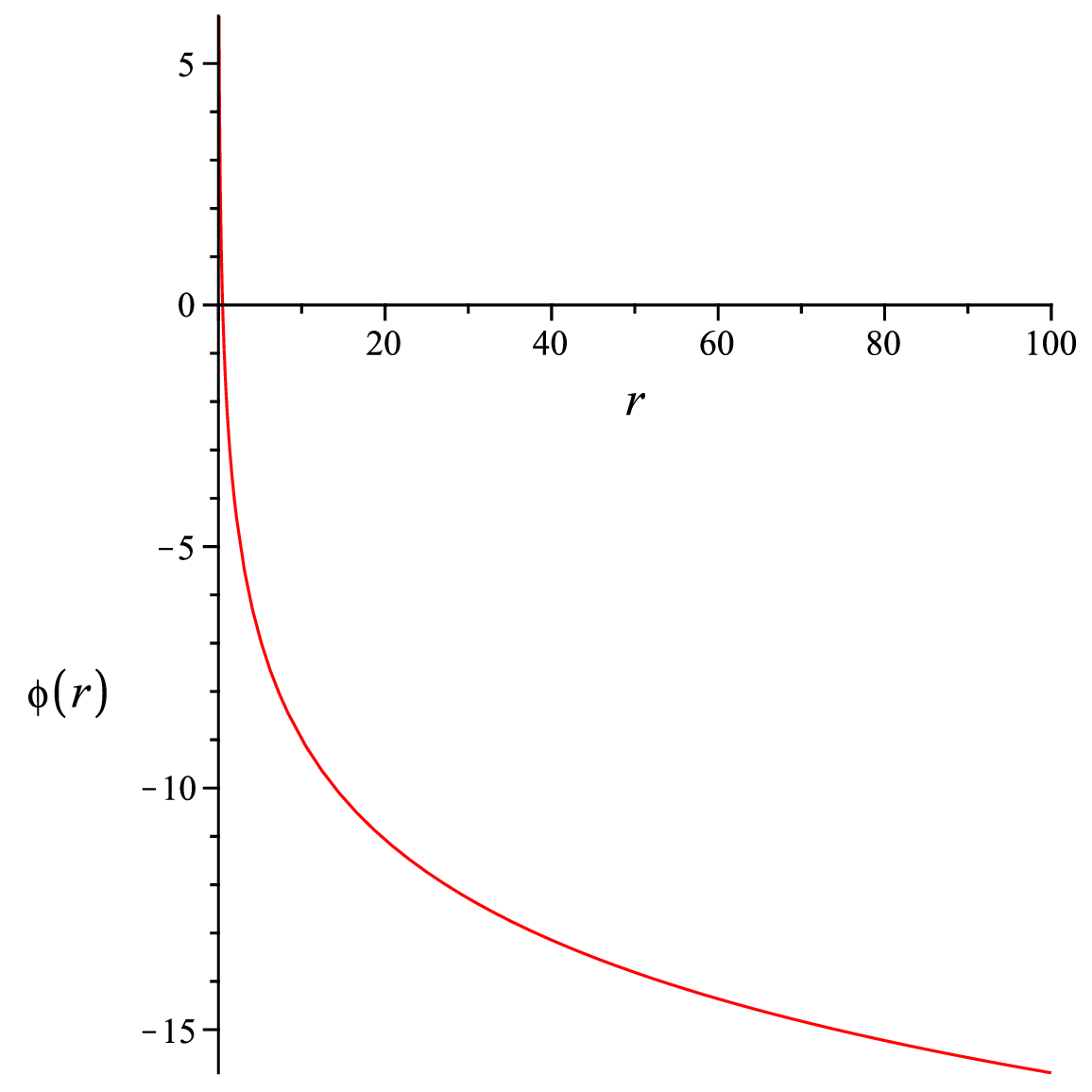}
		(a)
	\end{minipage}
	\begin{minipage}{.35\textwidth}
		\centering
		\includegraphics[width=.6\linewidth]{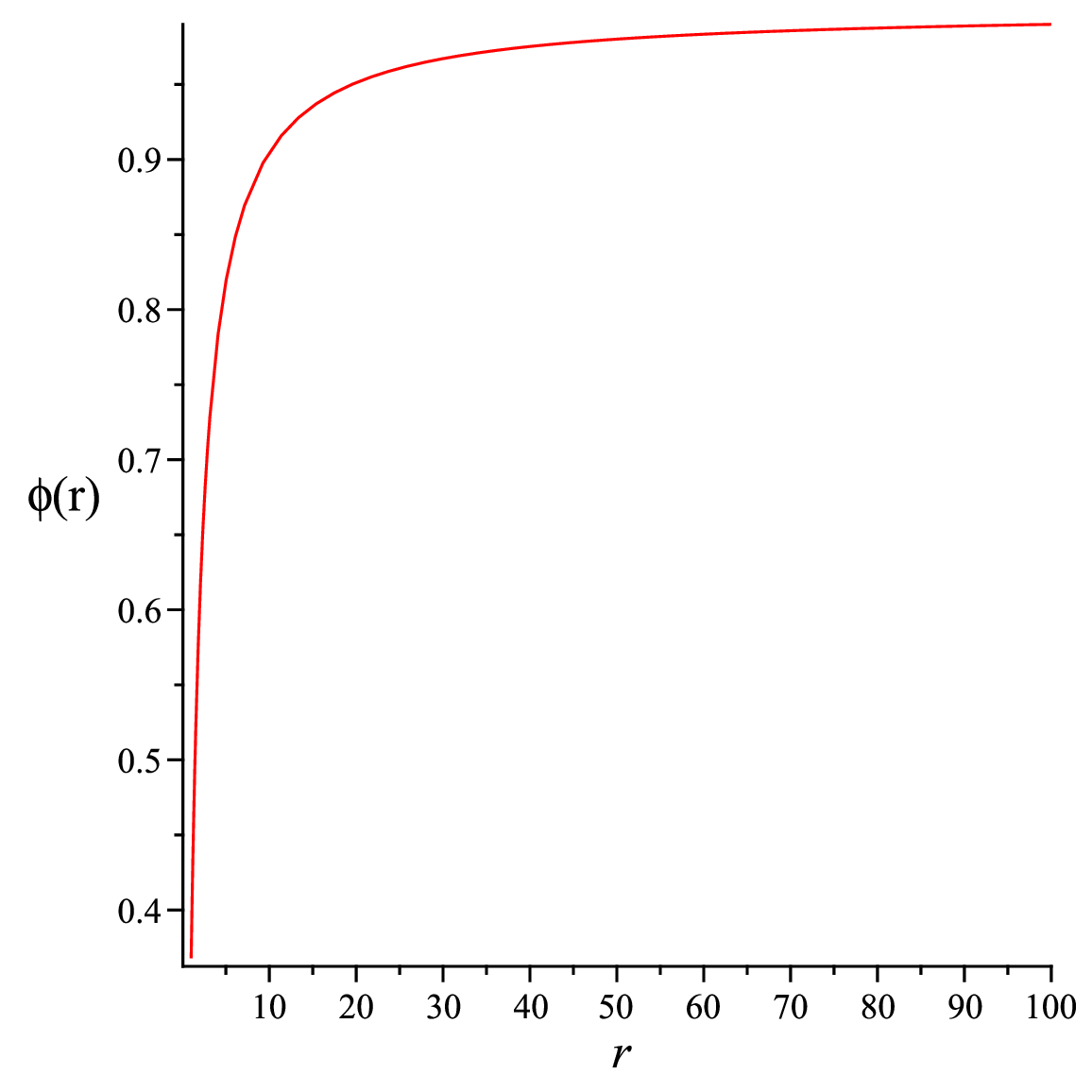}
		\centering (b)
	\end{minipage}
	\begin{minipage}{.35\textwidth}
		\centering
		\includegraphics[width=.6\linewidth]{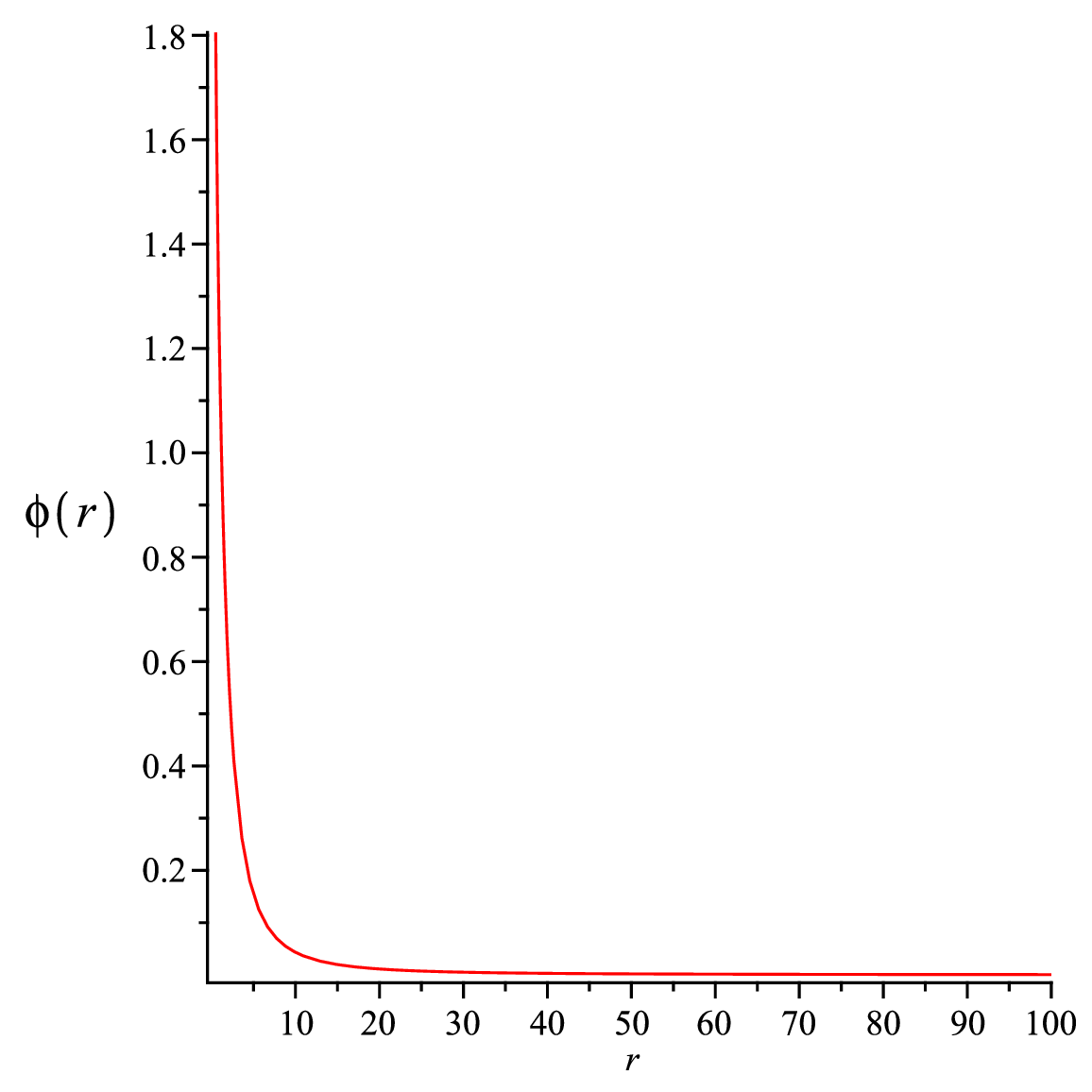}
		\centering (c)
	\end{minipage}
	\caption{Variation of $\phi(r)$ ((a) for $\phi(r)=j\ln(\frac{r}{r_0}),~ j=-3,~c=0.5$ (b) for $\phi(r)=e^{-\frac{r_0}{r}}$ and (c) for $\phi(r)=\ln\sqrt{1+\frac{\gamma^2}{r^2}},~\gamma=3,~r_0=0.5$) with radial co-ordinate `$r$'.}
	\label{figphis}
\end{figure}
\section{Validation of Energy conditions}
\label{sec5}
\begin{figure}[!]
	\centering
	\begin{minipage}{.45\textwidth}
		\centering
		\includegraphics[width=.6\linewidth]{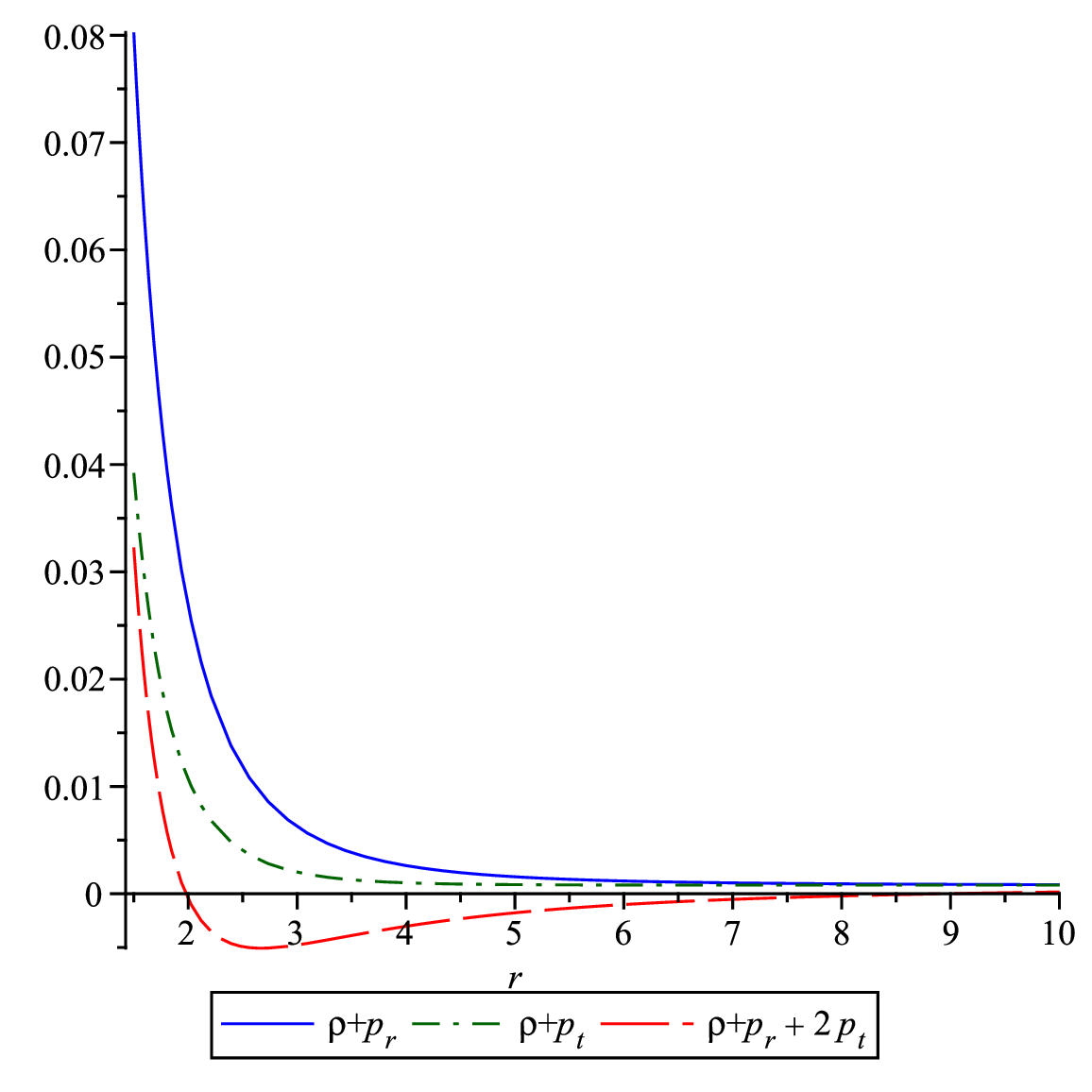}
		\centering (a)
	\end{minipage}
	\begin{minipage}{.45\textwidth}
		\centering
		\includegraphics[width=.6\linewidth]{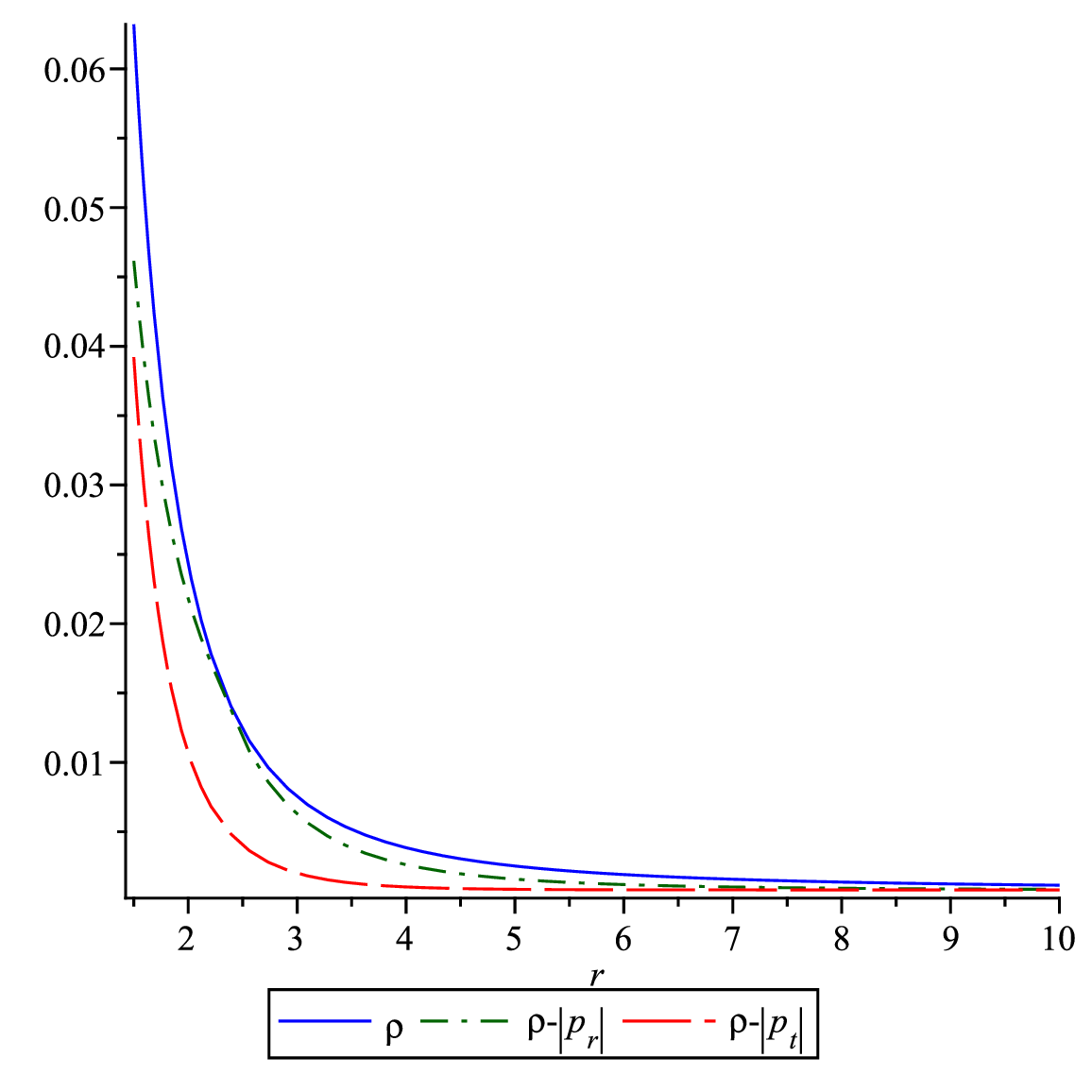}
		\centering (b)
	\end{minipage}
\caption{ Behavior of $\rho+p_r,~ \rho+p_t,~ \rho+p_r+2p_t$ (a) and $\rho,~\rho-|p_r|,~\rho-|p_t|$ diagrams (b) have been plotted for the obtained new shape function (\ref{obs}) with $\phi(r)=\ln r $ when $\lambda=1$ and $r_0=1.5$.}
\label{fig6}
\end{figure}

\begin{figure}[!]
	\centering
	\begin{minipage}{.45\textwidth}
		\centering
		\includegraphics[width=.6\linewidth]{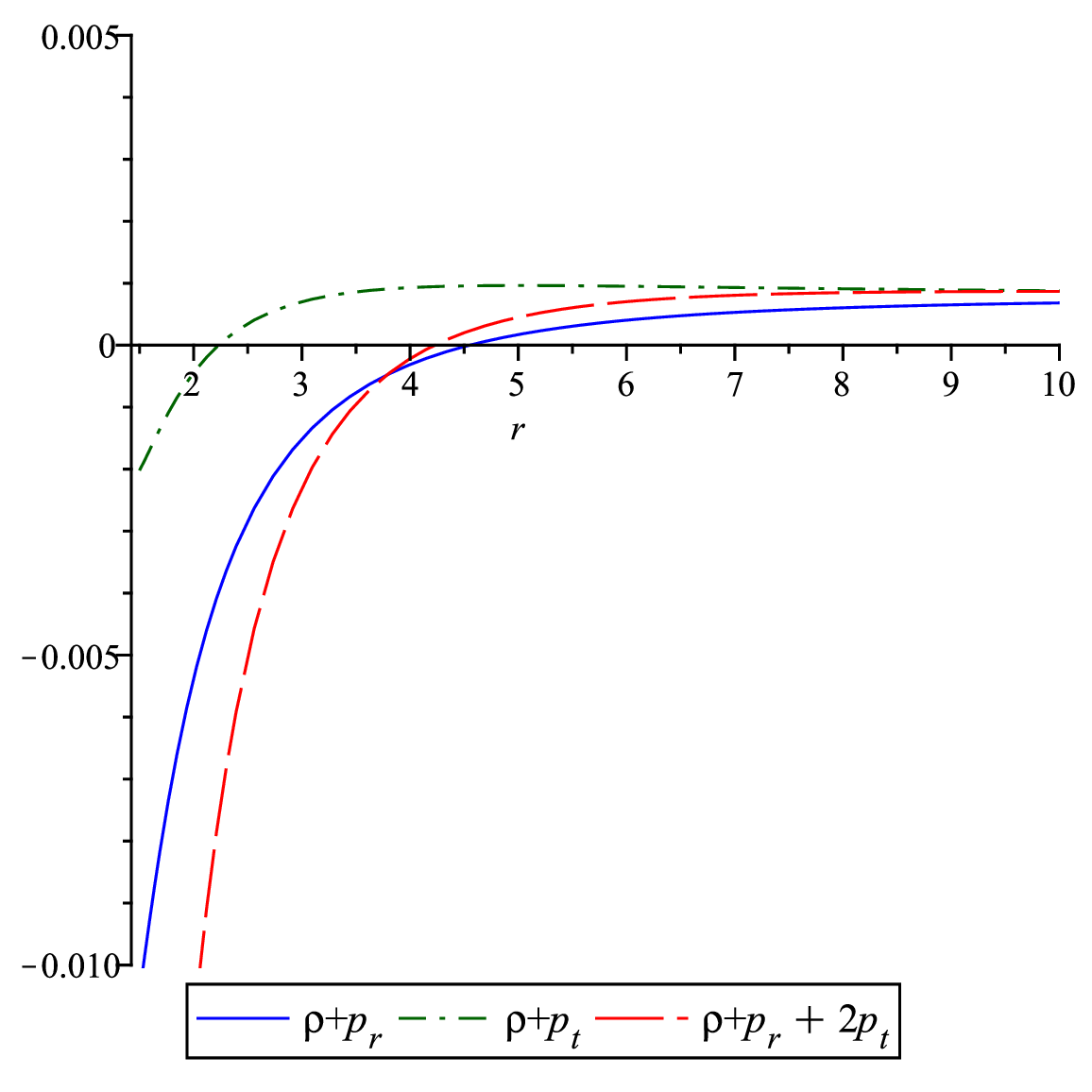}
		\centering (a)
	\end{minipage}
	\begin{minipage}{.45\textwidth}
		\centering
		\includegraphics[width=.6\linewidth]{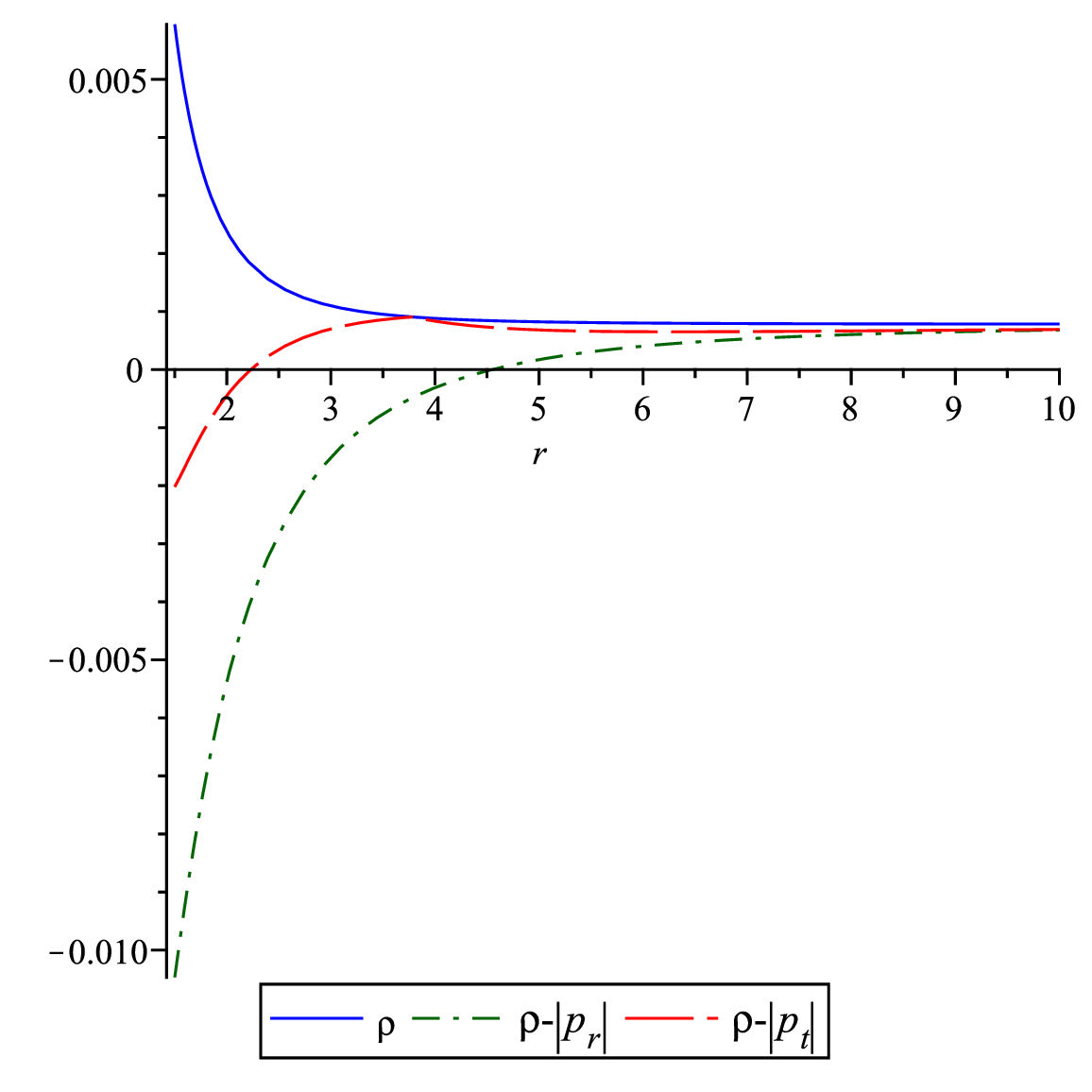}
		\centering (b)
	\end{minipage}
	\caption{  Behavior of $\rho+p_r,~ \rho+p_t,~ \rho+p_r+2p_t$ (a) and $\rho,~\rho-|p_r|,~\rho-|p_t|$ diagrams (b) have been plotted for the obtained new shape function (\ref{obs2}) with $\phi(r)=\text{Const.} $ when $\lambda=1$ and $r_0=1.5$.}
	\label{fig66}
\end{figure}

\begin{figure}[!]
	\centering
	\begin{minipage}{.45\textwidth}
		\centering
		\includegraphics[width=.6\linewidth]{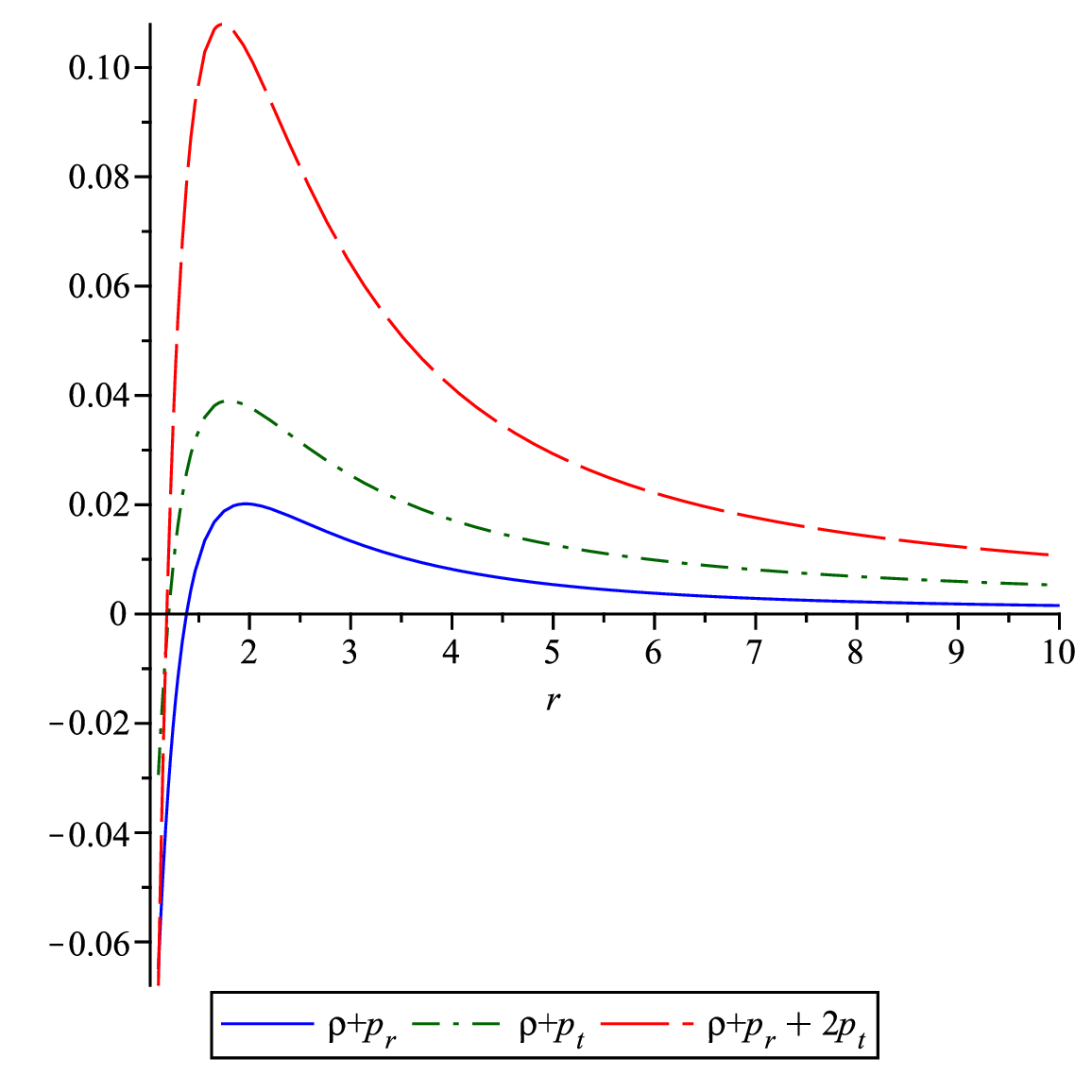}
		\centering (a)
	\end{minipage}
	\begin{minipage}{.45\textwidth}
		\centering
		\includegraphics[width=.6\linewidth]{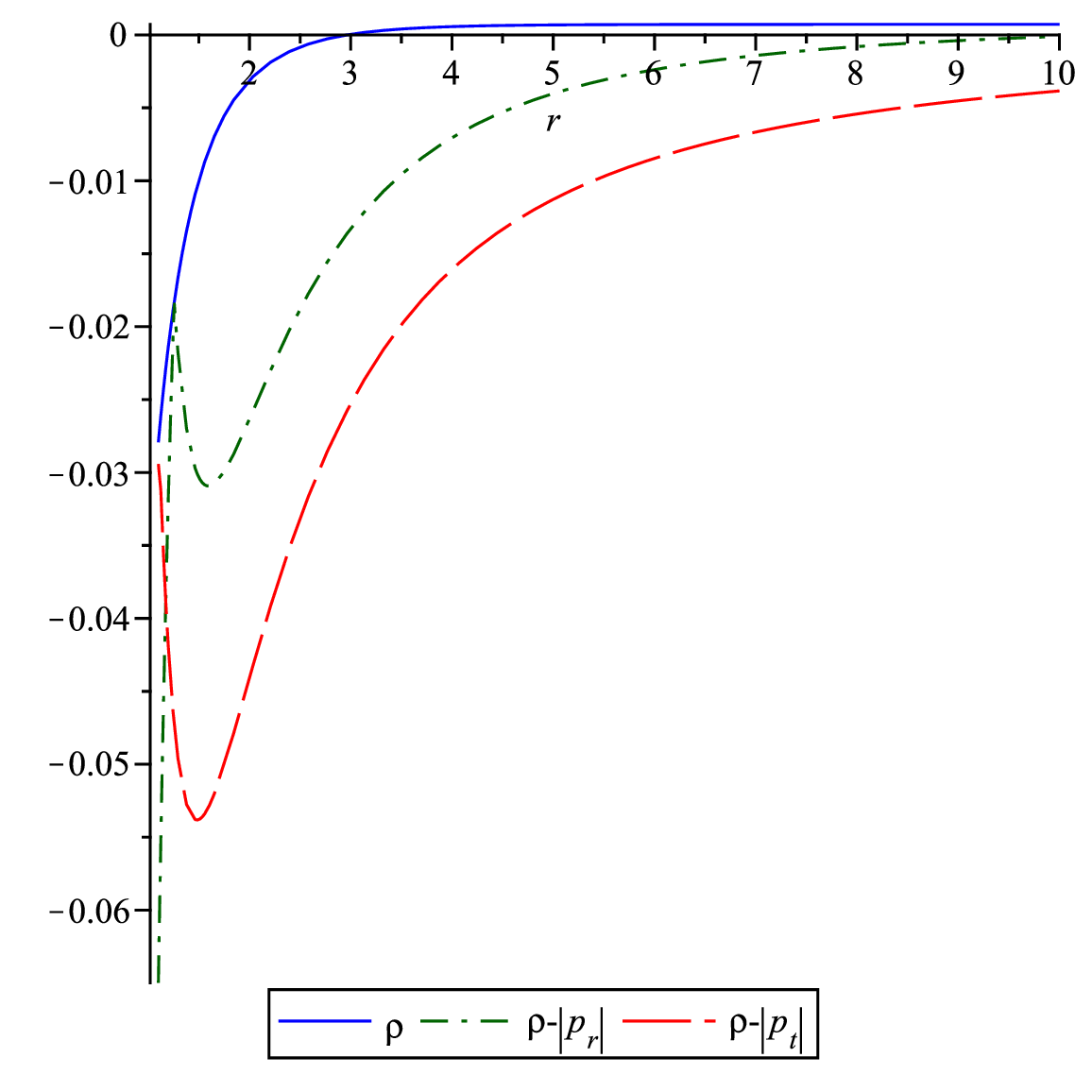}
		\centering (b)
	\end{minipage}
	\caption{ Behavior of $\rho+p_r,~ \rho+p_t,~ \rho+p_r+2p_t$ (a) and $\rho,~\rho-|p_r|,~\rho-|p_t|$ diagrams (b) have been plotted for the redshift function $\phi=j\ln\left(\frac{r}{r_0}\right)$ and shape function $b(r)=r_0e^{1-\frac{r}{r_0}}$ with the numerical values $j=2.2$, $r_0=1.1$ and $\lambda=0.9$}\label{fig7}
\end{figure}

\begin{figure}[!]
	\centering
	\begin{minipage}{.45\textwidth}
		\centering
		\includegraphics[width=.6\linewidth]{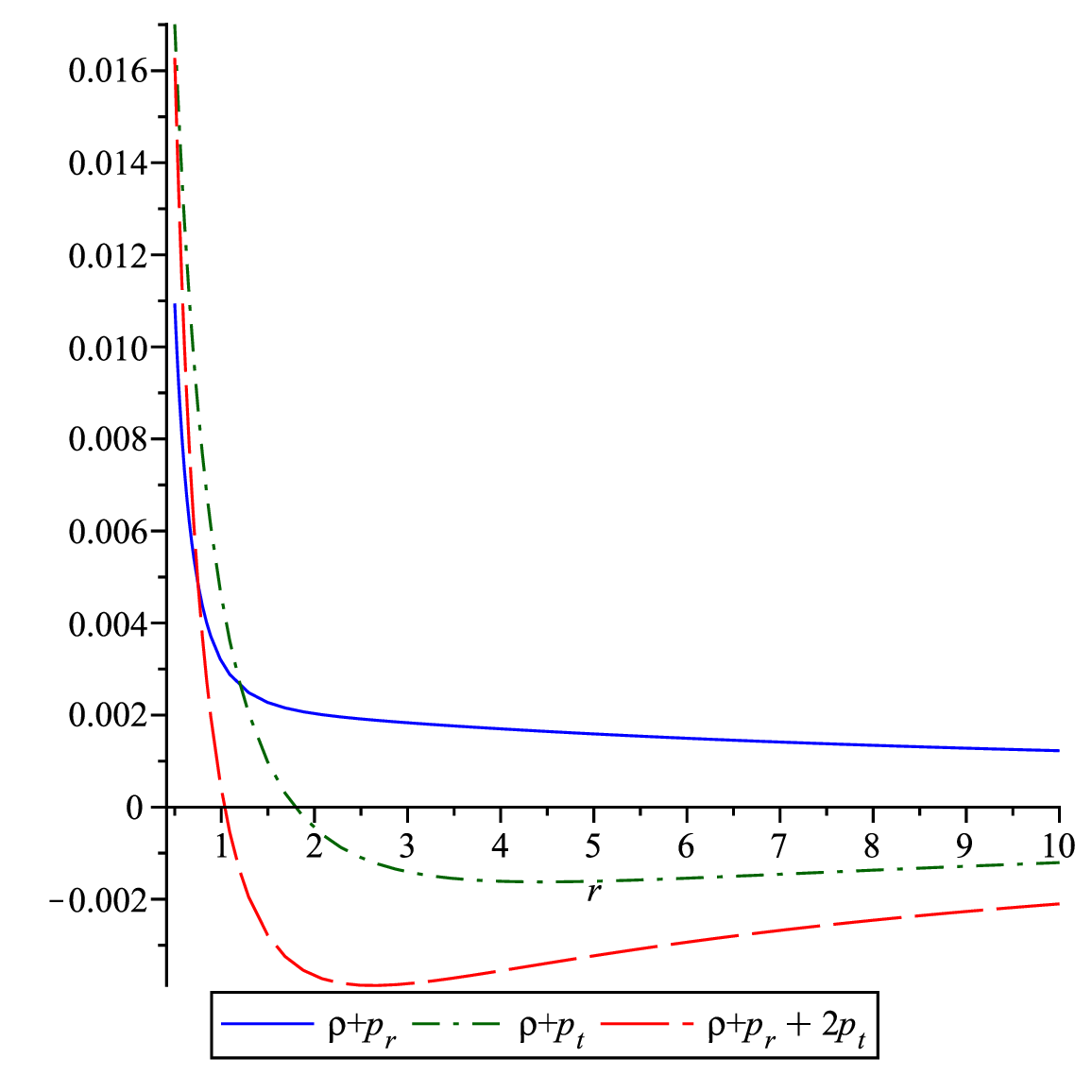}
		\centering (a)
	\end{minipage}
	\begin{minipage}{.45\textwidth}
		\centering
		\includegraphics[width=.6\linewidth]{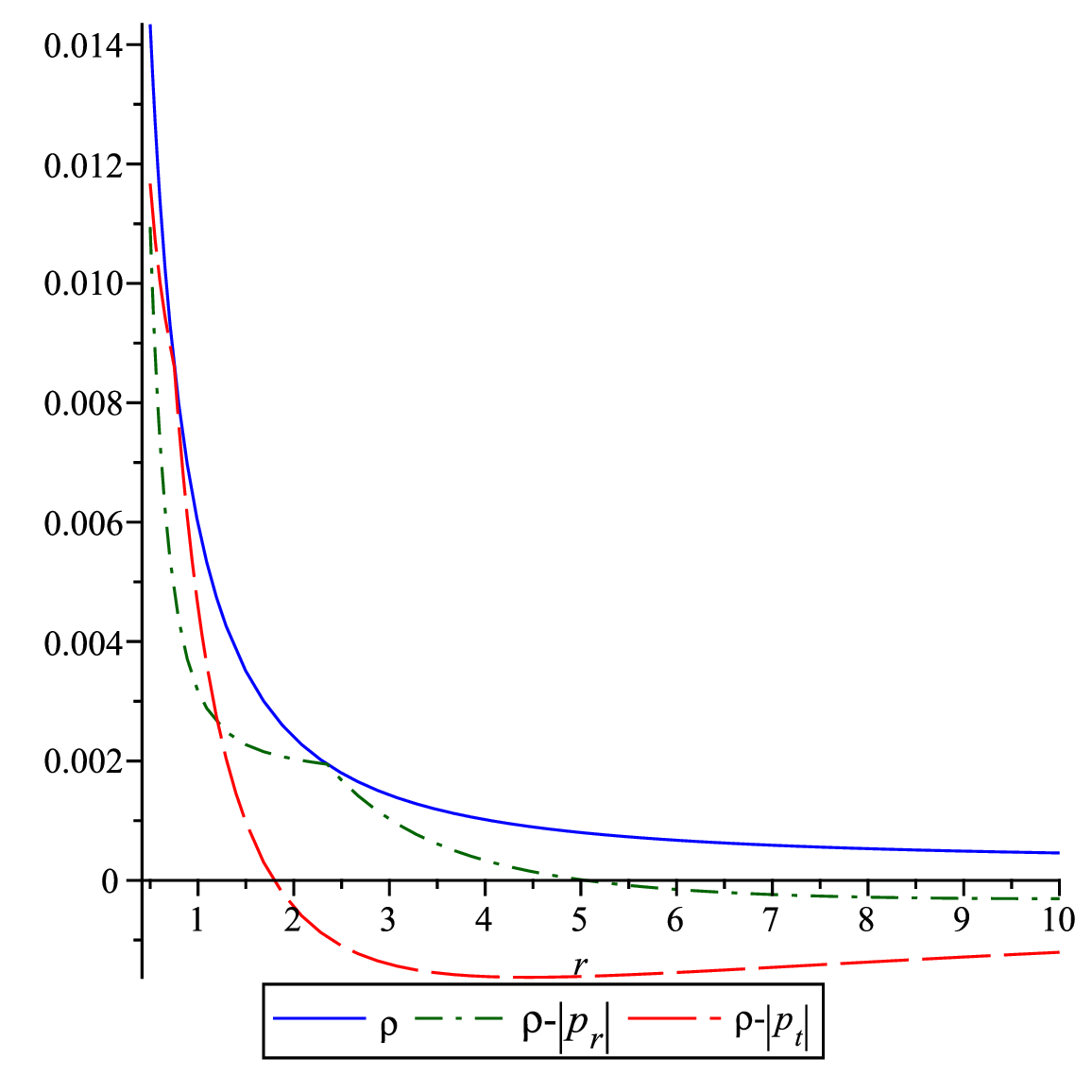}
		\centering (b)
	\end{minipage}
	\caption{  Behavior of $\rho+p_r,~ \rho+p_t,~ \rho+p_r+2p_t$ (a) and $\rho,~\rho-|p_r|,~\rho-|p_t|$ diagrams (b) have been plotted for the redshift function $\phi=j\ln\left(\frac{r}{r_0}\right)$ and shape $b(r)=r\frac{\ln(r+1)}{\ln(r_0+1)}$ with the numerical values $j=0.6$, $r_0=0.5$ and $\lambda=100$}\label{fig8}
\end{figure}

\begin{figure}[!]
	\centering
	\begin{minipage}{.45\textwidth}
		\centering
		\includegraphics[width=.6\linewidth]{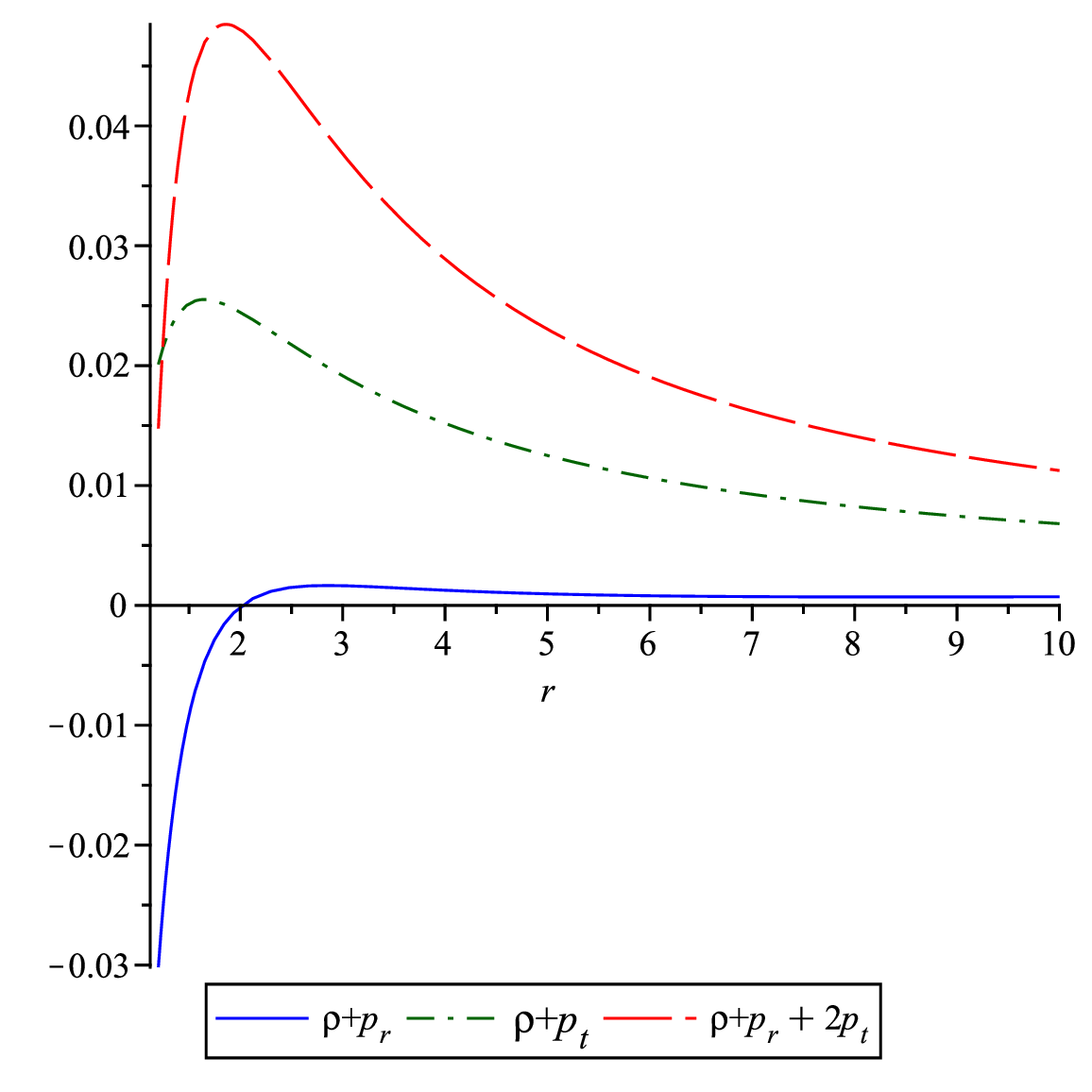}
		\centering (a)
	\end{minipage}
	\begin{minipage}{.45\textwidth}
		\centering
		\includegraphics[width=.6\linewidth]{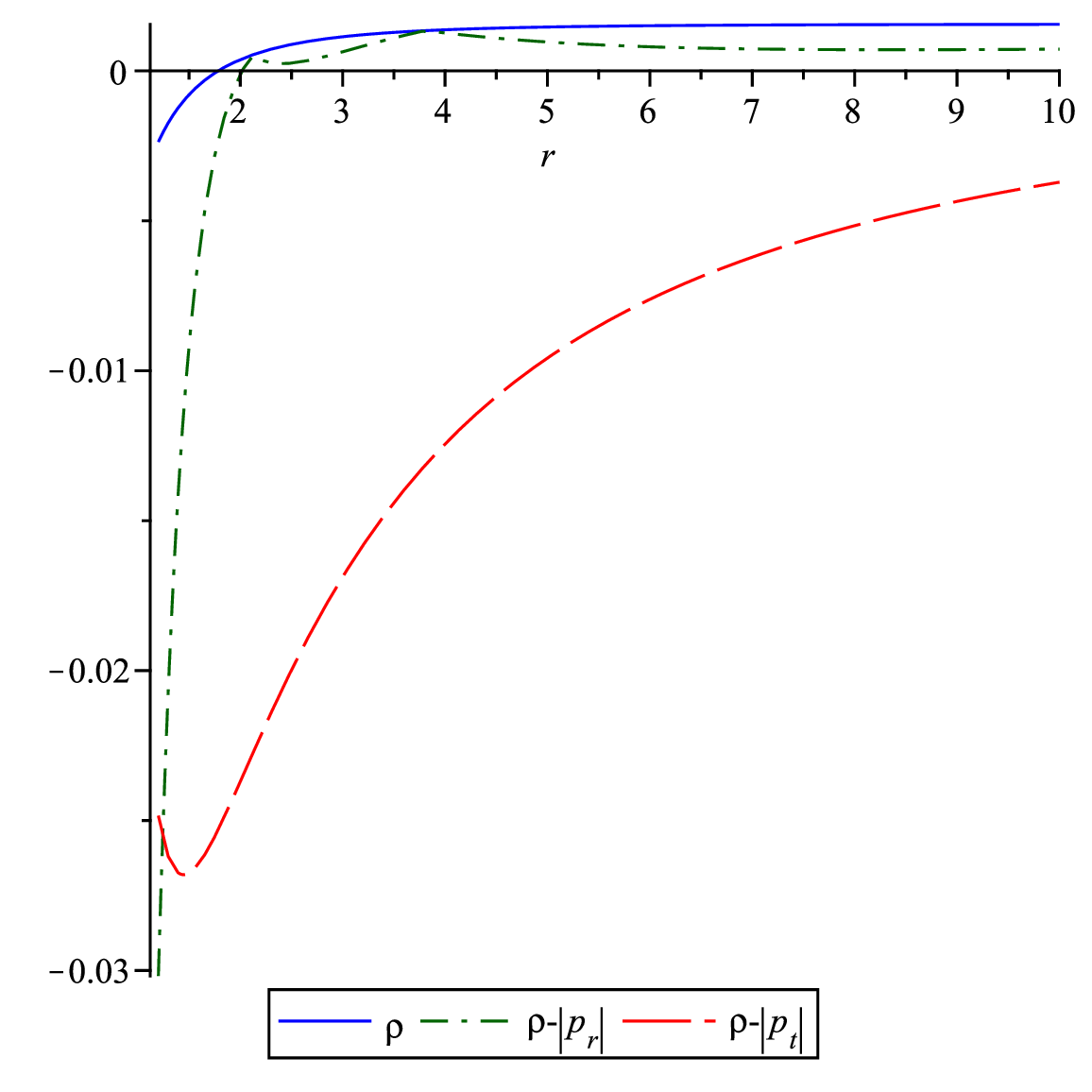}
		\centering (b)
	\end{minipage}
	\caption{ Behavior of $\rho+p_r,~ \rho+p_t,~ \rho+p_r+2p_t$ (a) and $\rho,~\rho-|p_r|,~\rho-|p_t|$ diagrams (b) have been plotted for the redshift function $\phi=j\ln\left(\frac{r}{r_0}\right)$ and shape function $b(r)=r_0\frac{a^r}{a^r_0}$ with the numerical values $j=1.6$, $r_0=1.2$, $a=0.8$ and $\lambda=5$}\label{fig9}
\end{figure}

\begin{figure}[!]
	\centering
	\begin{minipage}{.45\textwidth}
		\centering
		\includegraphics[width=.6\linewidth]{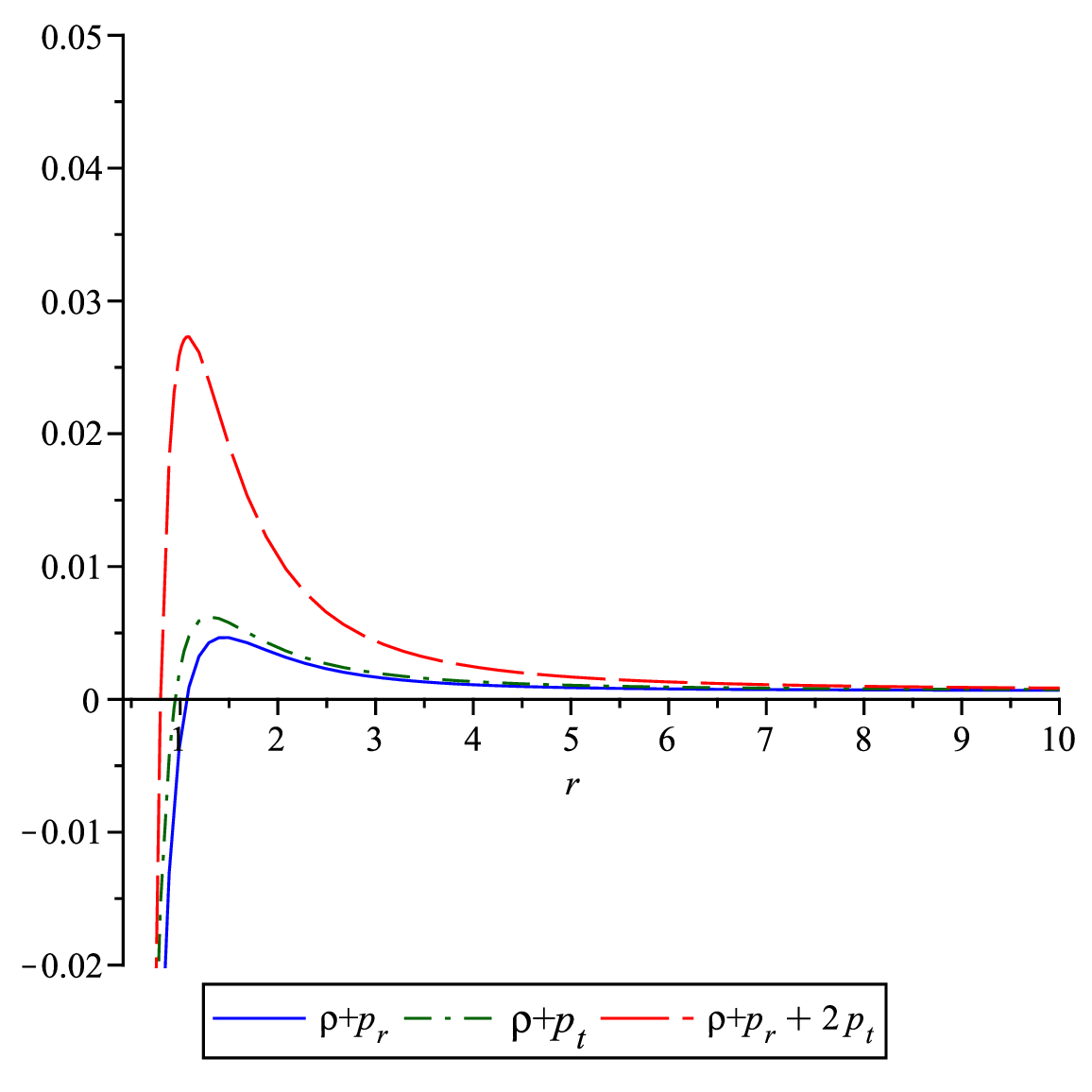}
		\centering (a)
	\end{minipage}
	\begin{minipage}{.45\textwidth}
		\centering
		\includegraphics[width=.6\linewidth]{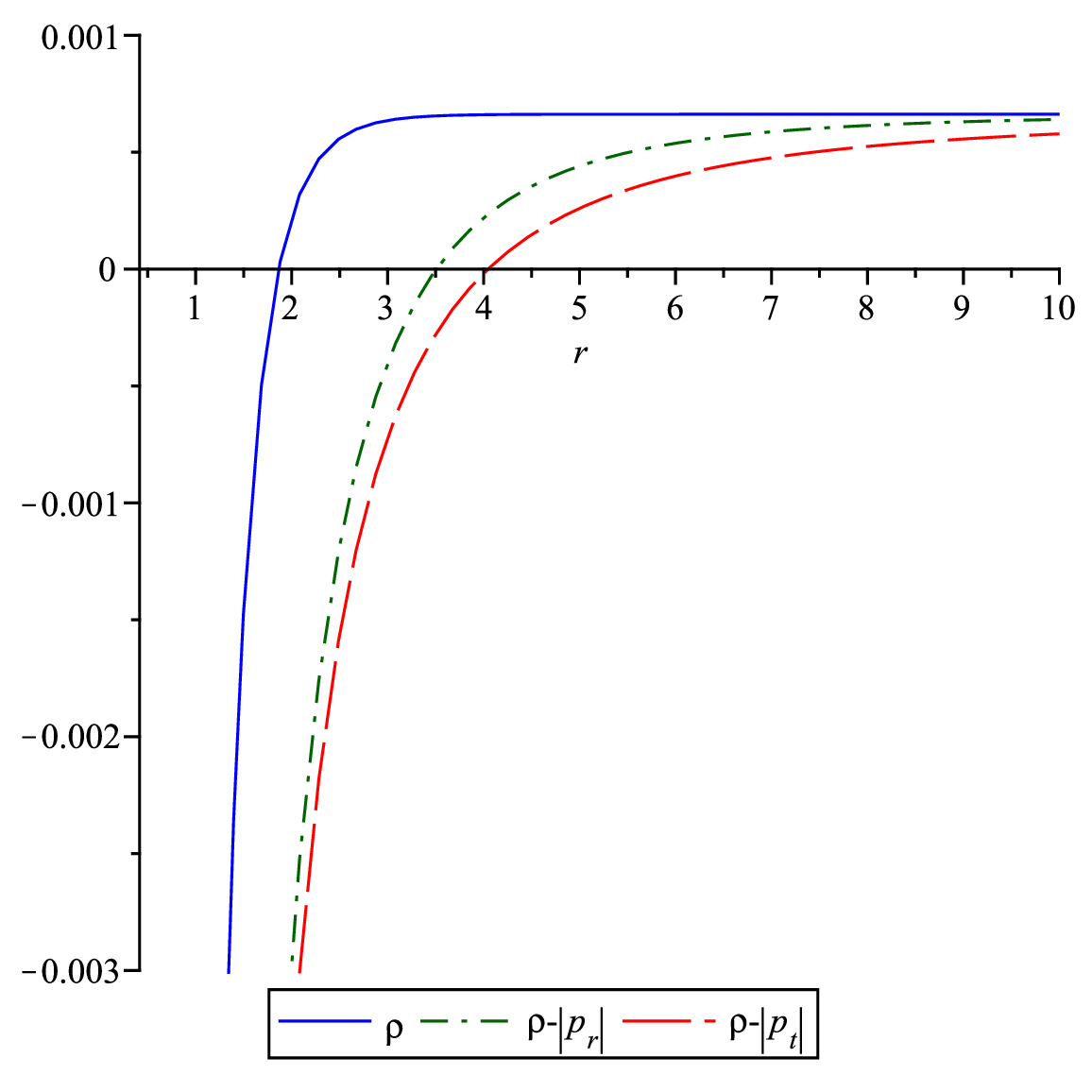}
		\centering (b)
	\end{minipage}
	\caption{Behavior of $\rho+p_r,~ \rho+p_t,~ \rho+p_r+2p_t$ (a) and $\rho,~\rho-|p_r|,~\rho-|p_t|$ diagrams (b) have been plotted for the redshift function $\phi=e^{-\frac{r_0}{r}}$ and shape function $b(r)=r_0e^{1-\frac{r}{r_0}}$ with the numerical values $r_0=0.5$ and $\lambda=0.8$}\label{fig10}
\end{figure}

\begin{figure}[!]
	\centering
	\begin{minipage}{.45\textwidth}
		\centering
		\includegraphics[width=.6\linewidth]{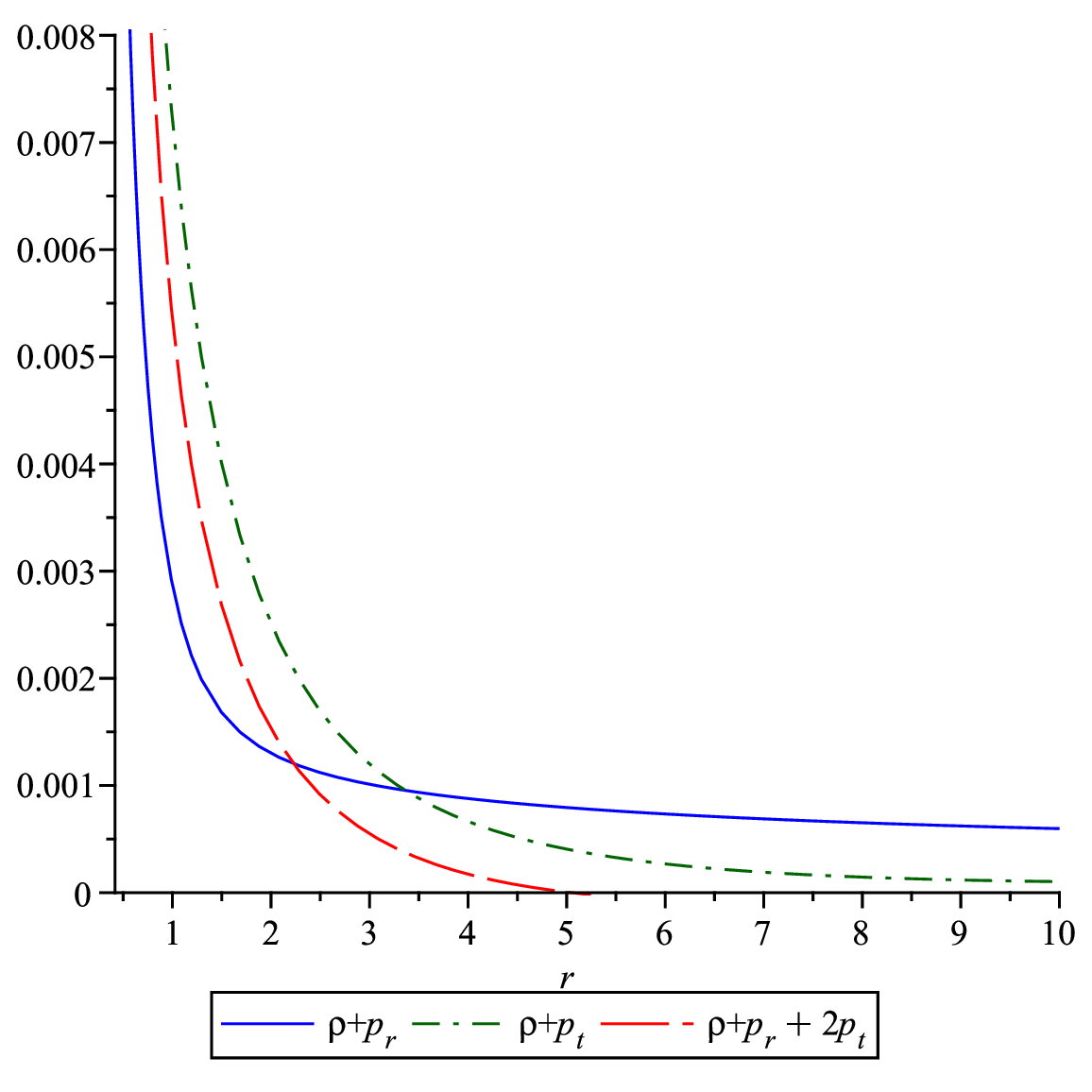}
		\centering (a)
	\end{minipage}
	\begin{minipage}{.45\textwidth}
		\centering
		\includegraphics[width=.6\linewidth]{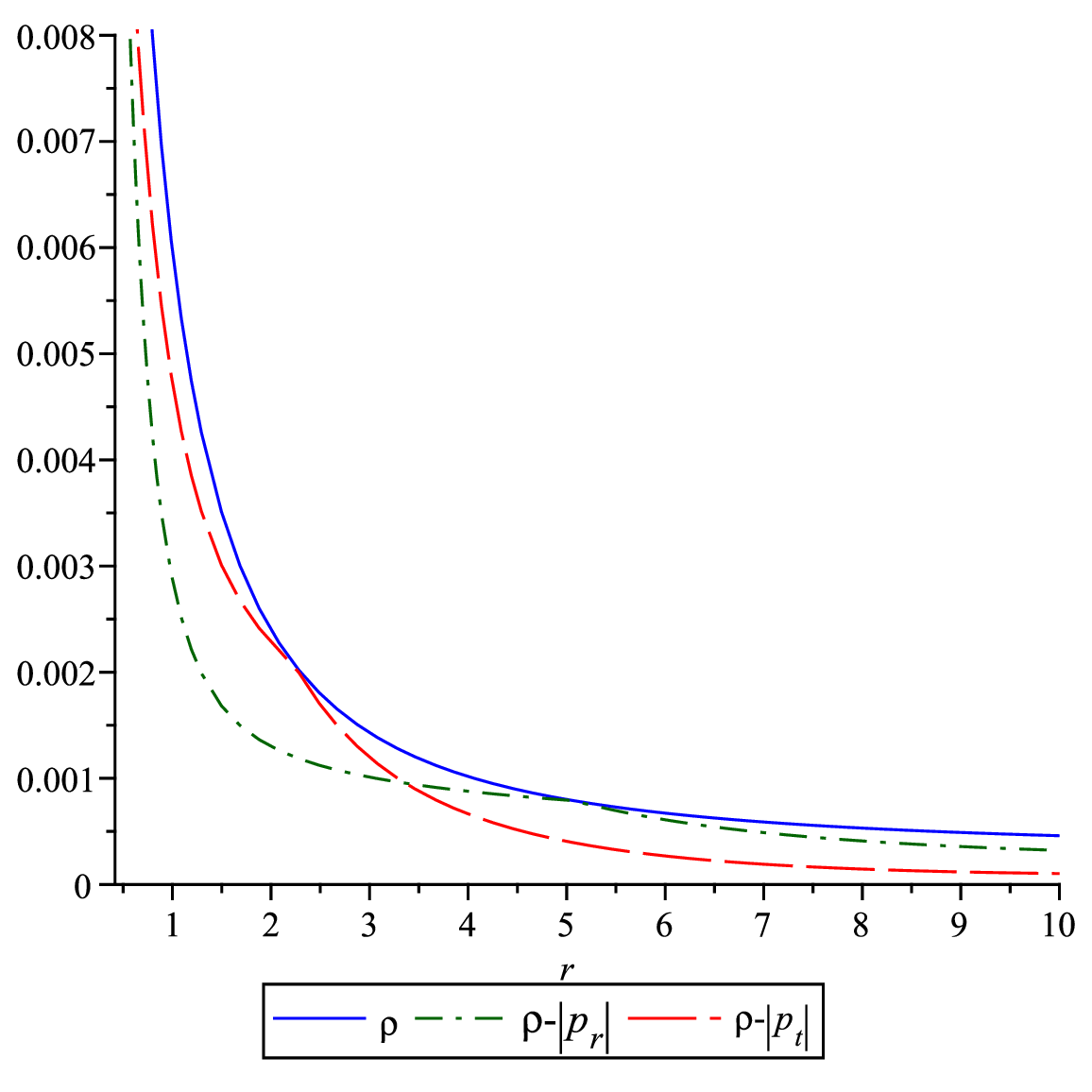}
		\centering (b)
	\end{minipage}
	\caption{  Behavior of $\rho+p_r,~ \rho+p_t,~ \rho+p_r+2p_t$ (a) and $\rho,~\rho-|p_r|,~\rho-|p_t|$ diagrams (b) have been plotted for the redshift function $\phi=e^{-\frac{r_0}{r}}$ and shape function $b(r)=r\frac{ln(r+1)}{ln(r_0+1)}$ with the numerical values $r_0=0.5$ and $\lambda=100$.}\label{fig11}
\end{figure}

\begin{figure}[!]
	\centering
	\begin{minipage}{.45\textwidth}
		\centering
		\includegraphics[width=.6\linewidth]{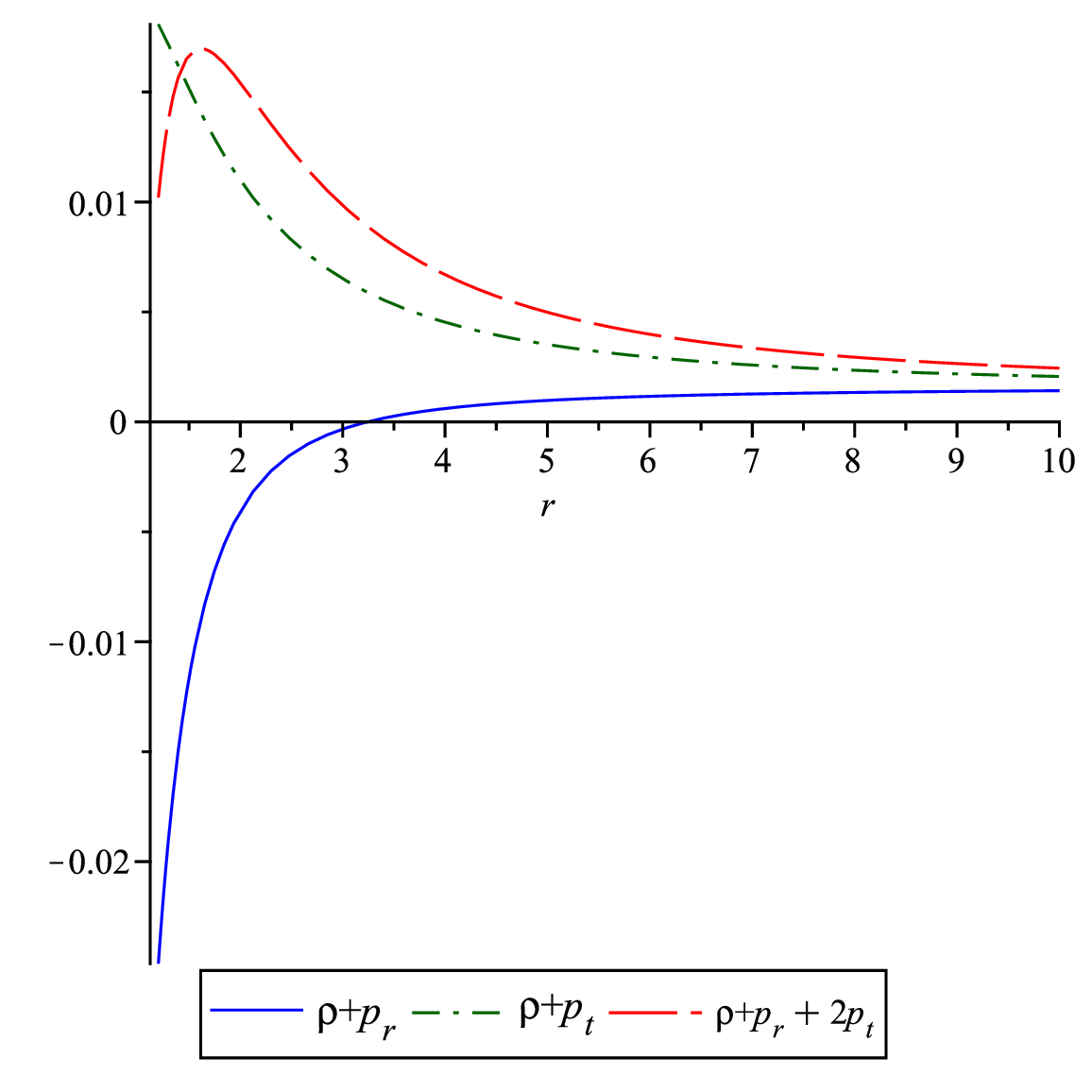}
		\centering (a)
	\end{minipage}
	\begin{minipage}{.45\textwidth}
		\centering
		\includegraphics[width=.6\linewidth]{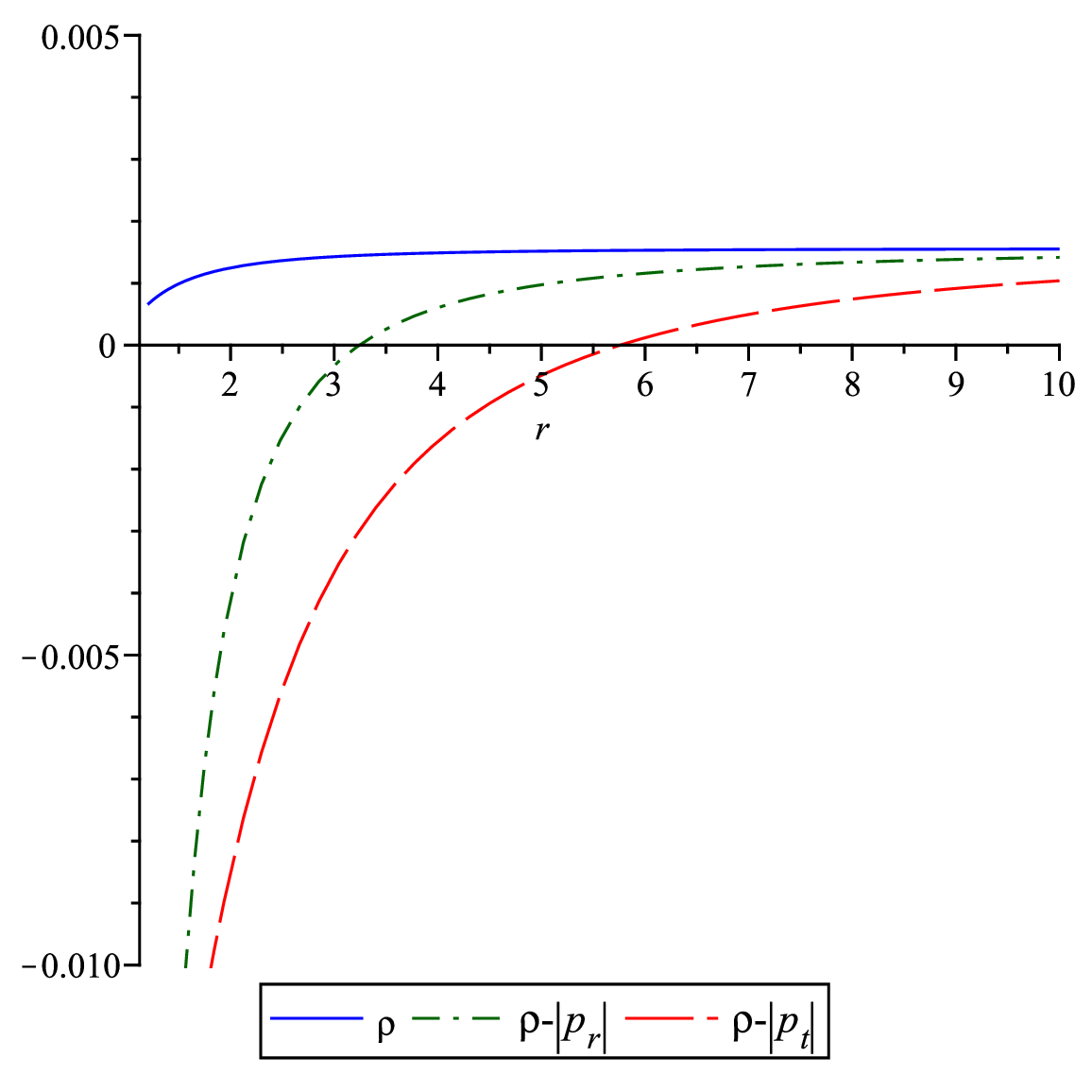}
		\centering (b)
	\end{minipage}
	\caption{ Behavior of $\rho+p_r,~ \rho+p_t,~ \rho+p_r+2p_t$ (a) and $\rho,~\rho-|p_r|,~\rho-|p_t|$ diagrams (b) have been plotted for the redshift function $\phi=e^{-\frac{r_0}{r}}$ and shape function $b(r)=r_0\frac{a^r}{a^r_0}$ with the numerical values $a=0.95$, $r_0=1.2$ and $\lambda=5$}\label{fig12}
\end{figure}

\begin{figure}[!]
	\centering
	\begin{minipage}{.45\textwidth}
		\centering
		\includegraphics[width=.6\linewidth]{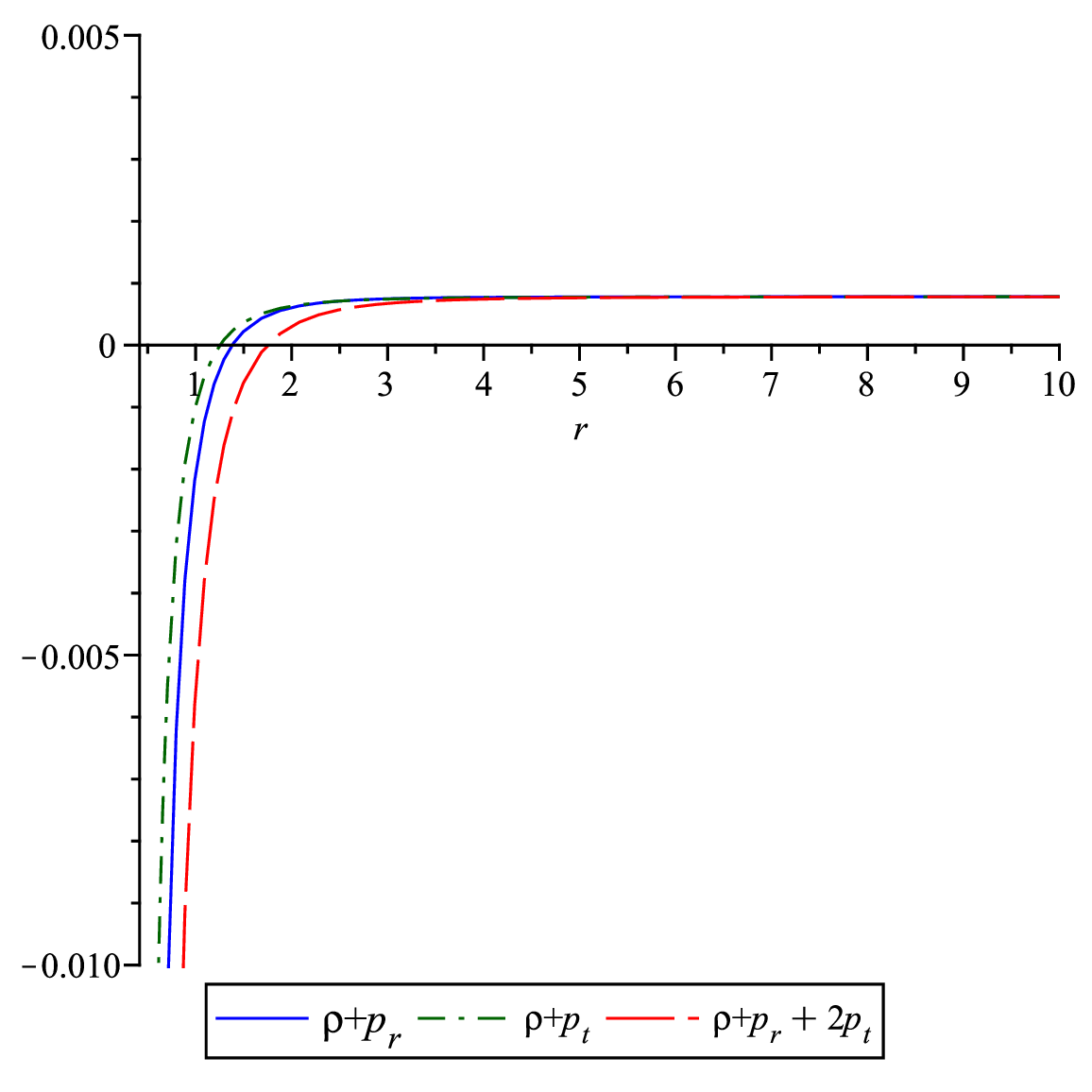}
		\centering (a)
	\end{minipage}
	\begin{minipage}{.45\textwidth}
		\centering
		\includegraphics[width=.6\linewidth]{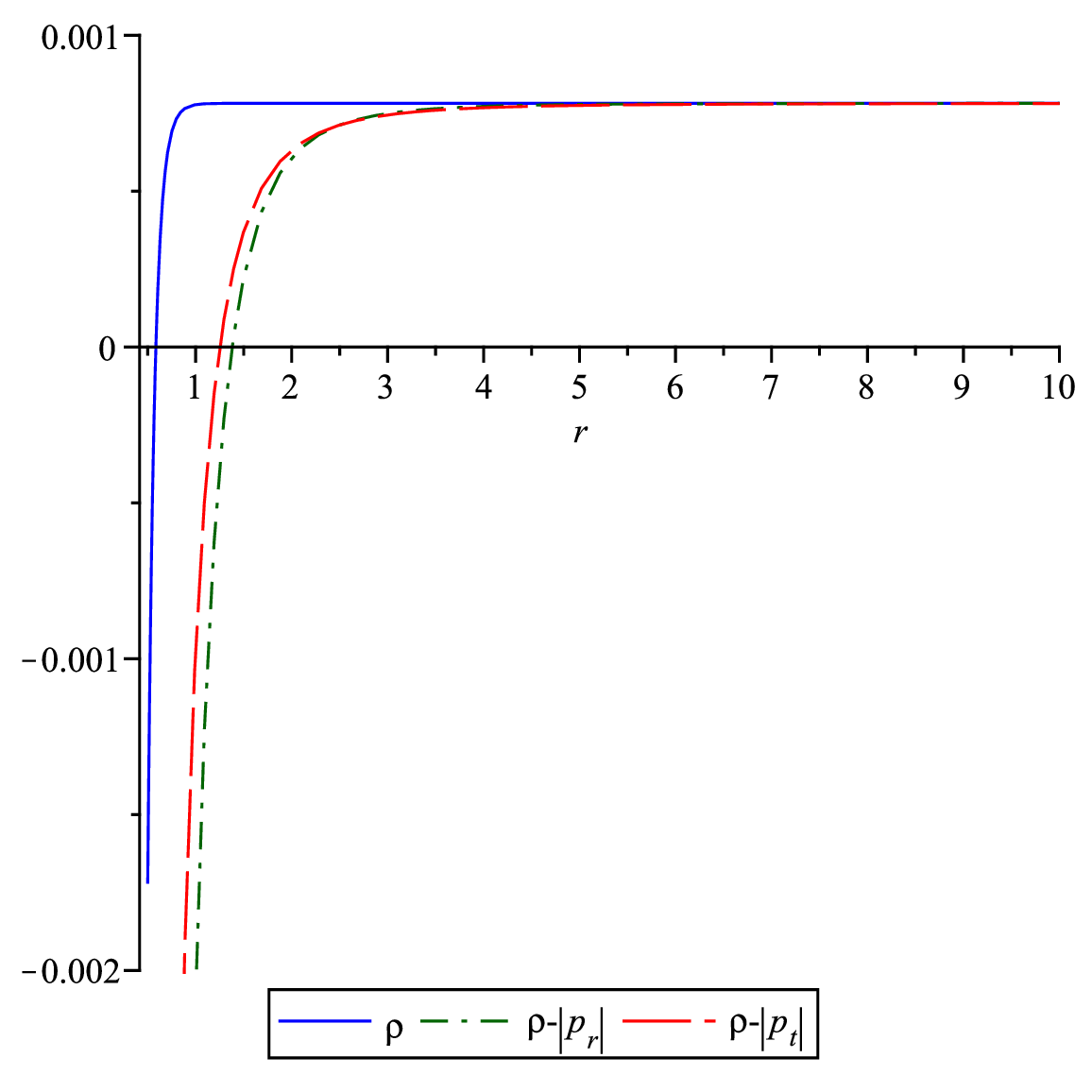}
		\centering (b)
	\end{minipage}
	\caption{ Behavior of $\rho+p_r,~ \rho+p_t,~ \rho+p_r+2p_t$ (a) and $\rho,~\rho-|p_r|,~\rho-|p_t|$ diagrams (b) have been plotted for the redshift function $\phi=\ln\left(\sqrt{1+\frac{\gamma^2}{r^2}}\right)$ and shape function $b(r)=r_0e^{1-\frac{r}{r_0}}$ with the numerical values $\gamma=0.2$, $r_0=0.1$ and $\lambda=1$}\label{fig13}
\end{figure}

\begin{figure}[!]
	\centering
	\begin{minipage}{.45\textwidth}
		\centering
		\includegraphics[width=.6\linewidth]{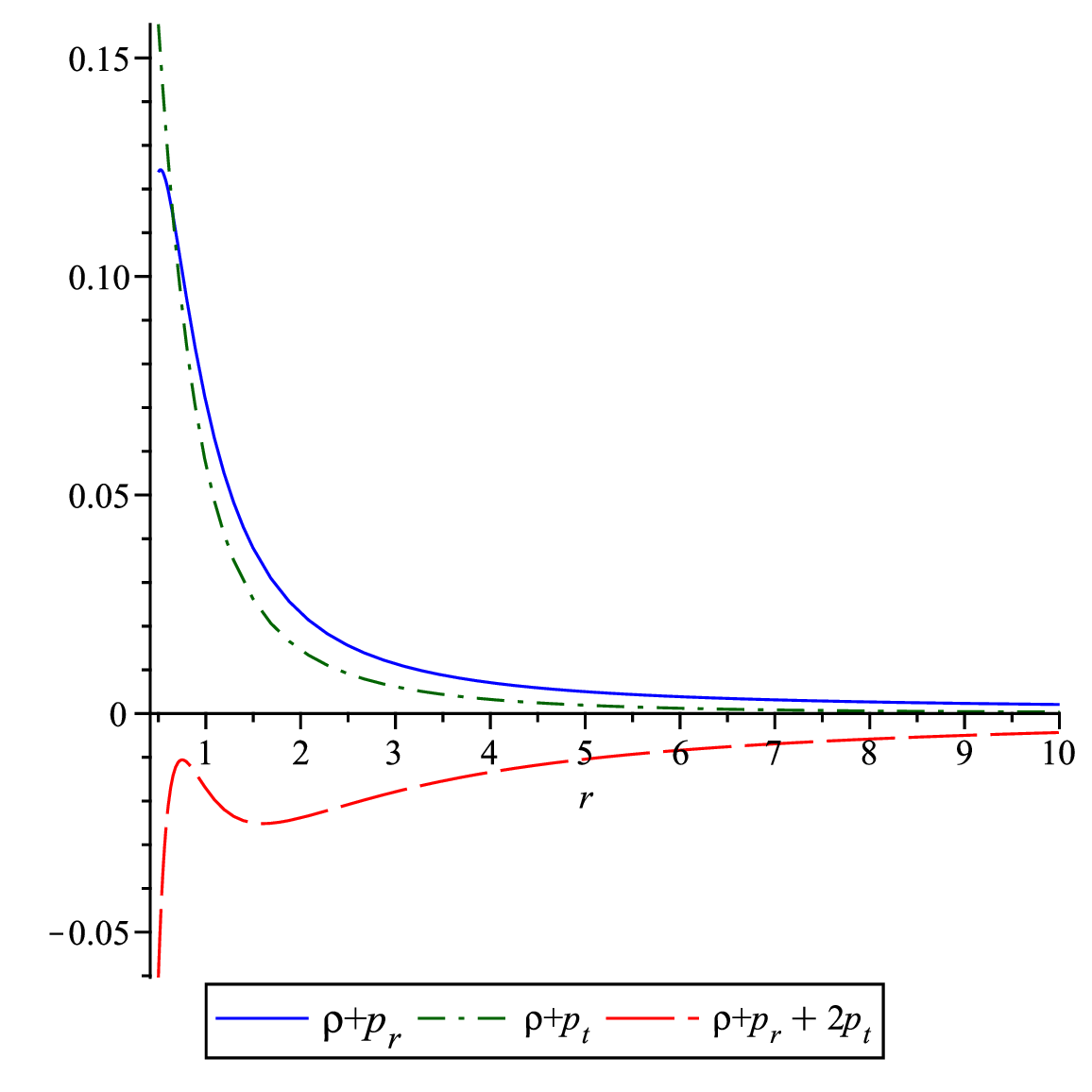}
		\centering (a)
	\end{minipage}
	\begin{minipage}{.45\textwidth}
		\centering
		\includegraphics[width=.6\linewidth]{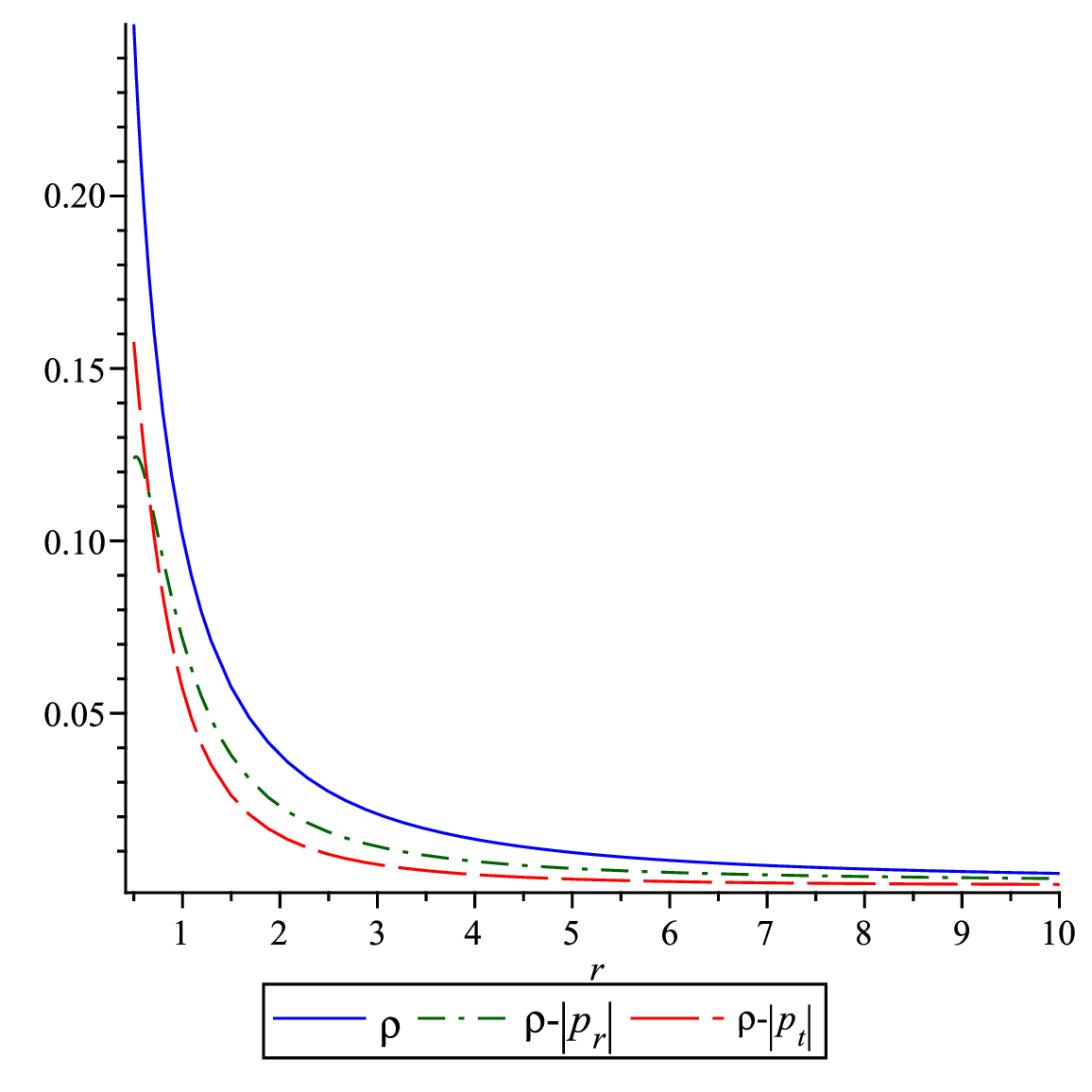}
		\centering (b)
	\end{minipage}
	\caption{ Behavior of $\rho+p_r,~ \rho+p_t,~ \rho+p_r+2p_t$ (a) and $\rho,~\rho-|p_r|,~\rho-|p_t|$ diagrams (b) have been plotted for the redshift function $\phi=\ln\left(\sqrt{1+\frac{\gamma^2}{r^2}}\right)$ and shape function $b(r)=r\frac{ln(r+1)}{ln(r_0+1)}$ with the numerical values $\gamma=0.8$, $r_0=0.5$ and $\lambda=1$}\label{fig14}
\end{figure}

\begin{figure}[htb]
	\centering
	\begin{minipage}{.45\textwidth}
		\centering
		\includegraphics[width=.6\linewidth]{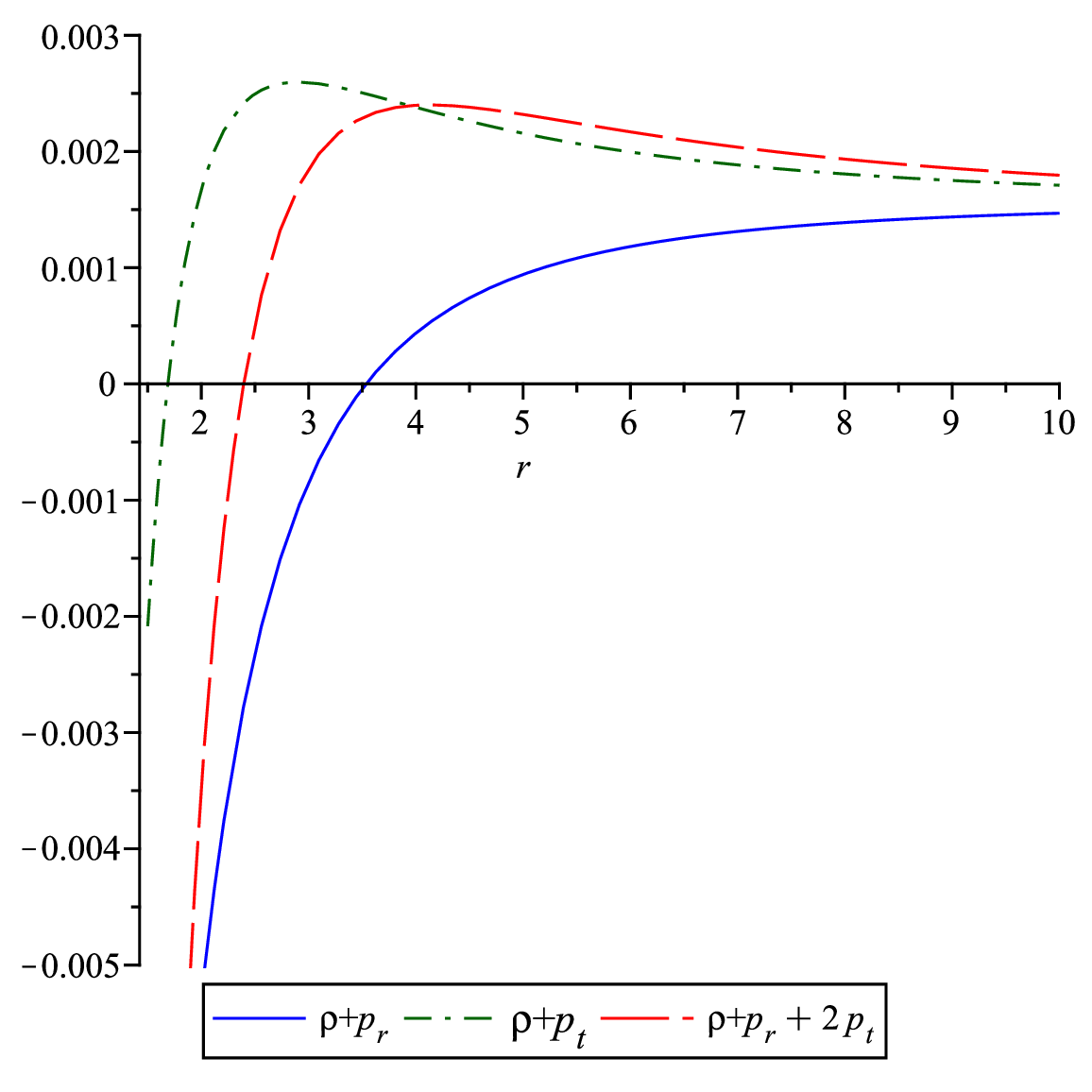}
		\centering (a)
	\end{minipage}
	\begin{minipage}{.45\textwidth}
		\centering
		\includegraphics[width=.6\linewidth]{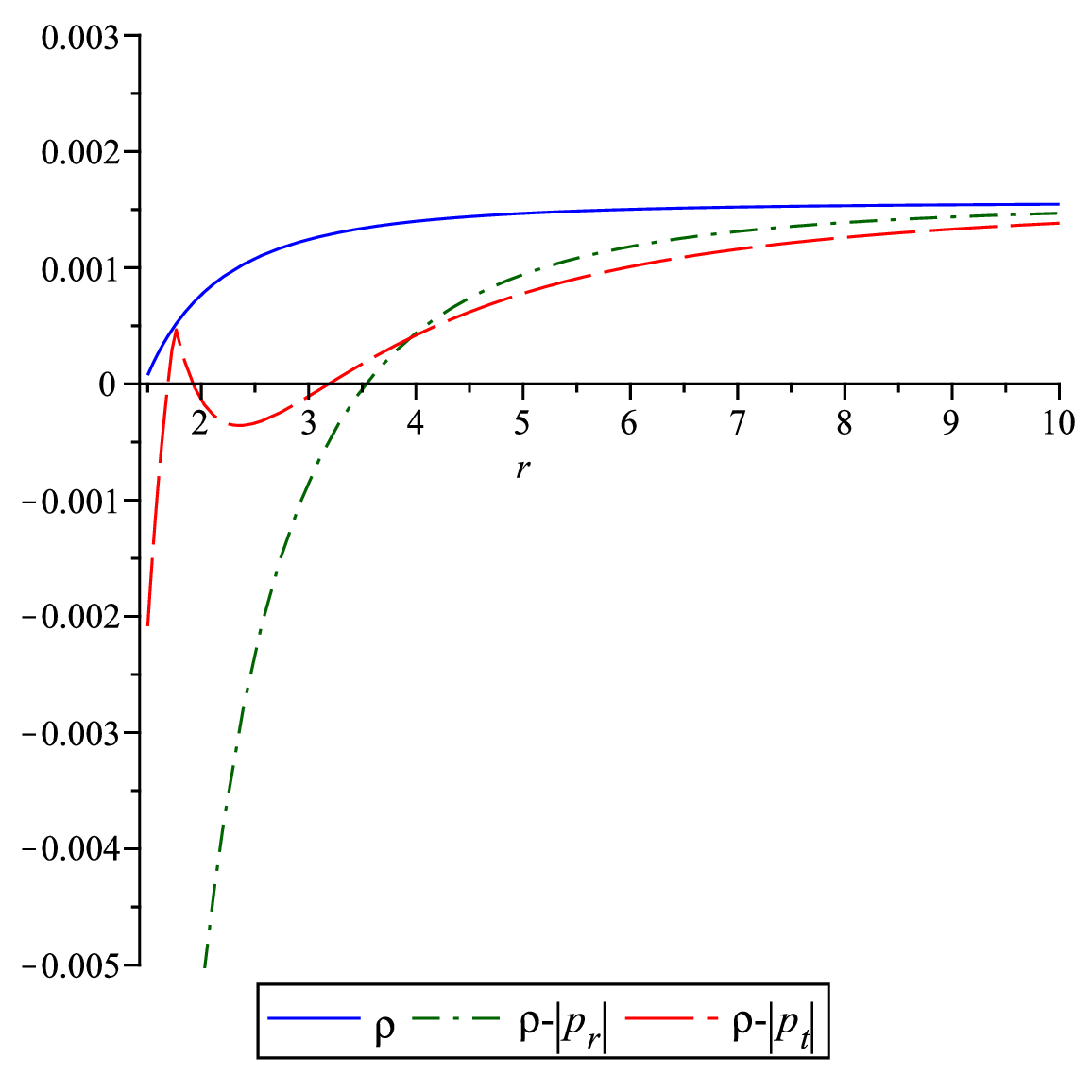}
		\centering (b)
	\end{minipage}
	\caption{ Behavior of $\rho+p_r,~ \rho+p_t,~ \rho+p_r+2p_t$ (a) and $\rho,~\rho-|p_r|,~\rho-|p_t|$ diagrams (b) have been plotted for the redshift function $\phi=\ln\left(\sqrt{1+\frac{\gamma^2}{r^2}}\right)$ and shape function  $b(r)=r_0\frac{a^r}{a^r_0}$ with the numerical values $\gamma=0.8$, $r_0=1.5$, $a=0.9$ and $\lambda=5$}\label{fig15}
\end{figure}

In this section, we continue our discussions with the validation of energy conditions and make some regional plots to check the validity of all energy conditions. In this work, we consider null energy condition (NEC), weak energy condition (WEC), strong energy condition (SEC) and dominant energy condition (DEC) to examine the wormholes. Mathematically, the above energy conditions can be written as NEC: $T_{\alpha\beta}K^\alpha K^\beta\geq0$, WEC: $T_{\alpha\beta}V^\alpha V^\beta\geq0$, SEC: $\left(T_{\alpha\beta}-\frac{1}{2}Tg_{\alpha\beta}\right)V^\alpha V^\beta\geq0$, DEC: $-T_\beta^\alpha V^\beta$ is future directed. Here $V^\alpha$ is a unit time-like vector while $K^\alpha$ is a null vector. So for anisotropic fluid the above energy conditions given as\cite{r4}:$(i)$ NEC: $\rho+p_r\geq0, ~\rho+p_t\geq0$, $(ii)$ WEC: $\rho\geq0, ~\rho+p_r\geq0, ~\rho+p_t\geq0$, $(iii)$ SEC: $\rho+p_r\geq0,~ \rho+p_t\geq0,~ \rho+p_r+2p_t\geq0$, $(iv)$ DEC: $\rho\geq0,~\rho-|p_r|\geq0,~ \rho-|p_t|\geq0$.
We have examined the energy conditions for all wormhole solutions in figures (\ref{fig6})-(\ref{fig15}). 


\section{Embedding diagrams}
\label{sec6}
One may use embedding diagrams to visualize a wormhole and extract some useful information for the choice of the shape function $b(r)$. In order to produce embeddings of two dimensional space slices (or hypersurface) of the wormhole in $\scriptsize{R}^3$, we make the restriction $\theta=\pi/2$. Here , we consider a fixed moment of time, $t=$constant and respectively the wormhole metric reduces 
\begin{equation}\label{eq24}
ds^2=\left(1-\frac{b(r)}{r}\right)^{-1}dr^2+r^2d\phi^2
\end{equation}
In the embedding space we introduce cylindrical coordinates $z$, $r$ and $\phi$. Then Euclidean metric of the embedding space has the form \cite{r2},
\begin{equation}
ds^2=dz^2+dr^2+r^2d\phi^2.
\end{equation}
The embedded surface will be axially symmetric , and hence can be described by the single function $z=z(r)$. On that surface the line element can be written as
\begin{equation}\label{eq26}
 ds^2=\Bigg[1+\left(\frac{dz}{dr}\right)^2\Bigg]dr^2+r^2d\phi^2.
\end{equation}
Now, comparing equation(\ref{eq24}) and equation (\ref{eq26}) , we acquire the expression for embedding function as 
\begin{equation}
z(r)=\pm\int_{r_0}^{r}\left(\frac{r}{b(r)}-1\right)^{-\frac{1}{2}}dr
\end{equation}
\begin{figure}[htb]
	\centering
	\begin{minipage}{.45\textwidth}
		\centering
		\includegraphics[width=.7\linewidth]{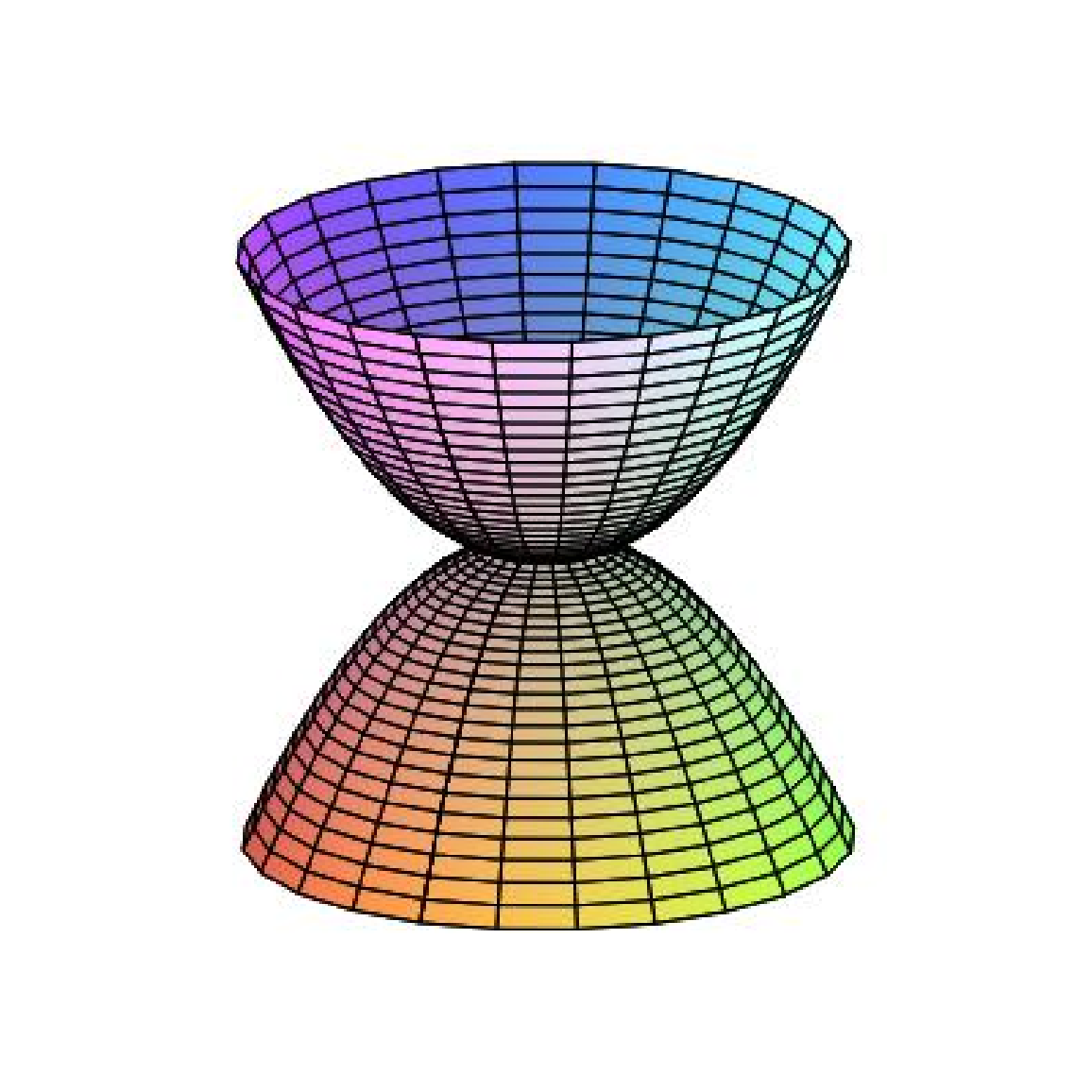}
		\centering (a) 
	\end{minipage}
\begin{minipage}{.45\textwidth}
	\centering
	\includegraphics[width=.7\linewidth]{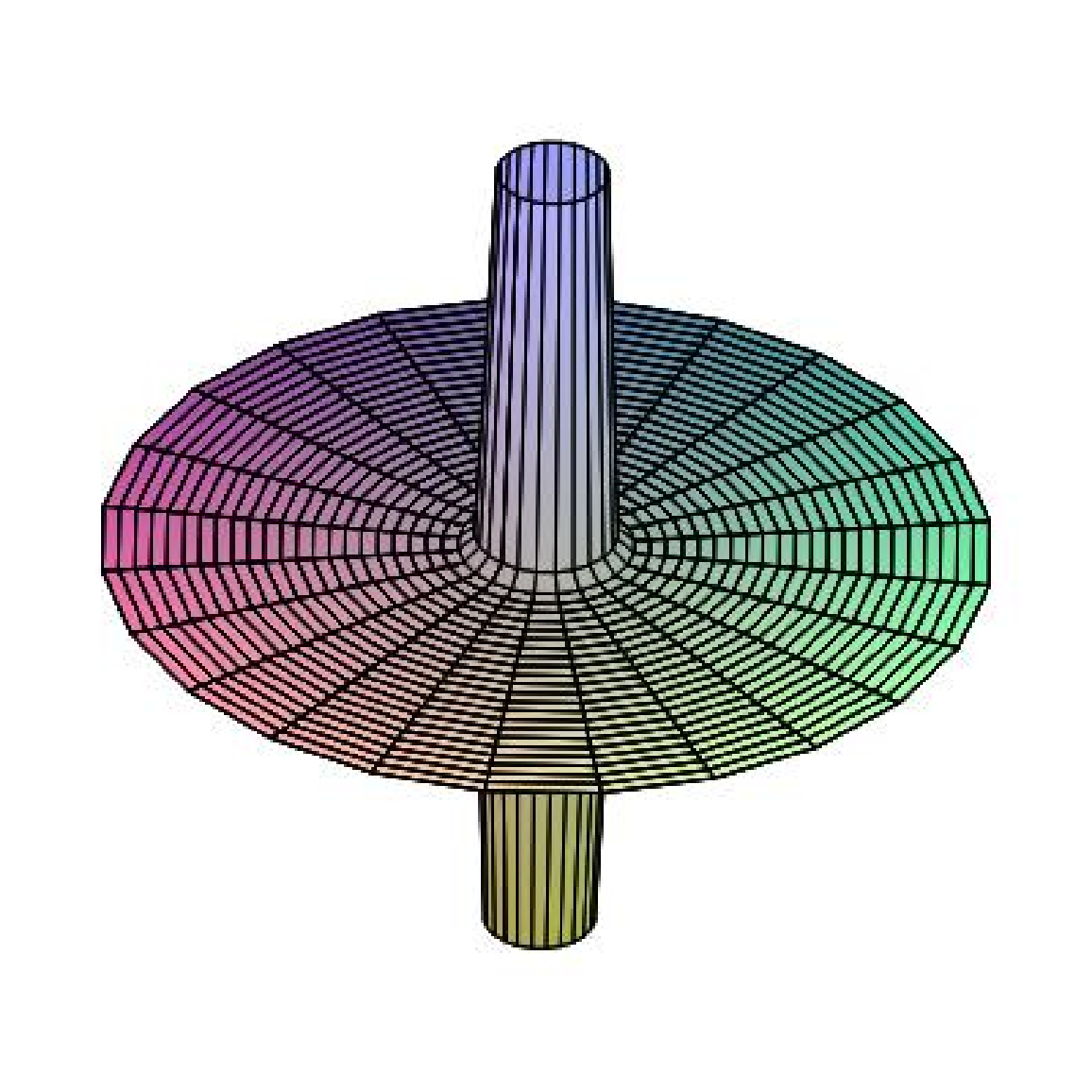}
	\centering (b)
\end{minipage}
	\caption{Embedding diagram (a) for the obtained new shape function (\ref{obs}) with $r_0=1.5$, (b) for the obtained new shape function (\ref{obs2}) when $r_0=1.5$ }
	\label{fig16}
\end{figure}
\begin{figure}[htb]
	\centering
	\begin{minipage}{.45\textwidth}
		\centering
		\includegraphics[width=.7\linewidth]{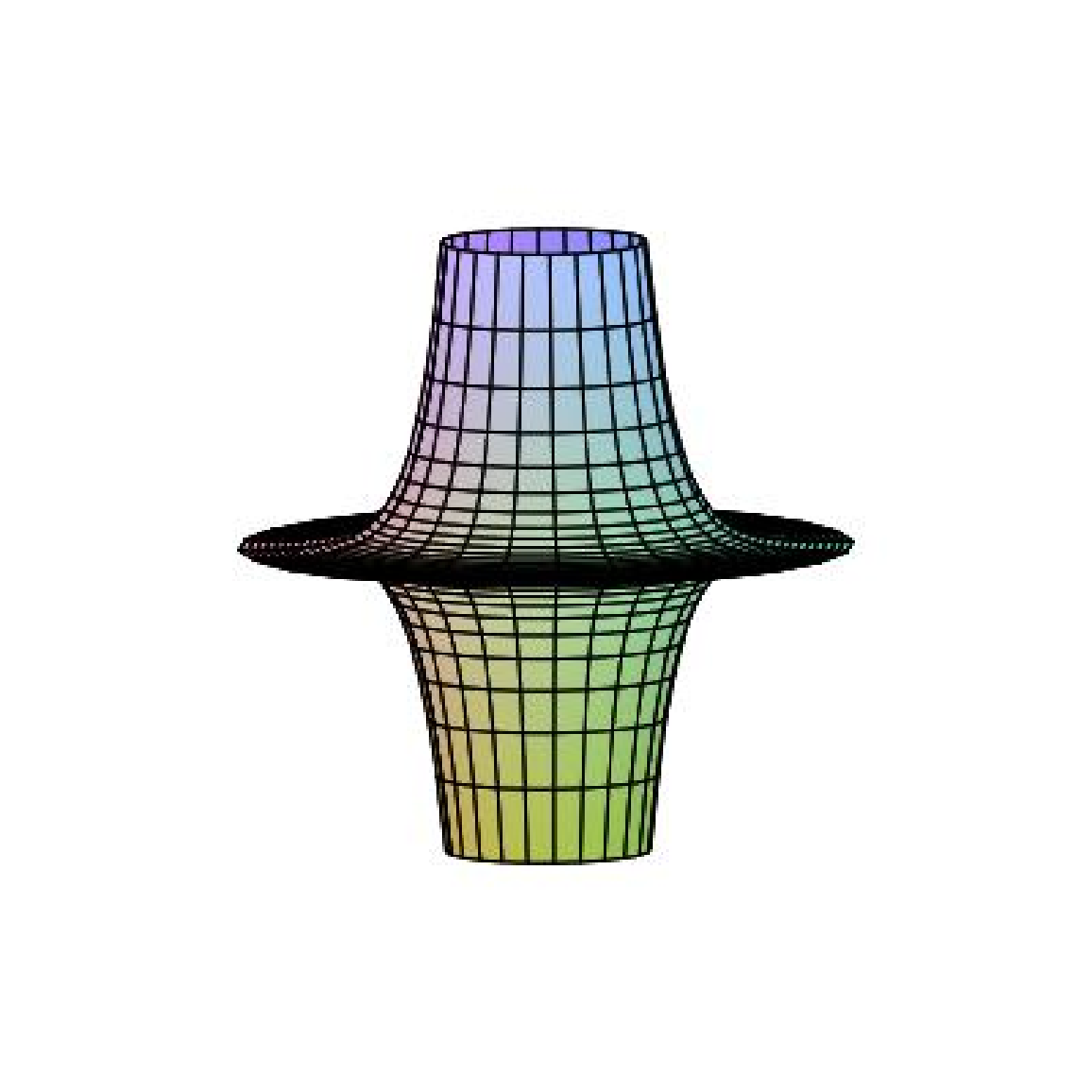}
		\centering (a) 
	\end{minipage}
	\begin{minipage}{.45\textwidth}
		\centering
		\includegraphics[width=.7\linewidth]{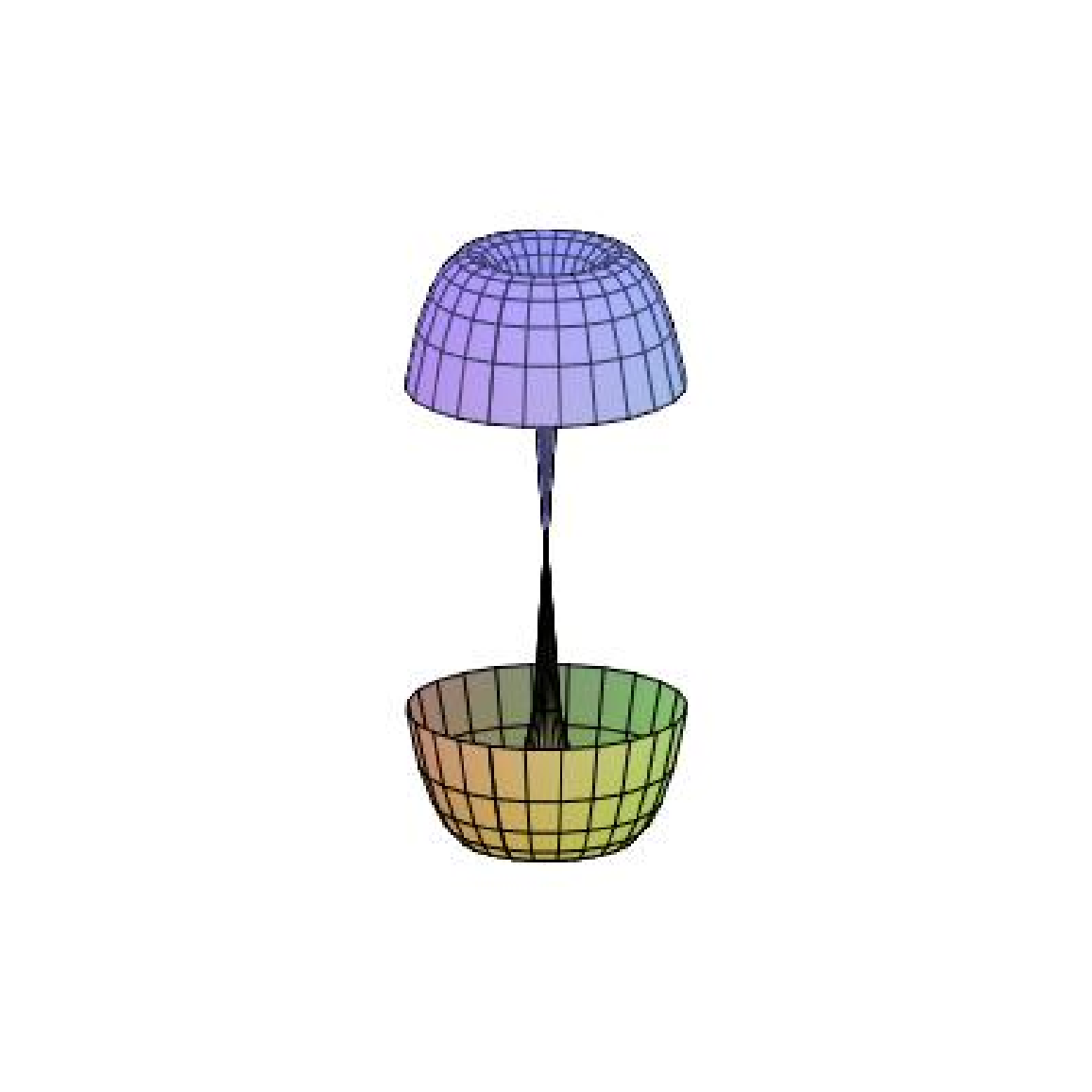}
		\centering (b)
	\end{minipage}
	\begin{minipage}{.45\textwidth}
		\centering
		\includegraphics[width=.7\linewidth]{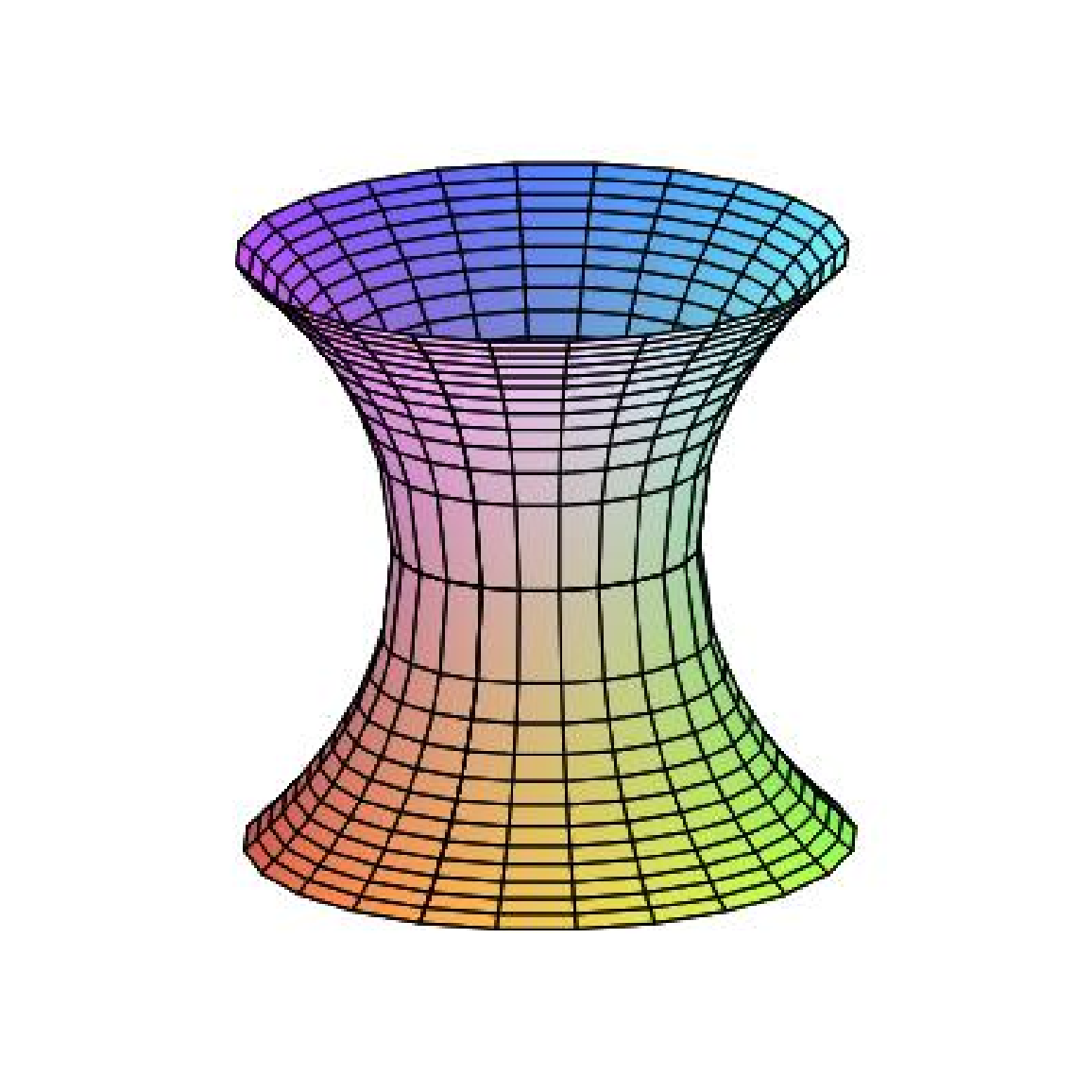}
		\centering (c)
	\end{minipage}
	\caption{Embedding diagram (a) for the shape function 1 with $r_0=0.5$, (b) for the shape function 2 with $r_0=0.5$ and (c) for the shape function 3 when $r_0=0.5$, $a=0.8$.}
	\label{fig17}
\end{figure}
The graphical representation of the embedding function is shown in figs (\ref{fig16})-(\ref{fig17}) for the discussed wormhole geometries.
\section{Results and Discussions}
\label{sec7}
%
\begin{table}[htb]
\centering
\caption{Range of radial coordinate `$r$' where the energy conditions are satisfied:}\label{T2}
\begin{tabular}{|c|c|c|c|c|c|c|}
	\hline	$\phi(r)$&$b(r)$& NEC
	
	 &WEC&SEC&DEC\\
	\hline
\multirow{3}{*}{$\phi(r)=j\ln\left(\frac{r}{r_0}\right)$} & {$r_0e^{1-\frac{r}{r_0}}$} & %
{(1.5, 10)}&(3.25, 10)&{(1.5, 10)} &$\times$\\
\cline{2-6}
& {$\frac{r\ln(r+1)}{\ln(r_0+1)}$} &{($r_0$, 1.75)}&{($r_0$, 1.75)}&{($r_0$, 1)}&{$r_0$, 1.75}\\
\cline{2-6}
& {$\frac{r_0a^r}{a^{r_0}}$}&{(2, 10)}&{(2, 10)}&{(2, 10)}&{$\times$} \\
	\hline
\multirow{3}{*}{$\phi(r)=e^{-\frac{r_0}{r}}$} & {$r_0e^{1-\frac{r}{r_0}}$} & %
{(1.25, 10)}&(2, 10)&{(1.25, 10)} &(4, 10)\\
\cline{2-6}
& {$\frac{r\ln(r+1)}{\ln(r_0+1)}$} &{($r_0$, 10)}&{($r_0$, 10)}&{($r_0$, 10)}&{$r_0$, 10}\\
\cline{2-6}
& {$\frac{r_0a^r}{a^{r_0}}$}&{(3.5, 10)}&{(3.5, 10)}&{(3.5, 10)}&{(6, 10)} \\	
	\hline
\multirow{3}{*}{$\phi(r)=ln\sqrt{1+\frac{\gamma^2}{r^2}}$} & {$r_0e^{1-\frac{r}{r_0}}$} & %
{(2, 10)}&(2, 10)&{(2, 10)} &(1.5, 10)\\
\cline{2-6}
& {$\frac{r\ln(r+1)}{\ln(r_0+1)}$} &{($r_0$, 10)}&{($r_0$, 10)}&{$\times$}&{$r_0$, 10}\\
\cline{2-6}
& {$\frac{r_0a^r}{a^{r_0}}$}&{(3.75, 10)}&{(3.75, 10)}&{(3.75, 10)}&{(3.75, 10)} \\	

	\hline
\end{tabular}
\end{table}
In 4-$D$ spacetime, a new shape function of the wormhole was obtained in the present article. We made this by choosing two generating functions and examined the energy conditions for that. In this case, all the energy conditions can be satisfied in at least a small region near the wormhole throat. From figure (\ref{fig6}), it is clear that all the energy conditions are satisfied in a region $r\in(r_0,2)$ for the obtained shape function (\ref{obs}). Also for the asymptotically flat wormhole, from figure (\ref{fig66}) it is clear that all the energy conditions are satisfied in a region $r\in(5,10)$ for the obtained shape function (\ref{obs2}).
\par 
It is shown from table (\ref{T2}), for most of the cases the wormhole satisfy all energy conditions in a region of `$r$' (see figures (\ref{fig8}), (\ref{fig10})-(\ref{fig13}), (\ref{fig15})). In some cases, $\rho$ is negative ( See figures (\ref{fig7}),(\ref{fig9}),(\ref{fig10}),(\ref{fig13})) in the neighbourhood of $r_0$ so for the existence of the traversable wormhole they need exotic matter. From figures (\ref{fig7}) and (\ref{fig8}) we can conclude two types of wormhole solutions satisfy all the energy conditions except DEC for some particular choices of parameter, one type did not satisfy SEC (See figure (\ref{fig14})). So from the above discussion we can conclude that some of the presented solutions violate the energy conditions and most
of them are not asymptotically flat (from figures (\ref{figss}) and (\ref{figphis})).
\par 
Figures (\ref{figgen})--(\ref{fig5}) show the behaviors of all generating functions $G(r)$ and $H(r)$. In \cite{21}, authors observed that the generating function $G(r)$ related to the redshift function is always positive and decreasing of `$r$' and the second generating function $H(r)$ is always negative and increasing in nature in Einstein gravity. Though, this observation regarding generating functions
 does not hold in $f(R,T)$ gravity (from figures \ref{fig3}(b), \ref{fig3}(c), \ref{fig3}(d), \ref{fig4}(d), \ref{fig5}(a) and \ref{fig5}(c)). If we choose $\lambda=0$ in the form of $f(R, T)$ then we will get the same $H(r)$ which we have got in Einstein gravity. Hence, comparing the work of Rahaman et al.\cite{21}, we can conclude that generating functions depend upon different gravity theories.
\par 
In this present work, to obtain a shape function using generating functions the following algorithm can be considered: At first, two generating functions $G(r)$ and $H(r)$ have to be considered. Secondly, from equation (\ref{G}) the redshift function is obtained using $G(r)$ (provided the integral part is integrable). Shape function $b(r)$ is found by using equation (\ref{b})(provided the integration exists) and the throat condition $b(r_0)=r_0$ (to obtain the integration constant). Finally, the equations (\ref{eqc2})-(\ref{eqc4}) have to be verified by the obtained $b(r)$. If all the  conditions are satisfied then $b(r)$ will be termed as a shape function.  

\end{document}